\documentclass[aps,prd,superscriptaddress,amsmath,amssymb,twocolumn,fleqn,epsf,epsfig,showpacs,nofootinbib]{revtex4-2} 
\usepackage{dcolumn}
\usepackage{bm}
\usepackage{amsfonts}
\usepackage{color}
\usepackage{graphicx}
\usepackage{colordvi}
\usepackage{mathtools}
\usepackage{empheq}
\usepackage[utf8]{inputenc}
\usepackage{hyperref}

\def\be{\begin{equation}}
\def\ee{\end{equation}}
\def\ba{\begin{eqnarray}}
\def\ea{\end{eqnarray}}

\begin{document}

\title{Redshift evolution of cosmic birefringence in CMB anisotropies}

\author{Matteo Galaverni}\email{matteo.galaverni@gmail.com} 
\affiliation{Specola Vaticana (Vatican Observatory), V-00120, Vatican City State}
\affiliation{INAF/OAS Bologna, Via Gobetti 101, I-40129 Bologna, Italy}
\author{Fabio Finelli}\email{fabio.finelli@inaf.it} 
\affiliation{INAF/OAS Bologna, via Gobetti 101, I-40129 Bologna, Italy}
\affiliation{INFN, Sezione di Bologna, Via Irnerio 46, I-40127 Bologna, Italy} 
\author{Daniela Paoletti}\email{daniela.paoletti@inaf.it}
\affiliation{INAF/OAS Bologna, Via Gobetti 101, I-40129 Bologna, Italy}
\affiliation{INFN, Sezione di Bologna, Via Irnerio 46, I-40127 Bologna, Italy}

\date{\today}

\begin{abstract}
We study the imprints of a cosmological redshift-dependent pseudoscalar field $\phi$ on
the rotation of cosmic microwave background
(CMB) linear polarization generated by a coupling $ \phi F^{\mu\nu} \tilde F_{\mu \nu}$.
We show how either phenomenological
or theoretically motivated redshift dependence of the pseudoscalar field, such as those in models of Early Dark Energy,
Quintessence or axion-like dark matter, lead to
CMB polarization and temperature-polarization power spectra which exhibit a multipole dependence which goes beyond the
widely adopted approximation in which the redshift dependence of the linear polarization angle is neglected.
Because of this multipole dependence, the isotropic birefringence effect due to a general coupling 
$\phi F^{\mu\nu} \tilde F_{\mu \nu}$ is not degenerate with 
a systematic calibration angle uncertainty.
By taking this multipole dependence into account, we calculate the parameters of these phenomenological 
and theoretical redshift dependence of the pseudoscalar
field which can be detected by future CMB polarization experiments 
on the basis of a $\chi^2$ analysis for a Wishart likelihood.
As a final example of our approach, we compute by Markov Chain MonteCarlo (MCMC) the minimal coupling $g_\phi$ in Early Dark Energy which could be detected by future experiments,
with or without marginalizing on a systematic rotation angle uncertainty.
\end{abstract}



\maketitle

\section{Introduction}

When the electromagnetic tensor $F_{\mu\nu}$ is coupled to
a pseudoscalar field $\phi(x)$ a new term appears in the Lagrangian density:
\begin{equation}
\mathcal{L}\supset  -\frac{g_\phi}{4}\phi F_{\mu\nu}\,\widetilde F^{\mu\nu}\,,
\end{equation}
where  
$g_\phi$ is a model dependent coupling constant
and $\widetilde F^{\mu\nu}\equiv\frac{1}{2}\epsilon^{\mu\nu\rho\sigma}F_{\rho\sigma}$ is the dual of the electromagnetic tensor.
The plane of linear polarization of a single photon
propagating in this evolving cosmological pseudoscalar field background undergoes a rotation given by \cite{Carroll:1989vb,Harari:1992ea}:
\begin{equation}
\frac{g_\phi}{2}\left[\phi(x)-\phi(x_\mathrm{em})\right]\,,
\end{equation}
where $\phi(x_\mathrm{em})$ is the value of the pseudoscalar field when light is emitted.
This effect is called {\em cosmological birefringence}.

First upper limits on the coupling constant $g_\phi$ were based on optical
imaging polarimetry of radio galaxies \cite{Carroll:1997tc,Carroll:1991zs,Cimatti:1993yc,Harari:1992ea,Carroll:1998zi,Alighieri:2010eu}.
Soon after it was realized that also CMB polarization could be used to study this interaction
which induces a rotation of the plane of linear polarization
\cite{Lue:1998mq,Feng:2004mq,Feng:2006dp} to leading order in $g_\phi$ as well as circular polarization
to the next-to-leading order \cite{Finelli:2008jv,Alexander:2008fp}.

Either the redshift dependence 
\cite{Finelli:2008jv,Liu:2006uh,Gubitosi:2014cua,Chigusa:2019rra,Sherwin:2021vgb,
Nakatsuka:2022epj,Greco:2022xwj} 
and the inhomogeneities \cite{Greco:2022xwj,Li:2008tma,Pospelov:2008gg,Kamionkowski:2008fp,Cai:2021zbb,Greco:2022ufo}
of the cosmological pseudo-scalar field contribute
to cosmological birefringence. In this paper we study the imprints of the
isotropic
redshift dependence
of $\phi$ along the line-of-sight from last scattering surface to the observer
into CMB parity even and odd power spectra. We consider either phenomenological
or theoretically motivated redshift dependence of the pseudoscalar field, such as those in models of Early Dark Energy (EDE),
Quintessence (DE) or axion-like dark matter (DM) \cite{Liu:2006uh,Gubitosi:2009eu,Gubitosi:2014cua,Liu:2016dcg,Sigl:2018fba,Capparelli:2019rtn,Fujita:2020aqt,Fujita:2020ecn,Fedderke:2019ajk,Murai:2022zur}.

As already pointed out \cite{Liu:2006uh,Finelli:2008jv,Gubitosi:2014cua,Sherwin:2021vgb,Nakatsuka:2022epj,Greco:2022xwj}, this redshift dependence induces a multipole dependence
of the cosmological birefringence effect in the CMB power spectra
which goes beyond the widely used approximation for which the rotation angle is assumed constant in redshift
\cite{Lue:1998mq}.
Although this approximation is a key working assumption for deriving constraints
from CMB polarization data \cite{Komatsu:2011,Gruppuso:2011ci,Finelli:2012wu,Gruppuso:2015xza,Planck:2016soo,POLARBEAR:2017beh,Wu:2019hek,
Gruppuso:2020kfy,BICEPKeck:2020hhe,BICEPKeck:2021sbt,Namikawa:2020ffr,Bortolami:2022whx}
and forecast the capabilities of future experiments
\cite{CMB-S4:2016ple,Molinari:2016xsy,NASAPICO:2019thw,Pogosian:2019jbt}, we believe it is timely to fully exploit
the theoretical predictions of isotropic cosmological birefringence for two main reasons.

Firstly, 
by assuming the isotropic birefringence
angle as independent on the multipoles an exact degeneracy between the  
cosmological birefringence effect and the uncertainty in the calibration angle which would be otherwise absent opens up.
We explicitly show how taking into account the redshift dependence of cosmological birefringence
mitigate this degeneracy (see also \cite{Sherwin:2021vgb,Nakatsuka:2022epj}).

As a second point, we stress that the advance in data analysis and in the increasingly precision of CMB polarization data
shrank error bars approximately by a factor 3 from the \textit{Planck} analysis 
on data release 2 \cite{Planck:2016soo}: hints of isotropic cosmic birefringence within the constant angle approximation
were claimed with  \textit{Planck} data release 3 (DR3) \cite{Minami:2020odp},  \textit{Planck} data release 4 (DR4) \cite{Diego-Palazuelos:2022dsq}, and more recently with
\textit{WMAP 9-year} and \textit{Planck} data-processing pipeline called NPIPE \cite{Planck:2020olo}
($\alpha=0.342^{+0.094}_{-0.091}$~deg \cite{Eskilt:2022cff}).
For a recent review see \cite{Komatsu:2022nvu}. 
We will indeed show that the differences between a physical model and the constant angle approximation are important and within the reach of future CMB polarization experiments. 

The paper is organized as follows: in Sect. \ref{Sect:II}
we review the Boltzmann equation
in presence of an isotropic redshift-dependent birefringence.
We compare the power spectra for some phenomenological models
with the widely used approximation where the time dependence of the linear polarization
angle is neglected.
The study of theoretically motivated redshift dependence of the pseudoscalar field is presented in Sect. \ref{Sect:III}:
Early Dark Energy, Quintessence and axion-like dark matter.
In Sect. \ref{Sect:IV}
we present the  forecasts for CMB experiments, focusing in particular on \textit{LiteBIRD}
on the basis of a $\chi^2$ analysis for a Wishart likelihood 
and we perform few exploratory runs exploring the whole cosmological and birefringence parameter space using the Markov Chain MonteCarlo code \texttt{cosmomc}.
We conclude in Sect. \ref{Sect:V}.

In this work, we use natural units, $\hbar=c= 1$, and assume 
flat $\Lambda CDM$ cosmological model with \textit{Planck} 2018 estimates of cosmological parameters \cite{Planck:2018vyg}:
$\Omega_b \, h^2 = 0.02237$,
$\Omega_c \, h^2 = 0.120$, 
$\tau=0.0544$, 
$n_s=0.9649$, $\ln \left(10^{10} A_s\right)=3.044$, 
$H_0 = 100 \, h \, {\rm km \, s}^{-1} \, \mathrm{Mpc}^{-1}=67.36\, \mathrm{km \, s}^{-1} \, \mathrm{Mpc}^{-1}$.

\section{Effects of redshift evolution of the birefringence field}
\label{Sect:II}

The linear polarization rotation for a of 
a CMB photon is described by:
\begin{eqnarray}
Q&=& Q_\mathrm{rec} \cos \left(2\beta\right) + U_\mathrm{rec} \sin \left(2\beta\right)\,,\\
U&=& U_\mathrm{rec} \cos \left(2\beta\right) - Q_\mathrm{rec} \sin \left(2\beta\right)\,,
\end{eqnarray}
where $Q_\mathrm{rec}$ and $U_\mathrm{rec}$ are the Stokes parameters at recombination, when CMB photons are last scattered.\footnote{
We follow CMB-HEALPix coordinate conventions:  the linear
polarization angle increases clockwise looking toward the source \cite{Galaverni:2014gca,Galaverni:2018zcm,Sperello:2017}.}

In the case of isotropic time-dependent birefringence angle 
induced by a cosmological pseudoscalar field 
the Boltzmann equation for linear polarization contains an additional term proportional to $g_\phi \phi^{\prime}(\eta)$, 
where $\phi^{\prime}$ is the derivative of $\phi$ with respect conformal time $\eta$ \cite{Kosowsky:1996yc,Liu:2006uh,Finelli:2008jv,Gubitosi:2014cua}:
\begin{eqnarray}
\label{Boltz_rot}
&& \Delta^\prime_{Q\pm i U} (k,\eta)+i k \mu \Delta_{Q\pm i U} (k,\eta)
\nonumber \\
&&= -n_e \sigma_T a(\eta)\left[\Delta_{Q\pm i U} (k,\eta)\right.
\nonumber \\& & 
\left. +\sum_m\sqrt{\frac{6 \pi}{5}} {}_{\pm2}Y_2^m S_P^{(m)}(k,\eta)\right] \nonumber\\
& & \mp i g_\phi \phi^{\prime}(\eta) \Delta_{Q\pm iU}(k,\eta)\,.
\end{eqnarray}
The the cosine of the angle between the CMB photon direction 
and the Fourier wave vector is indicated by $\mu$, $n_e$ is the number density of free electrons, 
$\sigma_T$ is the Thomson cross section,
${}_{s}Y_2^m$ are spherical harmonics with spin-weight $s$, 
and $S_P^{(m)}(k,\eta)$ is the source term generating linear polarization.

In order to integrate along the line-of-sight we note  that \cite{Gubitosi:2014cua}:
\begin{eqnarray}
&&\Delta^\prime_{Q\pm i U} (k,\eta)+\left[i k \mu +\tau^\prime(\eta) \pm i g_\phi \phi^{\prime}(\eta)  \right]\Delta_{Q\pm i U} (k,\eta)\nonumber\\
&&=e^{-i k \mu \eta} e^{\tau(\eta)} e^{ \mp i 2\alpha(\eta)}\nonumber\\
&&\quad \frac{d}{d\eta}\left[e^{i k \mu \eta} e^{-\tau(\eta)} e^{ \pm i 2\alpha(\eta)}\Delta_{Q\pm i U} (k,\eta)\right]\,,
\end{eqnarray}
where we introduced the differential optical depth $\tau^\prime(\eta)\equiv n_e \sigma_T a(\eta)$
and we formally integrated $\phi^{\prime}(\eta)$
defining $\alpha(\eta)=\frac{g_\phi}{2}\phi(\eta)$, defined up to a constant.
Therefore, the Boltzmann Eq.~\eqref{Boltz_rot} can be re-written as:
\begin{eqnarray}
\label{Boltz_rot_2}
&& e^{-i k \mu \eta} e^{\tau(\eta)} e^{ \mp i 2\alpha(\eta)}\nonumber\\
&& \quad \frac{d}{d\eta}\left[e^{i k \mu \eta} e^{-\tau(\eta)} e^{ \pm i 2\alpha(\eta)}\Delta_{Q\pm i U} (k,\eta)\right]\nonumber\\
&&= - \tau^\prime(\eta) \sum_m\sqrt{\frac{6 \pi}{5}} {}_{\pm2}Y_2^m S_P^{(m)}(k,\eta)\,.
\end{eqnarray}
Following the integration along the line-of-sight methodology
\cite{Seljak:1996is}, 
we obtain these expressions for the polarization $C_\ell$ auto- and cross-spectra:
\begin{eqnarray}
\label{ClXY}
& & C_\ell^{XY}=\left(4 \pi\right)^2\frac{9\left(\ell+2\right)!}{16\left(\ell-2\right)!}
\nonumber \\
& &\qquad
\int k^2 dk\, \left[\Delta_{X,\ell}(k,\eta_0)\Delta_{Y,\ell}(k,\eta_0)\right]\,,\\
\label{ClTX}
& & C_\ell^{TX}=\left(4 \pi\right)^2\sqrt{\frac{9\left(\ell+2\right)!}{16\left(\ell-2\right)!}}
\nonumber \\
& &\qquad 
\int k^2 dk\, \Delta_{T,\ell}(k,\eta_0) \Delta_{X,\ell}(k,\eta_0)\,,
\end{eqnarray}
where $X,Y$ can be either $E$ or $B$.
The integrals defining the polarization scalar perturbations $\Delta_{T,\ell}$, $\Delta_{E,\ell}$ and $\Delta_{B,\ell}$ are:
\begin{eqnarray}
\Delta_{T, \ell}(k,\eta_0)&=&\int_{\eta_{\mathrm{rec}}}^{\eta_0}d\eta\, g(\eta)S_T(k,\eta)j_\ell(k\eta_0-k\eta)\,,\\
\label{source:E}
\Delta_{E, \ell}(k,\eta_0)&=&\int_{\eta_{\mathrm{rec}}}^{\eta_0}d\eta\, g(\eta)S_P^{(0)}(k,\eta)
\frac{j_\ell(k\eta_0-k\eta)}{\left(k\eta_0-k\eta\right)^2} \nonumber\\ 
& &\; \cos2\left[ \alpha (\eta)-\alpha(\eta_{0}) \right]\,,\\
\label{source:B}
\Delta_{B,\ell}(k,\eta_0)&=&\int_{\eta_{\mathrm{rec}}}^{\eta_0}d\eta\, g(\eta)S_P^{(0)}(k,\eta)
\frac{j_\ell(k\eta_0-k\eta)}{\left(k\eta_0-k\eta\right)^2} \nonumber\\ 
& &\;  \sin2\left[ \alpha (\eta)-\alpha(\eta_{0}) \right]\,.
\end{eqnarray}
here $S_T(k,\eta)$ [$S_P^{(0)}(k,\eta)$] is the source term for temperature [scalar polarization] anisotropies, 
and $j_\ell$ is the spherical Bessel function of order $\ell$.

Note that $\Delta_E$ and $\Delta_B$ are sensitive to cosmic birefringence through a term proportional to $\alpha (\eta)-\alpha(\eta_{0})$,
where $\alpha(\eta)$ describes linear polarization rotation from recombination ($\eta_{\mathrm{rec}}$) to time $\eta$: 
\begin{equation}
\alpha(\eta)=\int^\eta_{\eta_{\mathrm{rec}}}\alpha^\prime(\eta_1) d\eta_1\,.    
\end{equation}

\begin{figure}[!htb]
\includegraphics[width=\columnwidth]{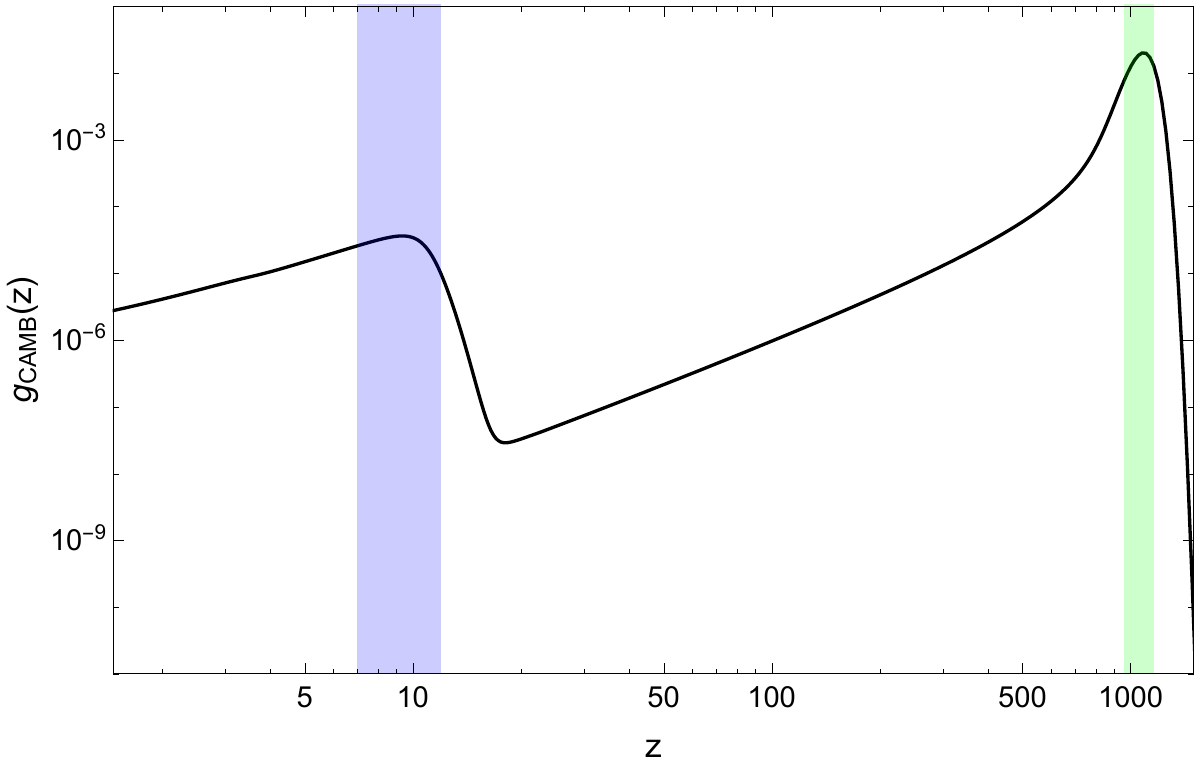}
\caption{\footnotesize\label{Fig:visibility} 
\texttt{CAMB} visibility  function ($g_\mathrm{CAMB}$) as a function of redshift $z$, blue band: $7<z<12$, green band: $z=z_\mathrm{rec}\pm 100$ ($z_\mathrm{rec}\simeq 1059$).}
\end{figure}

The visibility function $g(\eta)=\tau^\prime(\eta) e^{-\tau(\eta)}$ \cite{Seljak:1996is} is not constant 
for photons propagating from last scattering to nowadays: 
it reaches its maximum at recombination and a second peak is present at reionization epoch,
but it is several orders of magnitude smaller.  
The visibility function of the Boltzmann code \texttt{CAMB} \cite{Lewis:1999bs} 
is plotted in Fig.~\ref{Fig:visibility} as a function of redshift.

Since the visibility function is highly peaked at recombination 
a widely used approximation consists in evaluating the new term 
$\alpha (\eta)-\alpha(\eta_{0})$ appearing in Eqs.~\eqref{source:E}-\eqref{source:B} 
at recombination \cite{Gubitosi:2009eu}.
In this approximation:
\begin{equation}
\label{eq:baralpha}
\alpha (\eta)-\alpha(\eta_{0})\simeq\alpha (\eta_\mathrm{rec})-\alpha(\eta_{0})\equiv\bar{\alpha}\,,    
\end{equation}
the constant terms $\cos(2\bar{\alpha})$ and $\sin(2\bar{\alpha})$ exit integration over time in Eqs.~\eqref{source:E} - \eqref{source:B}
and the following expressions for the power spectra 
as a function of the power spectra at recombination (rec) are obtained (assuming at recombination both $C_\ell^{BB,\mathrm{rec}}=0$
and vanishing parity odd power spectra  $C_\ell^{TB,\mathrm{rec}}=C_\ell^{EB,\mathrm{rec}}=0$) 
\cite{Lue:1998mq,Feng:2004mq,Gubitosi:2014cua,Gruppuso:2016nhj,Minami:2019ruj}:
\begin{eqnarray}
\label{C_ll_TE_constant}
C_\ell^{TE,\mathrm{const}}&=&C_\ell^{TE,\mathrm{rec}} \cos(2 \bar{\alpha})
\,,\\
\label{C_ll_TB_constant}
C_\ell^{TB,\mathrm{const}}&=& C_\ell^{TE,\mathrm{rec}} \sin(2 \bar{\alpha})
\,,\\
\label{C_ll_EE_constant}
C_\ell^{EE,\mathrm{const}}&=& C_\ell^{EE,\mathrm{rec}} \cos^2(2 \bar{\alpha})
\,,\\
\label{C_ll_BB_constant}
C_\ell^{BB,\mathrm{const}}&=&
C_\ell^{EE,\mathrm{rec}} \sin^2(2 \bar{\alpha})
\,,\\
\label{C_ll_EB_constant}
C_\ell^{EB,\mathrm{const}}&=&\frac{1}{2}
C_\ell^{EE,\mathrm{rec}}
\sin(4 \bar{\alpha})
\,.
\end{eqnarray}
The main purpose of this paper is to compare 
the results of Eqs.~\eqref{ClXY}-\eqref{ClTX} 
with the constant approximation of 
Eqs.~\eqref{C_ll_TE_constant} - \eqref{C_ll_EB_constant}.

\begin{figure*}[!htb]
\includegraphics[width=168pt,height=4.5cm]{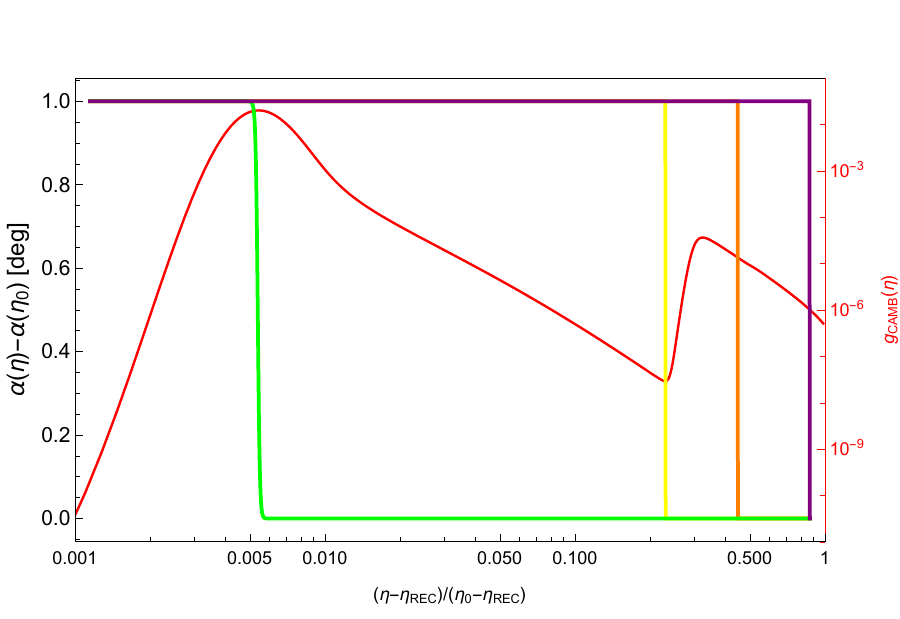}
\includegraphics[width=168pt,height=4cm]{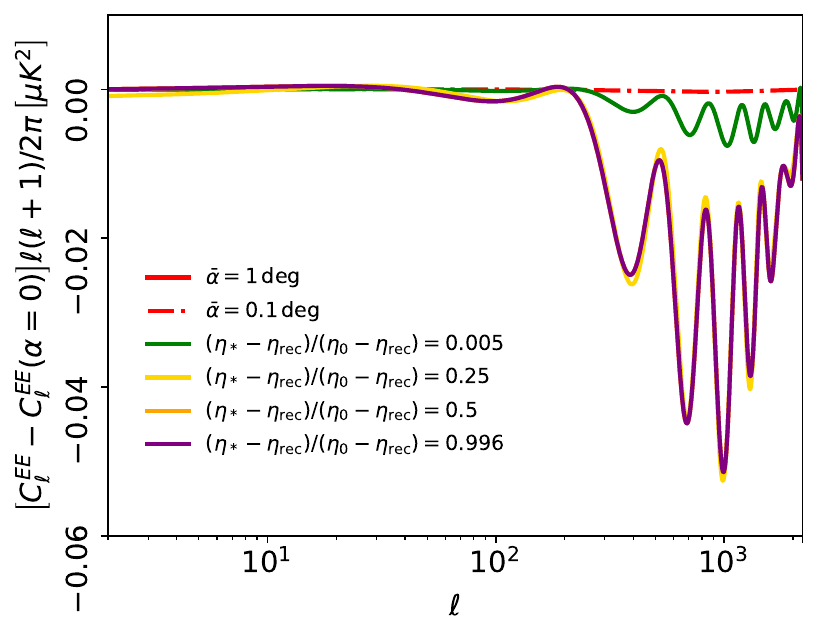}
\includegraphics[width=168pt,height=4cm]{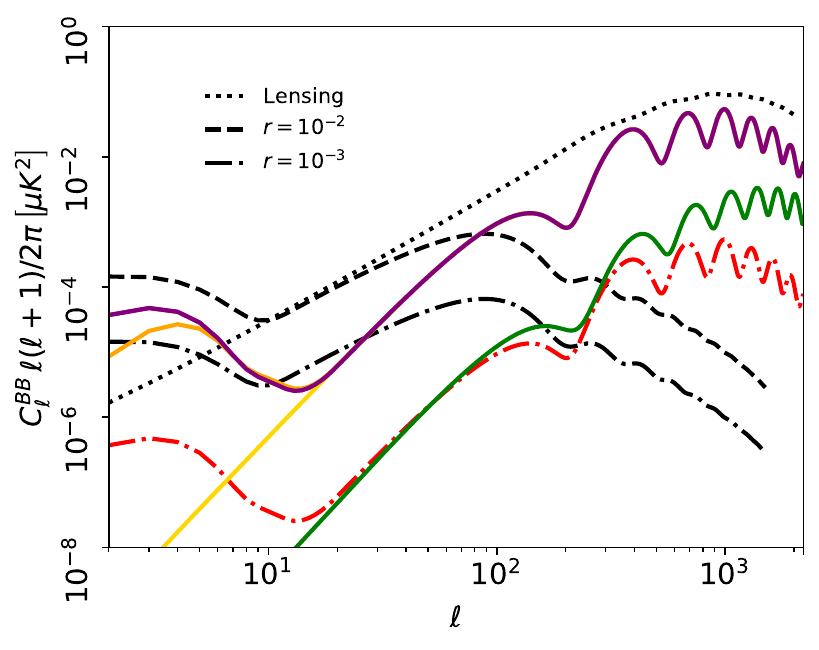}
\includegraphics[width=168pt,height=4cm]{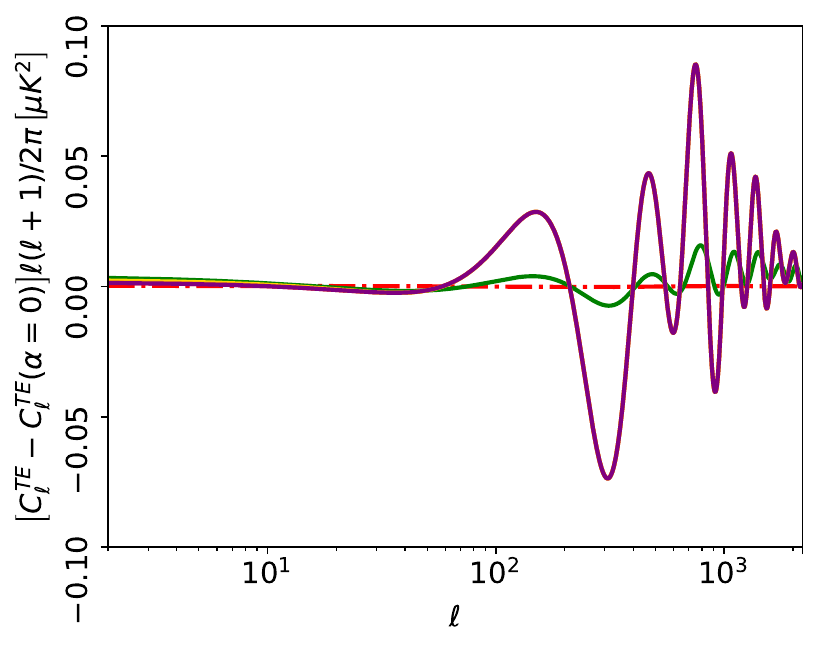}
\includegraphics[width=168pt,height=4cm]{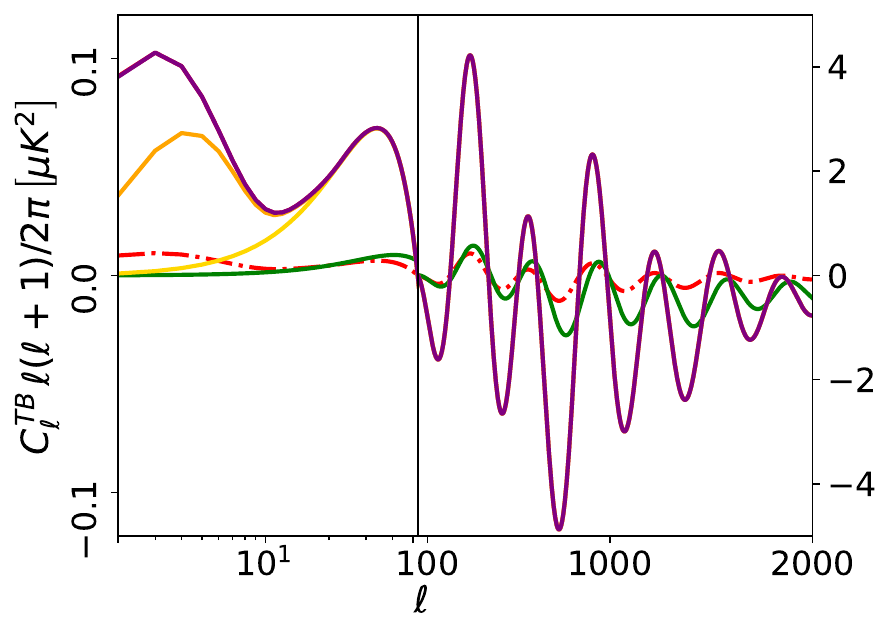}
\includegraphics[width=168pt,height=4cm]{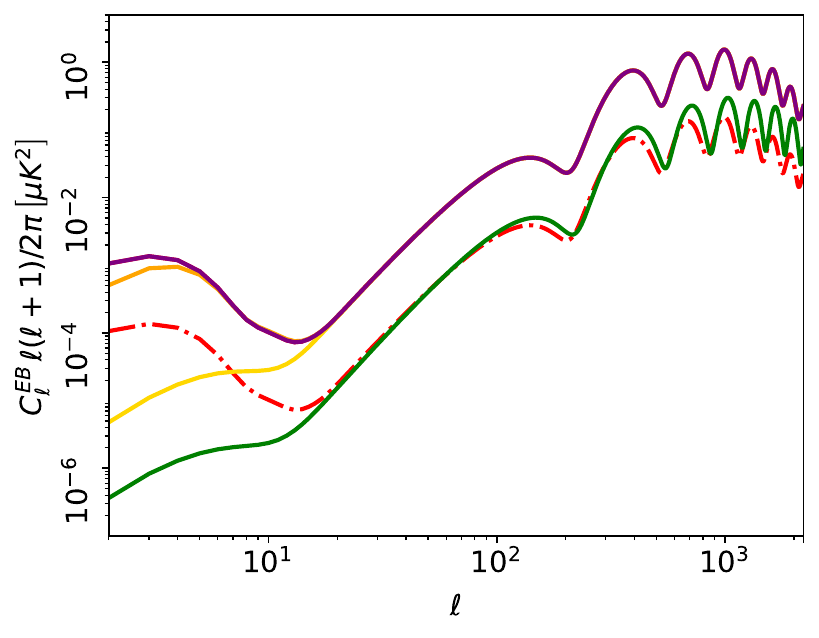}
\caption{\footnotesize\label{Fig:tanhevol:theta:120}  
(a) Evolution of $\alpha (\eta)-\alpha(\eta_{0})$ as a function of conformal time $\left(\eta-\eta_{\rm rec}\right)/\left(\eta_0-\eta_{\rm rec}\right)$; 
$\alpha$ always starts at $\alpha(\eta_{\rm rec})=0$~deg and ends at $\alpha(\eta_0)=-1$~deg,
but the ``sudden/instantaneous rotation'' happens at different times: 
purple line corresponds to a rotation near present time $x_*\equiv\left(\eta_*-\eta_{\rm rec}\right)/\left(\eta_0-\eta_{\rm rec}\right)=0.996$,
orange line line corresponds to a rotation occurring at $x_*=0.5$,
yellow line line corresponds to a rotation occurring at $x_*=0.25$,
and green line line corresponds to a rotation near last scattering surface $x_*=0.005$ 
- the \texttt{CAMB} visibility function $g_\mathrm{CAMB}$ is plotted in red (on a different scale); 
angular power spectra obtained with the modified version of \texttt{CAMB} are compared in
(b) $C_\ell^{EE} - C_\ell^{EE}(\bar{\alpha}=0)$, 
(c) $C_\ell^{BB}$,
here we plot for comparison  also the signal induced by gravitational lensing (black dotted line), 
primordial signal for $r=10^{-2}$ (black dashed line), and
primordial signal for $r=10^{-3}$ (black dot-dashed line)
(d)  $C_\ell^{TE} - C_\ell^{TE}(\bar{\alpha}=0)$, (e) $C_\ell^{TB}$, (f) $C_\ell^{EB}$.}
\end{figure*}

\begin{figure*}[!htb]
\includegraphics[width=168pt,height=4.5cm]{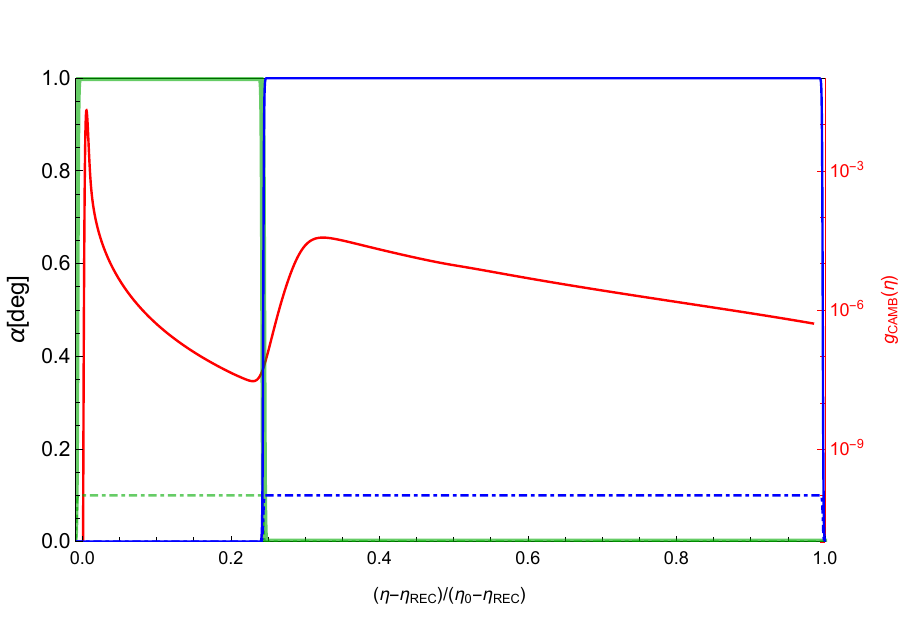}
\includegraphics[width=168pt,height=4cm]{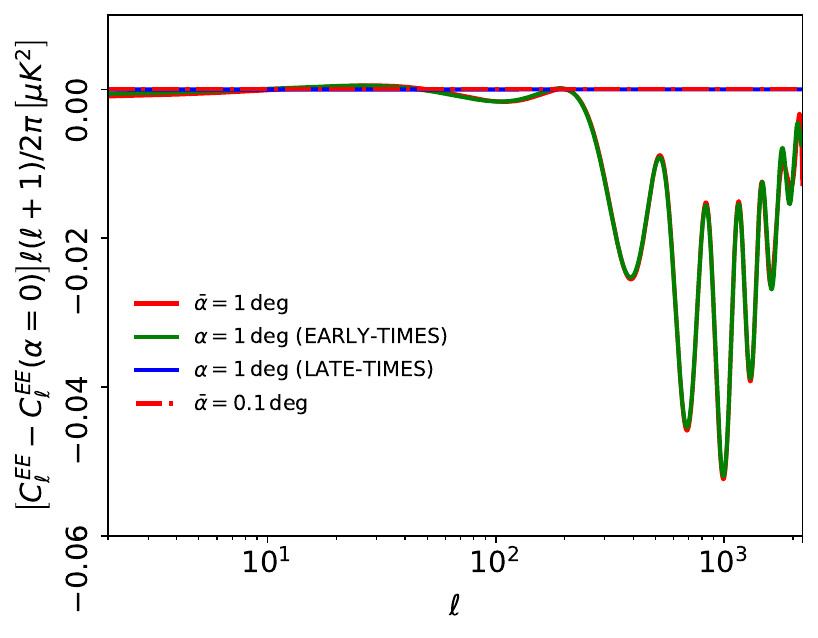}
\includegraphics[width=168pt,height=4cm]{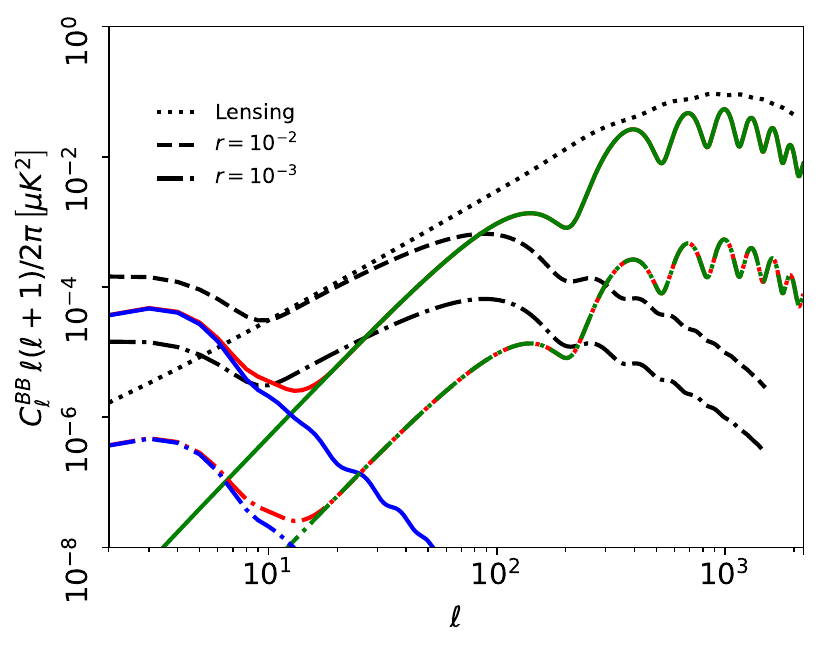}
\includegraphics[width=168pt,height=4cm]{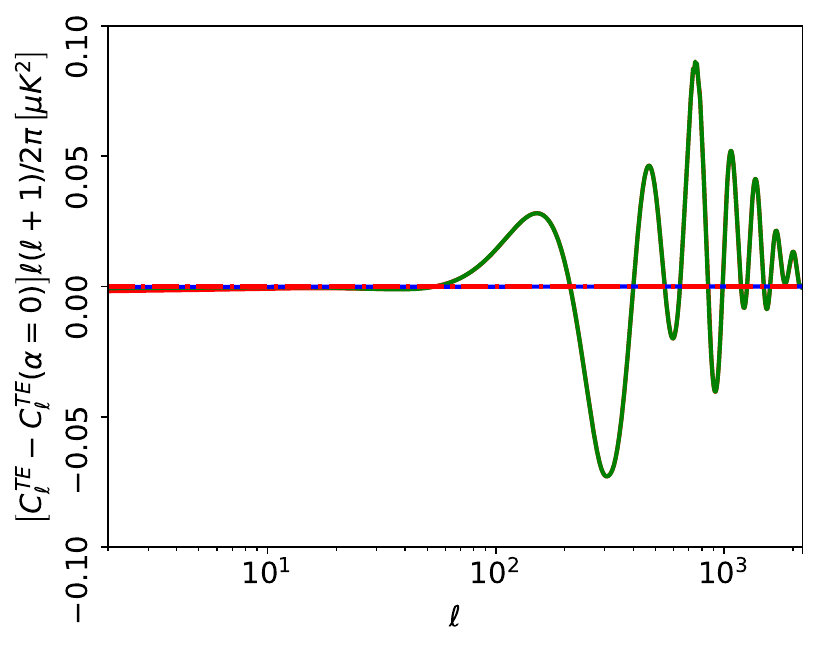}
\includegraphics[width=168pt,height=4cm]{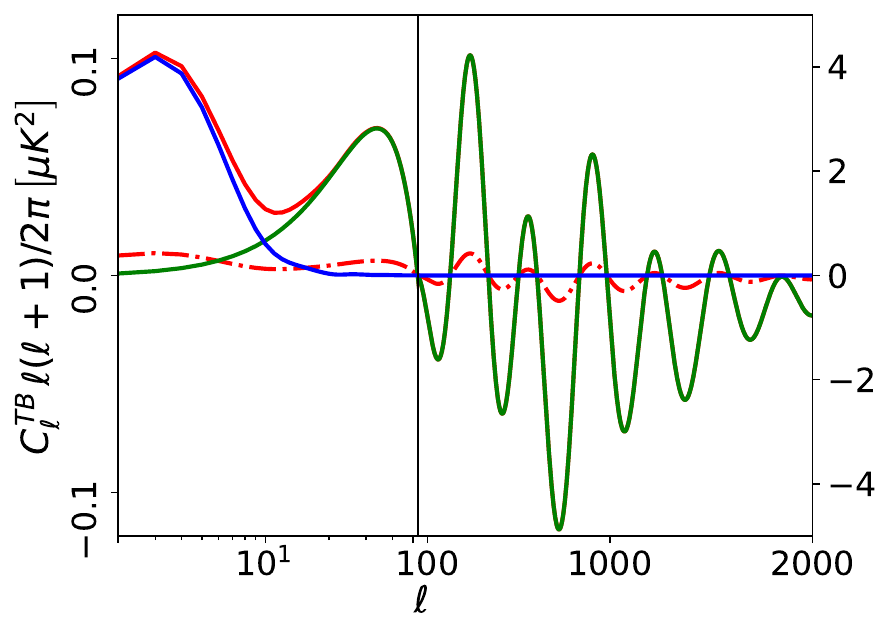}
\includegraphics[width=168pt,height=4cm]{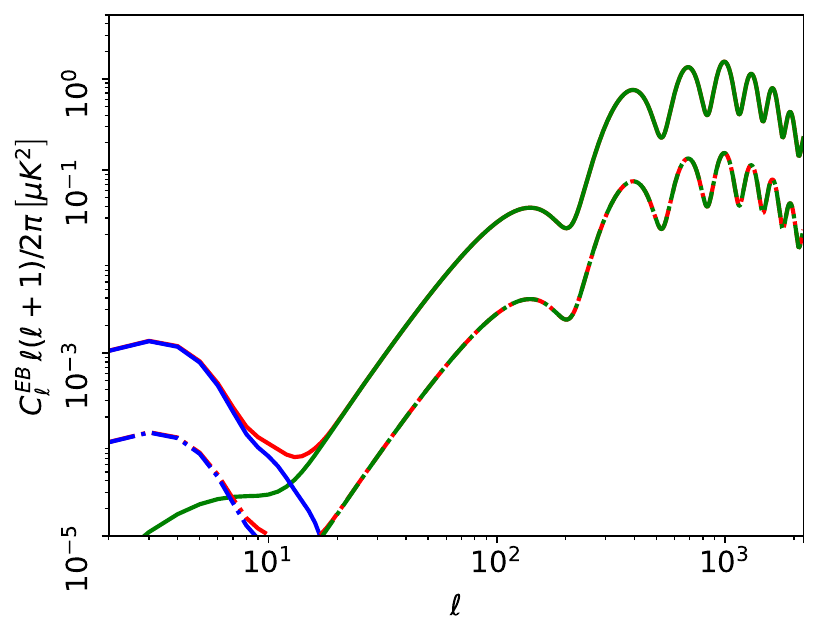}
\caption{\footnotesize\label{Fig:tanhevol}  (a) Evolution of the birefringence angle $\alpha$ as a function of conformal time $\left(\eta-\eta_{\rm rec}\right)/\left(\eta_0-\eta_{\rm rec}\right)$: 
rotation angle equal to 1~deg (0.1~deg) from last scattering to reionization green continuous (dash-dotted line) line,
rotation angle equal to 1~deg (0.1~deg) from reionization to nowadays blue continuous (dash-dotted line) line
- the \texttt{CAMB} visibility function $g_\mathrm{CAMB}$ is plotted in red (on a different scale); 
angular power spectra obtained with the modified version of \texttt{CAMB} are compared in
(b) $C_\ell^{EE} - C_\ell^{EE}(\bar{\alpha}=0)$, (c) $C_\ell^{BB}$,
(d)  $C_\ell^{TE} - C_\ell^{TE}(\bar{\alpha}=0)$, (e) $C_\ell^{TB}$, (f) $C_\ell^{EB}$.
}
\end{figure*}

In order to study in a more detailed way the effects of isotropic cosmic birefringence $\alpha(\eta)$
we modified source terms in the Boltzmann code \texttt{CAMB} \cite{Lewis:1999bs} following  Eqs.~\eqref{source:E} and \eqref{source:B}.
In Fig.~\ref{Fig:tanhevol:theta:120} we compare the effects on the power spectra of a sudden/instantaneous rotation  $\bar\alpha=1$~deg
occurring at different epochs. 
The initial value of the linear polarization angle is always 
$\alpha(\eta_\mathrm{rec})=0$~deg, 
then $\alpha$ drops 
to $-1$~deg,
but at different epochs.
We consider in particular 
$\alpha(\eta)=-1/2\left\{ 1+\tanh\left[ 10^4 (x-x_*) \right] \right\}$ deg,
where $x\equiv\left(\eta-\eta_{\rm rec}\right)/\left(\eta_0-\eta_{\rm rec}\right)$ and $x_*=\left\{0.005,0.25,0.5,0.996\right\}$.
To give an idea of the numbers involved in $\Lambda$CDM we have $x(\eta_{\rm rec})=0$, 
$x(\eta_{z=100})=0.08$, $x(\eta_{z=10})=0.31$, $x(\eta_{z=5})=0.44$, $x(\eta_0)=1$.
If the change of the linear polarization angle happens nowadays (at $\eta\simeq\eta_0$, or $x_*\simeq 1$) then we clearly have
$\alpha (\eta)-\alpha(\eta_{0})=1$~deg during all integration along the line-of-sight.
In this case the power spectra obtained using the modified Boltzmann code exactly coincide
with the analytic expressions of Eqs.~\eqref{C_ll_TE_constant} - \eqref{C_ll_EB_constant} fixed  $\bar{\alpha}=1$ deg.
Note that a miscalibration of the orientation of the detector is assimilable to a rotation at present time of the linear polarization vector
and gives an analogue effect on the power spectra \cite{Keating:2013,Minami:2019ruj,Minami:2020fin}. 
We clearly see that earlier in time the rotation happens, smaller are the effects on the power spectra 
(in particular the difference is larger at lower $\ell$).
For BB we also plot the power spectra induced by lensing (black dotted line) and tensor perturbations assuming 
a tensor-to-scalar ratio $r=10^{-2}$ (dashed black line) and $r=10^{-3}$ (dot-dashed black line). 
For a detailed discussion of the impact of lensing on cosmic birefringence we refer to  \cite{Namikawa:2021gbr}. 

\begin{figure*}[!htb]
\includegraphics[width=168pt,height=4.5cm]{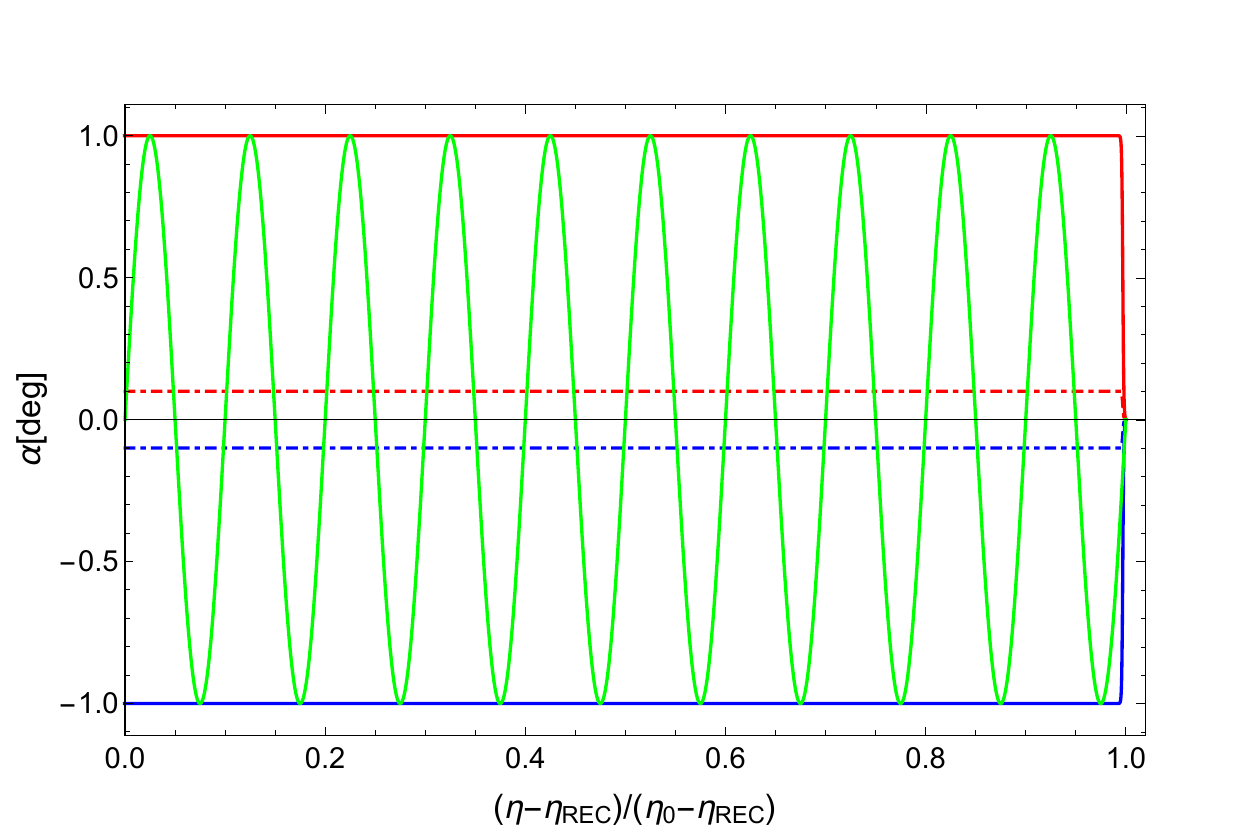}
\includegraphics[width=168pt,height=4cm]{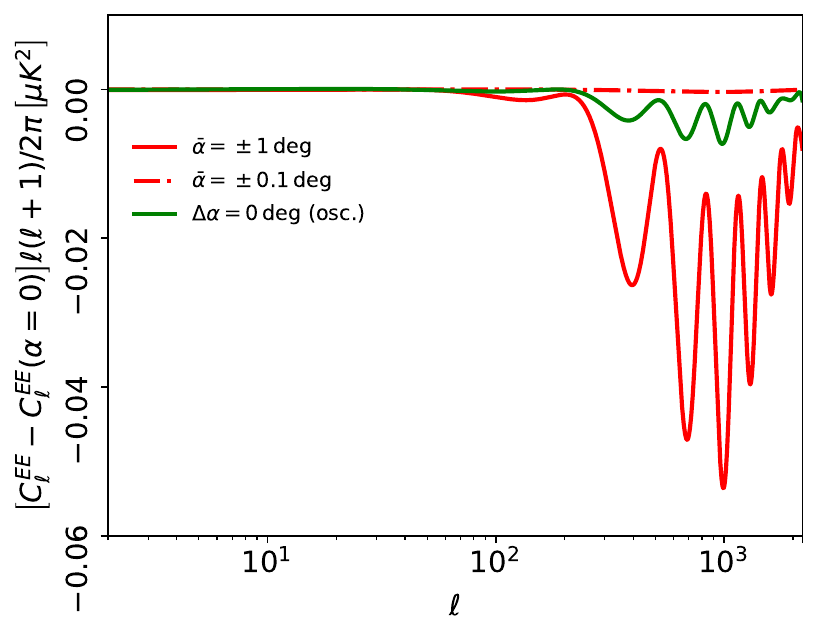}
\includegraphics[width=168pt,height=4cm]{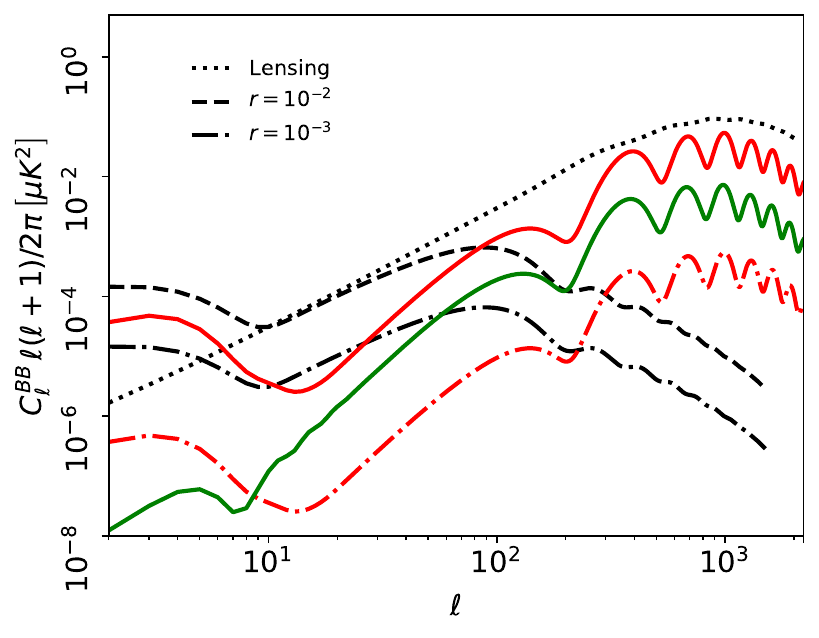}
\includegraphics[width=168pt,height=4cm]{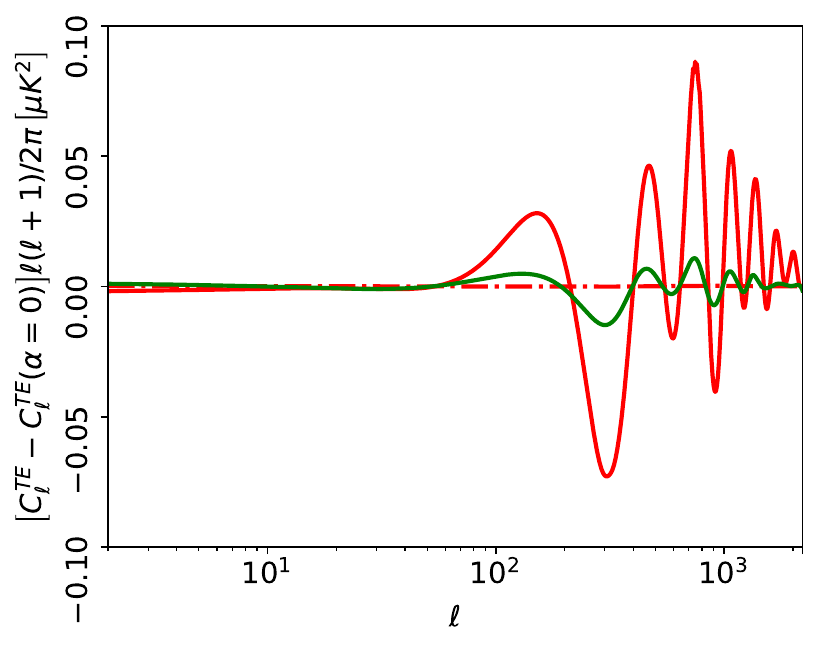}
\includegraphics[width=168pt,height=4cm]{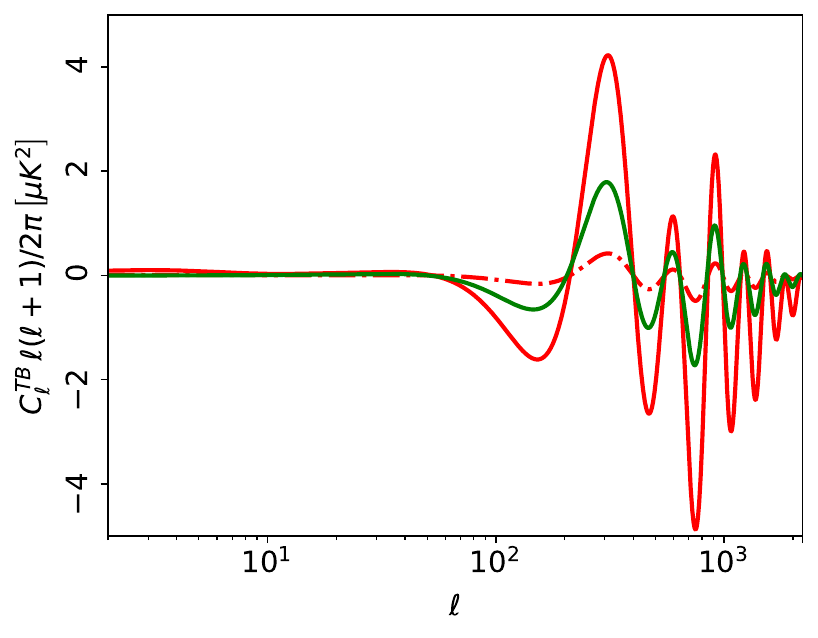}
\includegraphics[width=168pt,height=4cm]{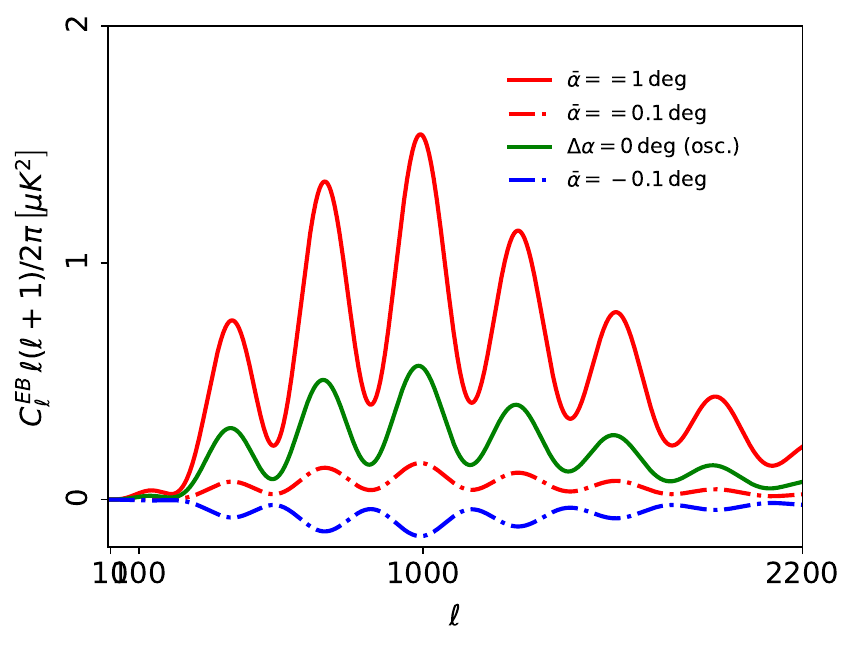}
\caption{\footnotesize\label{Fig:sin20}
(a) Oscillating birefringence angle with $\bar{\alpha}\equiv\alpha (\eta_\mathrm{rec})-\alpha(\eta_{0}) =0$~deg, 
the linear polarization angle oscillates between 1~deg and $-1$~deg: $\alpha(\eta)= \sin(20\pi x)$ (green line),
we plot for comparison also a the case of a sudden rotation of $+/-1$~deg ($+/-0.1$~deg) occurring at present time, 
see continuous (dot-dashed) red/blue line; 
(b) $C_\ell^{EE} - C_\ell^{EE}(\bar{\alpha}=0)$, (c) $C_\ell^{BB}$,
(d)  $C_\ell^{TE} - C_\ell^{TE}(\bar{\alpha}=0)$, (e) $C_\ell^{TB}$, (f) $C_\ell^{EB}$.}
\end{figure*}

In Fig.~\ref{Fig:tanhevol} 
we show the output of the modified Boltzmann code considering a rotation $\bar{\alpha}$ of linear polarization  
$(i)$ localized only at {\em early times} (from last scattering to reionization)
$\alpha(\eta)= \bar{\alpha} \left\{\frac{\tanh\left[10^4 (x-0.001)\right]+1}{2}-\frac{\tanh\left[10^4 (x-0.25)\right]+1}{2}\right\}$, 
or $(ii)$ only at {\em late times} (from reionization to nowadays)
$\alpha(\eta)= \bar{\alpha} \left\{\frac{\tanh\left[10^4 (x-0.25)\right]+1}{2}-\frac{\tanh\left[10^4 (x-0.996)\right]+1}{2}\right\}$.
Since the linear polarization rotation is not constant over time, 
the effects on the power spectra are different.
If $\alpha$ is rotated only at early times the effects on the power spectra are localized at
$\ell\gtrsim 10$. 
Otherwise if the linear polarization angle rotates after reionization (late times),
the effects are visible only at 
$\ell\lesssim 10$.

Interestingly we stress that birefringence effects on the power spectra can be present even if $\alpha (\eta_\mathrm{rec})=\alpha(\eta_{0})$,
differently from what stated in Ref. \cite{Fedderke:2019ajk}.
In this case according to the analytic expressions
of Eqs.~\eqref{C_ll_TE_constant}-\eqref{C_ll_EB_constant} there should be no effects
since $\bar{\alpha}=0$.
On the contrary there are evident effects on the power spectra using the modified \texttt{CAMB} code 
based on the Boltzmann equation for cosmic birefringence, 
see in particular Fig.~\ref{Fig:sin20}. 
See also Appendix~\ref{App_1} for other interesting phenomenological cases with 
$\alpha (\eta_\mathrm{rec})=\alpha(\eta_{0})$.

\section{Theory Modeling}
\label{Sect:III}

In an expanding universe a spatially homogeneous scalar field obeys:
\newpage 
\begin{equation}
\ddot{\phi}+3H\dot{\phi}-\frac{dV}{d\phi}=0\,,    
\end{equation}
where $\dot{~}$ denotes derivative respect to cosmic time $t$.
In this Section we specify the potential $V(\phi)$  for: 
(a) axion-like Early Dark Energy (Sect. \ref{Sect:III:EDE}),
(b) Quintessence (Sect. \ref{Sect:III:DE}), and
(c) axion-like dark matter (Sect. \ref{Sect:III:DM}).
From the evolution of $\phi(t)$ we estimate the effects on CMB power spectra using a modified version of \texttt{CAMB} \cite{Lewis:1999bs}.

\subsection{Axion-like as  Early Dark Energy}
\label{Sect:III:EDE}

\begin{figure}[!htb]
\includegraphics[width=\columnwidth]{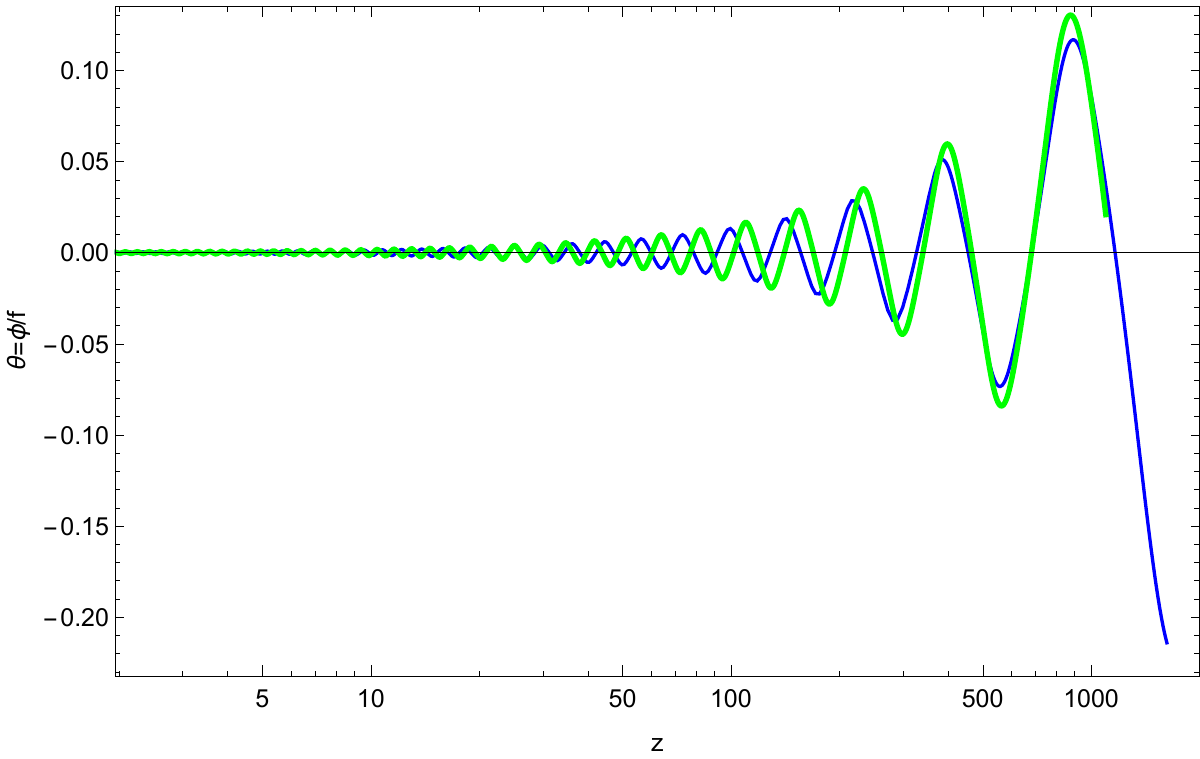}
\caption{\footnotesize\label{Fig:phi_approx} 
Early dark energy, evolution of $\Theta$ as a function of $z$ from recombination to nowadays 
fixed $n=2$, $\Lambda=0.417$ eV, $f=0.05\, M_\mathrm{pl}=1.22\times 10^{17}$ GeV,
$\Theta_i=1$ and $\dot{\Theta}_i=0$:
numerical fit of Eq.~\eqref{Theta:EDE_2} (green line) and the evolution provided by \texttt{CAMB-1.3.2} (blue line);
in this version of \texttt{CAMB} EDE is implemented following \cite{Smith:2019ihp}.
}
\end{figure}

Early Dark Energy was proposed in order to solve the tension between the local and the cosmological measurements 
of the Hubble parameter \cite{Poulin:2018dzj,Fujita:2020ecn,Poulin:2018cxd,Capparelli:2019rtn,Murai:2022zur}.
In this case we consider a potential of the form:
\begin{equation}
V\left(\phi\right) = \Lambda^4 \left(1-\cos\frac{\phi}{f}\right)^n\,,
\end{equation}
describing the spontaneous breaking of a continuous symmetry at scale $f$.
The evolution of the pseudoscalar field $\phi$ is determined by the following system of equations:
\begin{equation}
\left\{
\begin{array}{l}
\ddot{\phi}+3H\dot{\phi}+\frac{n\Lambda^4}{f}
\left(1-\cos\frac{\phi}{f}\right)^{n-1}\sin\frac{\phi}{f}=0\,,\\
H^2=\frac{1}{3 M_\mathrm{pl}^2}\left(\rho_\mathrm{RAD}+\rho_\mathrm{ MAT}+\rho_\Lambda+\rho_{\phi}\right)\,,
\end{array}
\right.
\end{equation}
where $ M_\mathrm{pl}=2.43\times10^{18}$ GeV is the reduced Planck mass.
Initially the field is frozen and acts as a cosmological constant, 
and it begins to oscillate when the effective mass becomes of the order of $H$.
In practice, we solve numerically this system in the new variable $x\equiv\ln t/t_i$,
from a fixed point $t_i$ in radiation dominated era to nowadays ($t_0$):
\begin{equation}
\left\{
\begin{array}{l}
\frac{d \Theta}{dx^2}+\left(\frac{3}{a}\frac{da}{dx}-1\right)\frac{d\Theta}{dx}
\\
\qquad+t_i^2 e^{2x}\frac{n \Lambda^4 }{f^2} \left(1-\cos \Theta\right)^{n-1} \sin\Theta=0\,,\\
\frac{da}{dx}=t_i e^x H_i a  \left[ \Omega_\mathrm{RAD,i}\left(\frac{a_i}{a}\right)^4
+\Omega_\mathrm{MAT,i}\left(\frac{a_i}{a}\right)^3\right.\\
\left.\qquad
+\Omega_\mathrm{ \Lambda,i}+\frac{1}{6}\frac{f^2}{H_i^2  M_\mathrm{ pl}^2 t_i^2}e^{-2 x}\left(\frac{d\Theta}{dx}\right)^2\right.\\
\left.\qquad +\frac{1}{3}\frac{\Lambda^4}{H_i^2 M_\mathrm{ pl}^2} \left(1-\cos \Theta\right)^n
\right]^{1/2}\,,
\end{array}
\right.
\label{eq_back_EDE}
\end{equation}
where $\Theta(t)\equiv\phi(t)/f$.
In the oscillating regime, for $n=2$,
we approximate the evolution of $\Theta$ as a function of cosmic time
with an elliptic sine  ($\mathrm{sn}$),
see \cite{Abramowitz,Greene:1997fu,Finelli:1998bu}.
In particular, fixed $\Lambda=0.417$ eV, $f=0.05\, M_\mathrm{pl}=1.22\times 10^{17}$ GeV,
$\Theta_i=1$ and $\dot{\Theta}_i=0$ the following numerical fit for
$\Theta$ is obtained \cite{EllipticSine}:
\begin{eqnarray}
\label{Theta:EDE_2}
\Theta(\eta)&\simeq& \left(-6.49\times 10^{-3}+2.15\times 10^{-3}\frac{\eta_0}{\eta}\right)\nonumber\\
&*&
\mathrm{sn}\left(6.35\times 10^{-1}
+5.18\times 10^{2}\frac{\eta}{\eta_0},\frac{1}{\sqrt{2}}\right)\,.
\end{eqnarray}
In Fig.~\ref{Fig:phi_approx} we plot this function for $\Theta$ as a function of redshift $z$, from recombination to $z=0$.

\begin{figure*}[!htb]
\includegraphics[width=168pt,height=4.5cm]{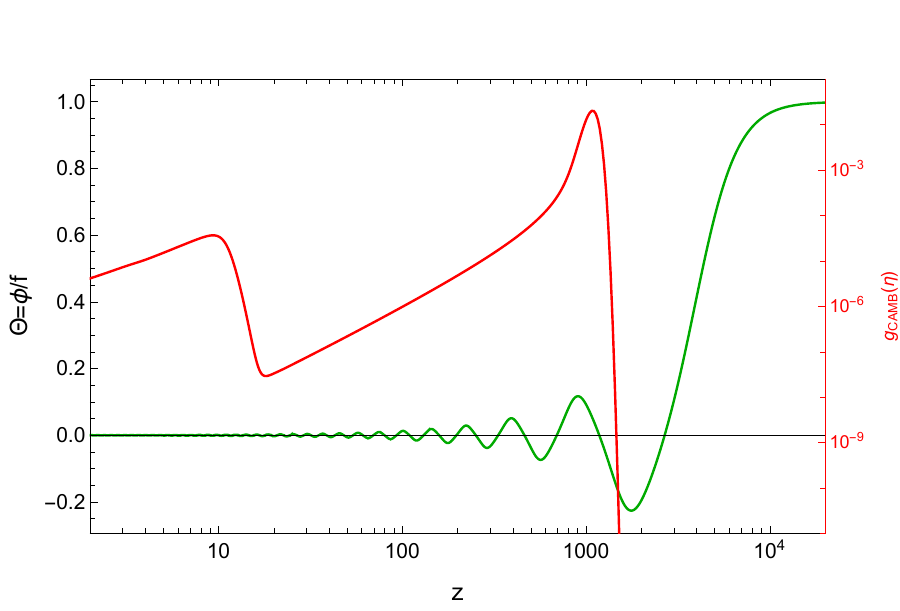}
\includegraphics[width=165pt,height=4cm]{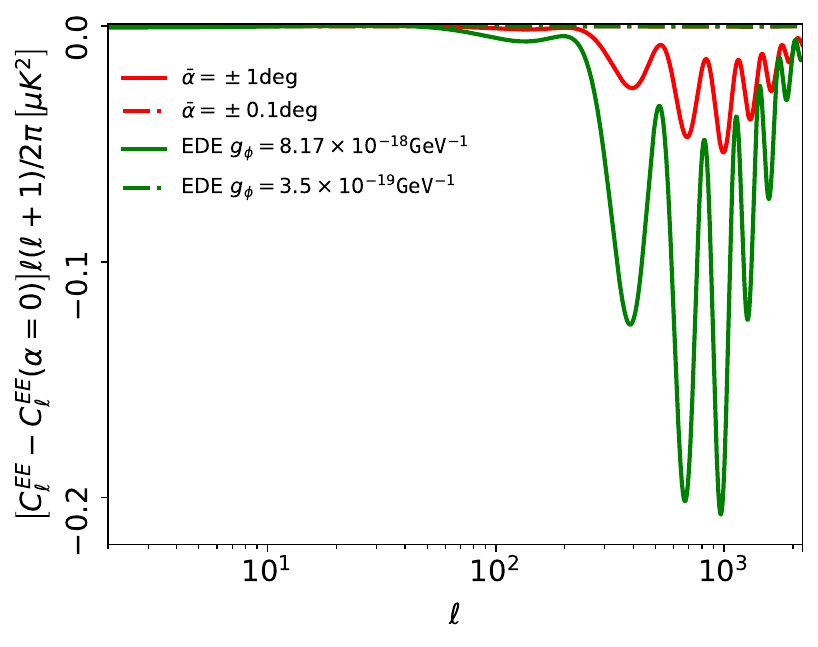}
\includegraphics[width=165pt,height=4cm]{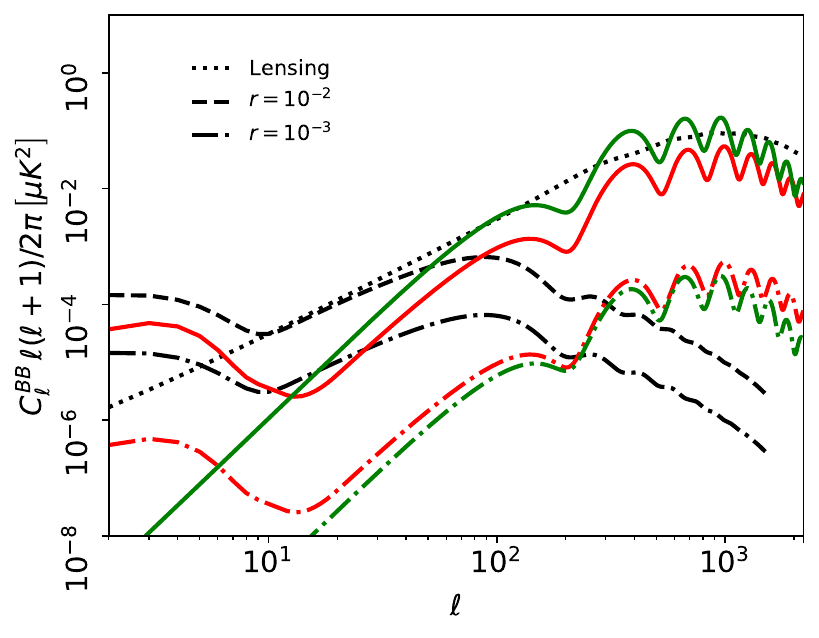}
\includegraphics[width=165pt,height=4cm]{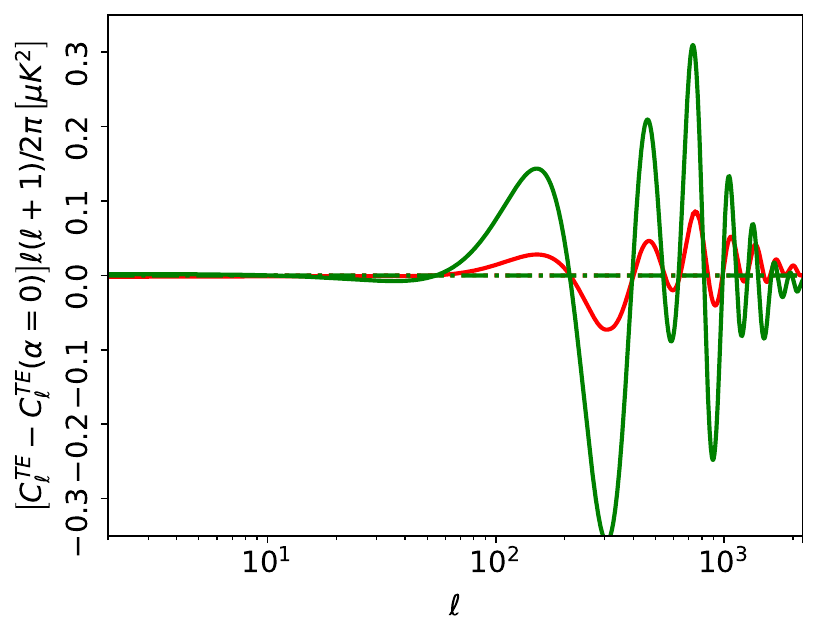}
\includegraphics[width=172pt,height=4cm]{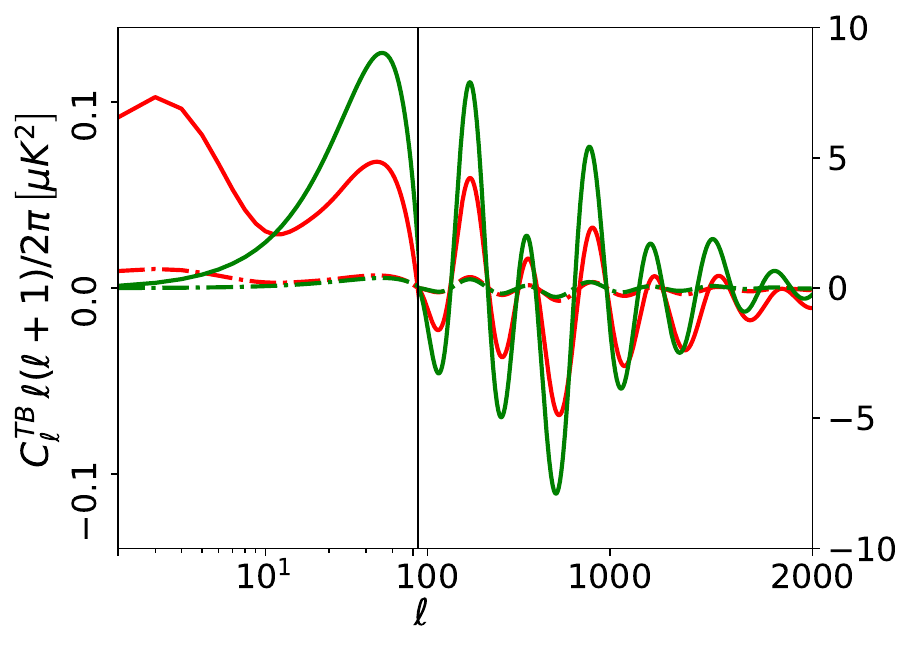}
\includegraphics[width=165pt,height=4cm]{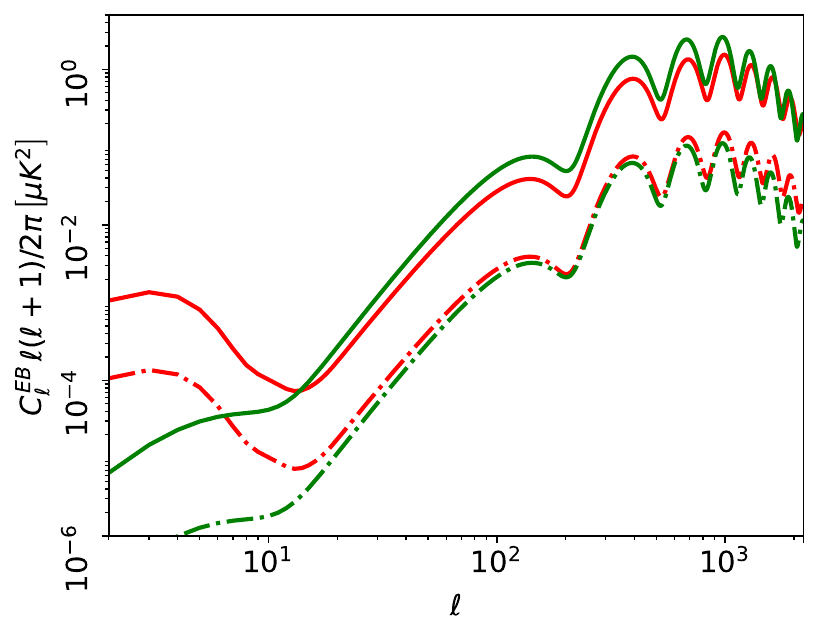}
\caption{\footnotesize\label{Fig:axionEDE_2_newCAMB_nostartphib} 
Early Dark Energy (a) evolution for $\Theta\equiv\phi/f$ as a function of redshift (green line)
provided by the code \texttt{CAMB} fixed $n=2$, 
$\Lambda=0.417$ eV, 
$f=0.05\, M_\mathrm{pl}=1.22\times 10^{17}$ GeV,
$\Theta_i=1$ and $\dot{\Theta}_i=0$; 
the CAMB visibility function $g_\mathrm{CAMB}$ is plotted in red (on a different scale).
Assuming $g_\phi=8.17 \times 10^{-18}\mathtt{GeV}^{-1}$ or $g_\phi=3.5\times10^{-19}\mathtt{GeV}^{-1}$ we plot the angular power spectra for  
(b) $C_\ell^{EE} - C_\ell^{EE}(\bar{\alpha}=0)$, (c) $C_\ell^{BB}$,
(d)  $C_\ell^{TE} - C_\ell^{TE}(\bar{\alpha}=0)$, (e) $C_\ell^{TB}$, (f) $C_\ell^{EB}$.
}
\end{figure*}

In Fig.~\ref{Fig:axionEDE_2_newCAMB_nostartphib} we plot the power spectra 
using the evolution of $\Theta(\eta)$ directly provided by \texttt{CAMB} code. 
Note that for this model $\Theta(\eta_0)$
is negligible 
compared to to the values at $\eta_{\rm rec}$,
since the field is quickly oscillating at $z=0$.
We consider two values of the coupling constant: 
$g_\phi=8.17\times10^{-18}\mathtt{GeV}^{-1}$ 
- corresponding to $\bar{\alpha}=1.15$~deg - and 
$g_\phi=3.5\times10^{-19}\mathtt{GeV}^{-1}$ 
- corresponding to $\bar{\alpha}=0.05$ deg.
For $EE$ and $TE$
we decided to plot the difference with the standard un-rotated spectra, 
$C_\ell^{EE} - C_\ell^{EE}(\bar{\alpha}=0)$ and $C_\ell^{TE} - C_\ell^{TE}(\bar{\alpha}=0)$, in order to underline the differences.

\subsection{Axion-like as dark energy}
\label{Sect:III:DE}

We consider dark energy driven by an axion-like pseudo-scalar, 
as suggested in \cite{Frieman:1995pm}, 
with a potential:
\begin{equation}
V\left(\phi\right) = M^4 \left(1+\cos\frac{\phi}{f}\right) \,.
\end{equation}

We solve numerically the system in the new variable $x$, as in the previous Subsection:
\begin{equation}
\left\{
\begin{array}{l}
\frac{d \Theta}{dx^2}+\left(\frac{3}{a}\frac{da}{dx}-1\right)\frac{d\Theta}{dx}
- t_i^2 e^{2x}\frac{M^4}{f^2}\sin\Theta=0\,,\\
\frac{da}{dx}=t_i e^x H_i a  \left[ \Omega_\mathrm{RAD,i}\left(\frac{a_i}{a}\right)^4
+\Omega_\mathrm{MAT,i}\left(\frac{a_i}{a}\right)^3\right.\\
\left.\qquad
+\frac{1}{6}\frac{f^2}{H_i^2  M_\mathrm{ pl}^2 t_i^2}e^{-2 x}\left(\frac{d\Theta}{dx}\right)^2\right.\\
\left.\qquad +\frac{1}{3}\frac{M^4}{H_i^2 M_\mathrm{ pl}^2} \left(1+\cos \Theta\right)
\right]^{1/2}\,.\nonumber
\end{array}
\right.
\end{equation}

For $M\sim10^{-3}$ eV and $f\lesssim M_\mathrm{ pl}$ the pseudoscalar field
mimics the cosmological constant contribution. 
There are indications from string theory that $f$ cannot be larger than 
$M_\mathrm{ pl}$ \cite{Dine:2001xh,Banks:2003sx}.
In the future, when the expansion rate of the universe will become smaller, 
the field will start to oscillate and 
the universe will become cold dark matter dominated.

\begin{figure*}[!htb]
\includegraphics[width=165pt,height=4.5cm]{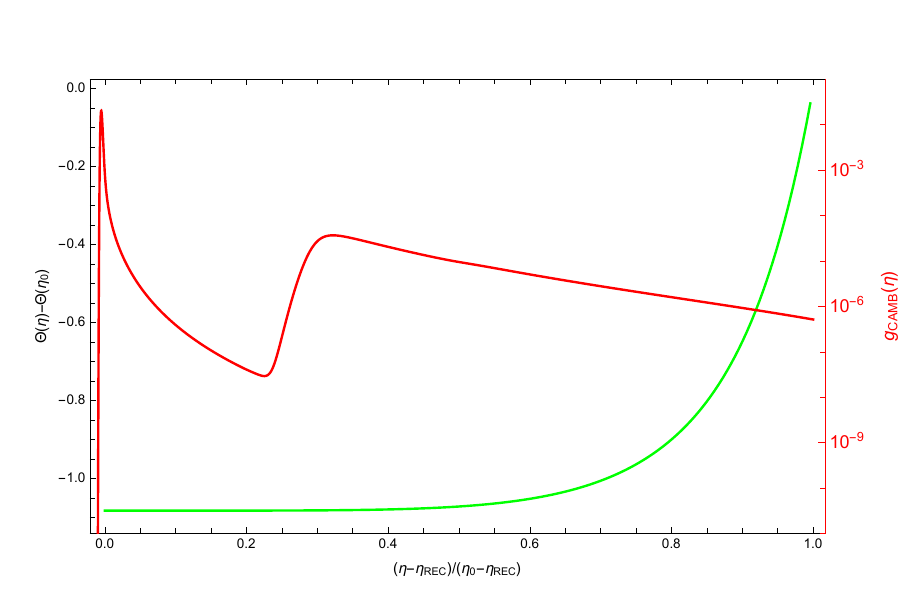}
\includegraphics[width=165pt,height=4cm]{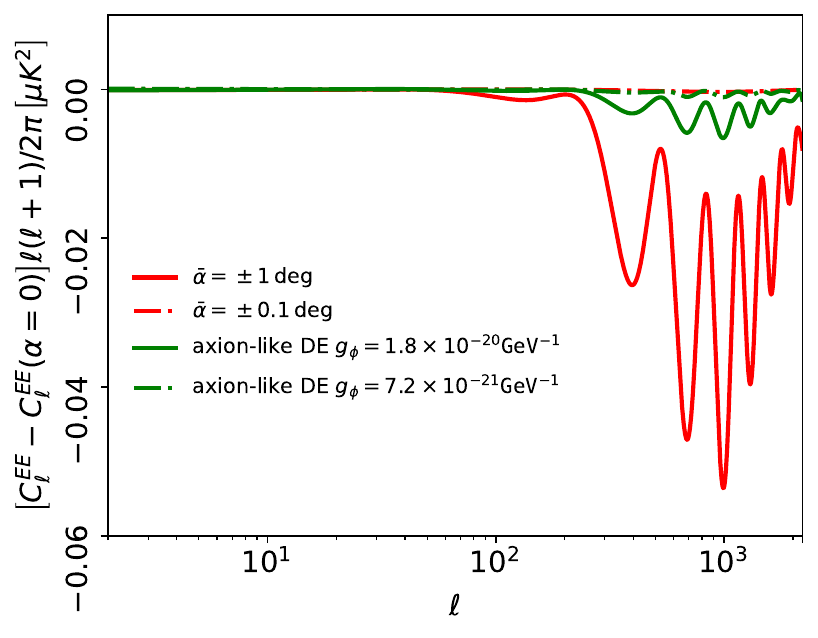}
\includegraphics[width=165pt,height=4cm]{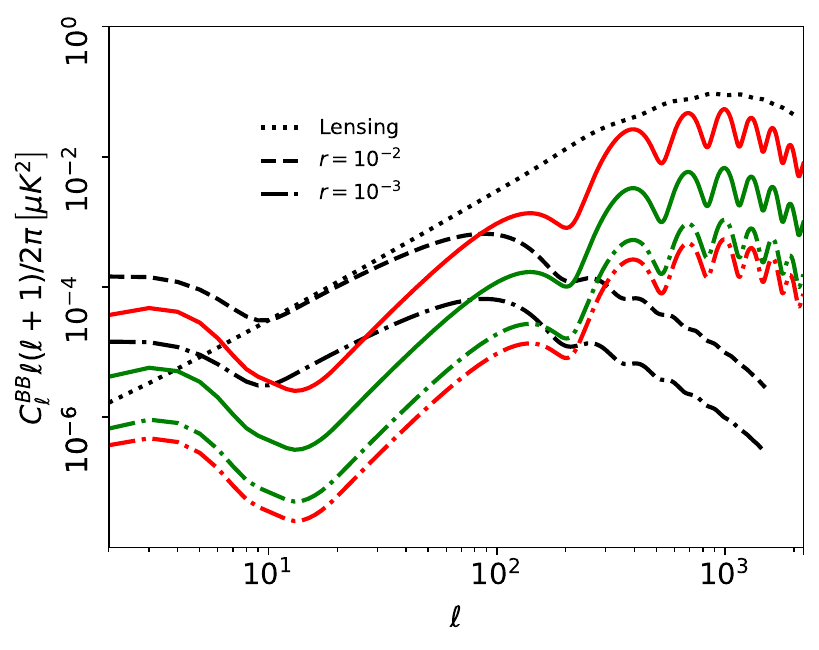}
\includegraphics[width=165pt,height=4cm]{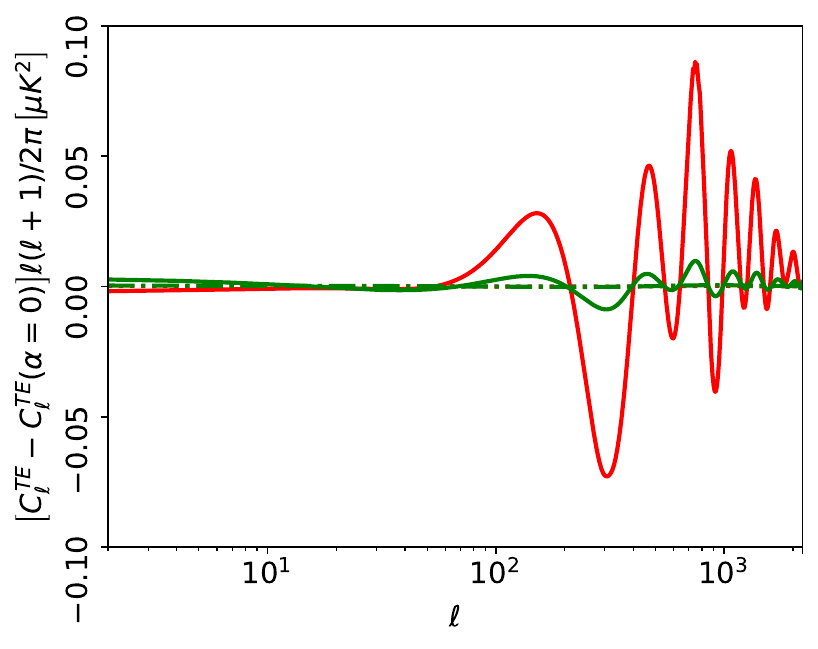}
\includegraphics[width=172pt,height=4cm]{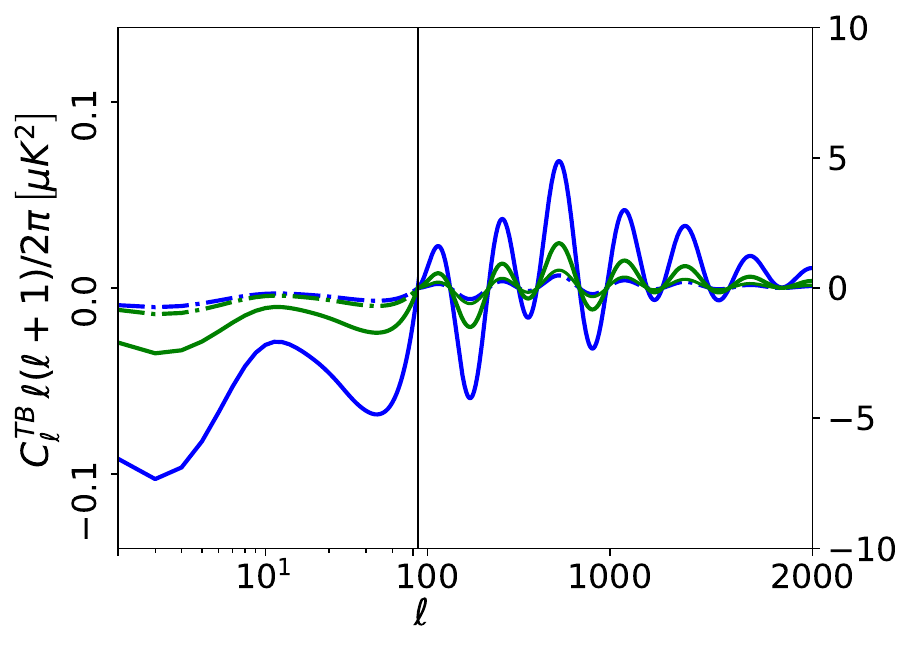}
\includegraphics[width=165pt,height=4cm]{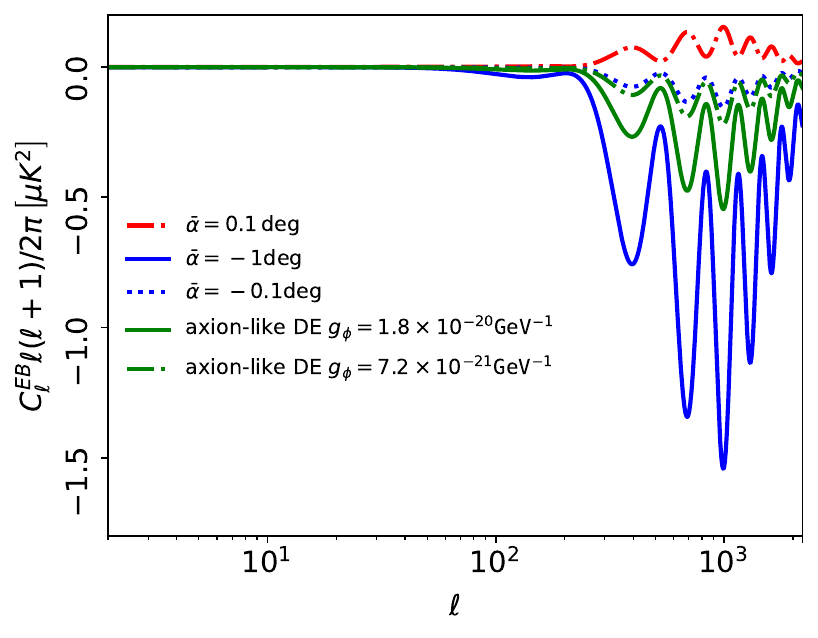}
\caption{\footnotesize\label{Fig:axionDE_1}
Axion-like dark energy (a) evolution of $\Theta(\eta)-\Theta(\eta_0)$
as a function of conformal time $\left(\eta-\eta_{\rm rec}\right)/\left(\eta_0-\eta_{\rm rec}\right)$
is plotted in green fixed $M=1.95\times10^{-3}$ eV, $f=0.265 M_\mathrm{pl}$,
$\Theta_i=0.25$ and $\dot{\Theta}_i=0$;
the \texttt{CAMB} visibility function $g_\mathrm{CAMB}$ is plotted in red (on a different scale).
Assuming $g_\phi=1.8\times10^{-20}\mathtt{GeV}^{-1} $ 
and $g_\phi=\frac{2\,\alpha_\mathrm{EM} }{f\,\pi}\simeq7.2\times10^{-21}\mathtt{GeV}^{-1}$, 
we plot the angular power spectra for  
(b) $C_\ell^{EE} - C_\ell^{EE}(\bar{\alpha}=0)$, (c) $C_\ell^{BB}$,
(d)  $C_\ell^{TE} - C_\ell^{TE}(\bar{\alpha}=0)$, (e) $C_\ell^{TB}$, (f) $C_\ell^{EB}$.
}
\end{figure*}

The pseudoscalar field becomes dynamical only recently.
By fixing $M=1.95\times10^{-3}$ eV, $f=0.265\, M_\mathrm{ pl}$,
$\Theta_i=0.25$ and $\dot{\Theta}_i=0$ we use the following numerical fit for
$\Theta(\eta)$:
\begin{equation}
\label{Theta:PNGB2}
\Theta(\eta)\simeq 0.25+1.468
\times 10^{-4} \exp\left[8.857\left(\frac{\eta-\eta_{\rm rec}}{\eta_0-\eta_{\rm rec}}\right)\right]\,.
\end{equation}

Differently from the Early Dark Energy model, discussed in the previous Subsection,
here the field is not oscillating at $z=0$ and it is important to consider $\Theta(\eta_0)$.
Using this numerical fit we evaluate the linear polarization angular power spectra 
for $g_\phi=1.8\times10^{-20}\mathtt{GeV}^{-1}$,
corresponding to a total rotation angle today $\alpha(\eta_0)=0.35$ deg. 
In some models the coupling constant between the pseudoscalar field 
is assumed to be proportional to the inverse of the energy breaking scale $f$
\cite{Sikivie:1983ip,Raffelt:1996wa}: 
\begin{equation}
\mathcal{L}\supset  -\frac{C\,\alpha_\mathrm{EM} }{2\pi f}\phi F_{\mu\nu}\,\widetilde F^{\mu\nu}\,,    
\end{equation}
where $C\simeq\mathcal{O}(1)$ is a model dependent constant.
Therefore we discuss also the case: $g_\phi=\frac{2\,\alpha_\mathrm{EM} }{f\,\pi}
\simeq 7.2\times10^{-21}\mathtt{GeV}^{-1}$ - corresponding to $\alpha(\eta_0)=0.14$ deg.

See Fig.~\ref{Fig:axionDE_1} for the power spectra $C_\ell^{EE} - C_\ell^{EE}(\bar{\alpha}=0)$, $C_\ell^{BB}$,
$C_\ell^{TE} - C_\ell^{TE}(\bar{\alpha}=0)$, $C_\ell^{TB}$, and $C_\ell^{EB}$ evaluated using \texttt{CAMB}.

\subsection{Axion-like as dark matter}
\label{Sect:III:DM}

For axion-like field acting as dark matter \cite{Raffelt:1996wa,Kolb:1990vq,Sikivie:2006ni}
we consider the potential: 
\begin{equation}
V\left(\phi\right) = m^2 \frac{f^2}{N^2} \left(1-\cos\frac{\phi N}{f}\right)\,,
\end{equation}
in the regime where the  pseudoscalar field  oscillates near the minimum. 
The field evolves according to \cite{Finelli:2008jv,Galaverni:2009zz}:
\begin{eqnarray}
\phi(t)
&=&\sqrt{6\Omega_\mathrm{ MAT}}\frac{H_0 M_\mathrm{ pl}}{m a^{3/2}(t)}\nonumber\\
\label{E:phi_t_1}
& &
\sin\left[m t \sqrt{1-\left(1-\Omega_\mathrm{ MAT}\right)\left(\frac{3H_0}{2m}\right)^2}\right]\,,  
\end{eqnarray}
where the evolution of the scale factor is \cite{Gruppuso:2005xy}:
\begin{eqnarray}
a(t)&=&\left(\frac{\Omega_\mathrm{ MAT}}{1-\Omega_\mathrm{ MAT}}\right)^\frac{1}{3}\nonumber\\
& &\left[\sinh \left(\frac{3}{2}\sqrt{1-\Omega_\mathrm{ MAT}} H_0 t\right)\right]^\frac{2}{3}\,.
\end{eqnarray}

\begin{figure*}[!htb]
\includegraphics[width=165pt,height=4.5cm]{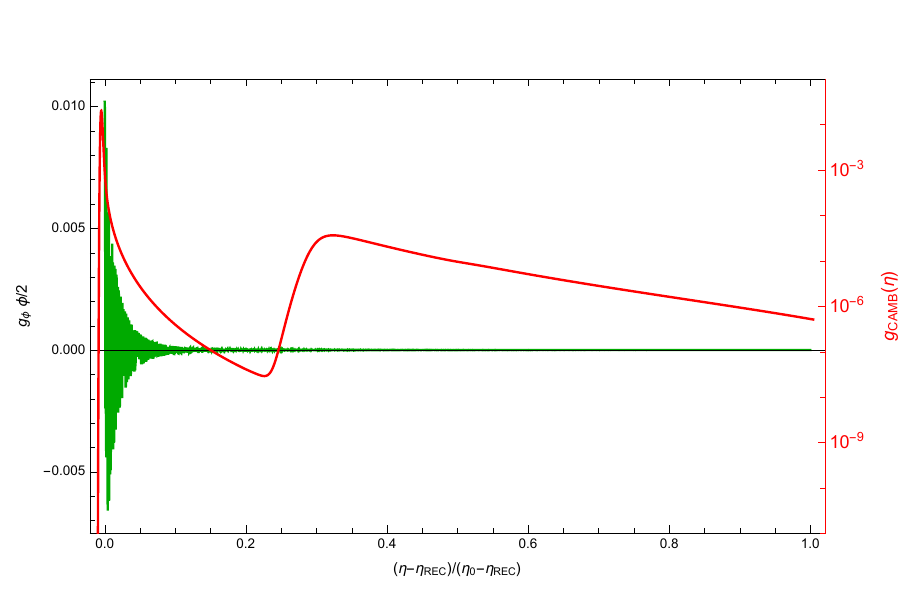}
\includegraphics[width=165pt,height=4cm]{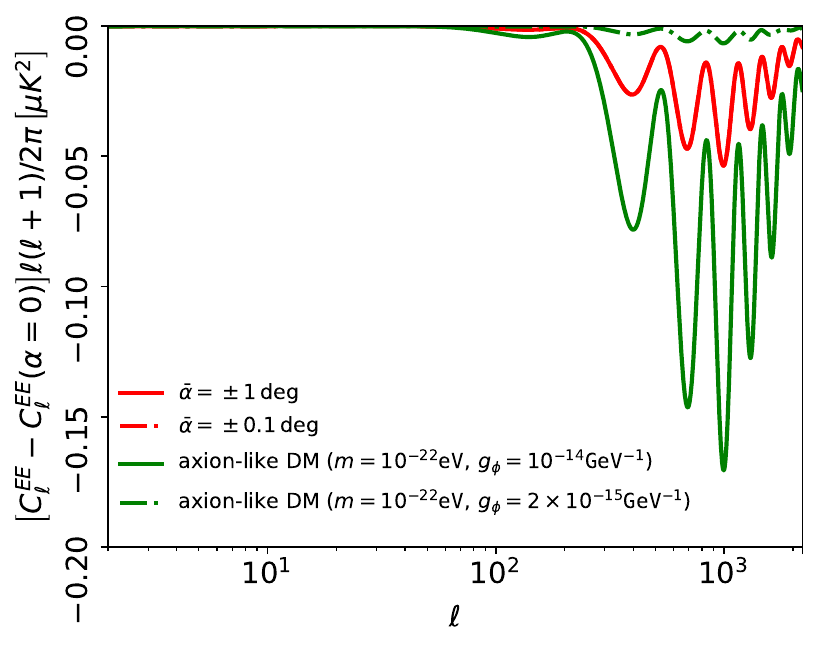}
\includegraphics[width=165pt,height=4cm]{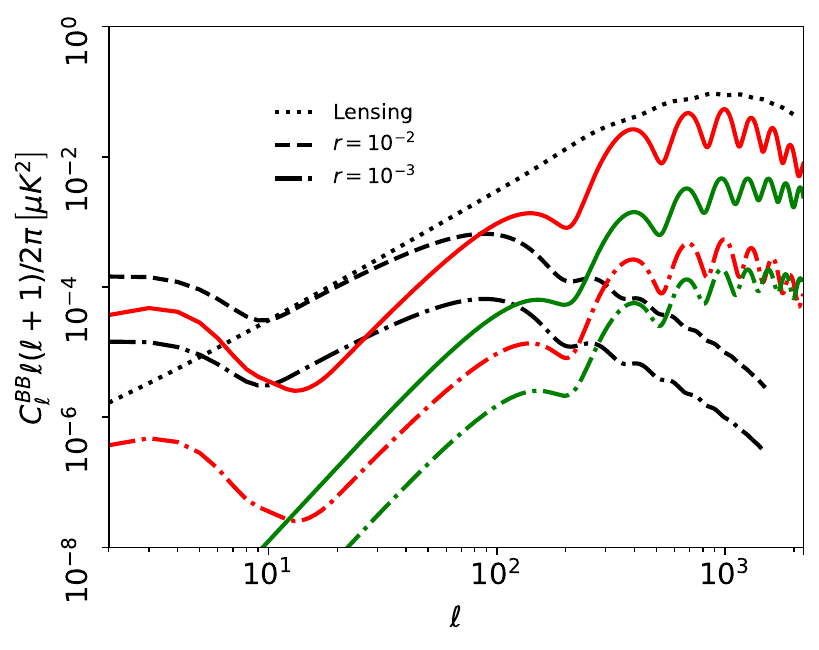}
\includegraphics[width=165pt,height=4cm]{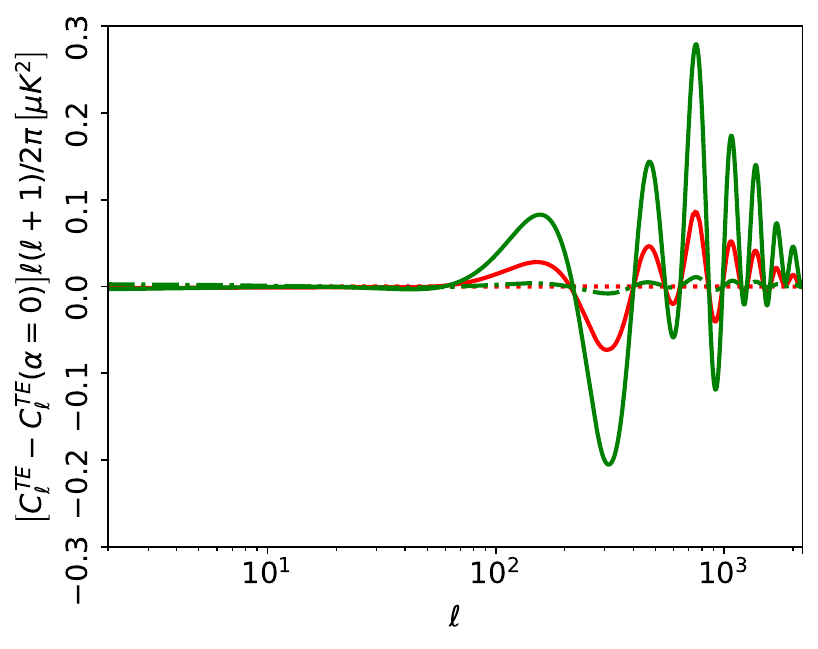}
\includegraphics[width=172pt,height=4cm]{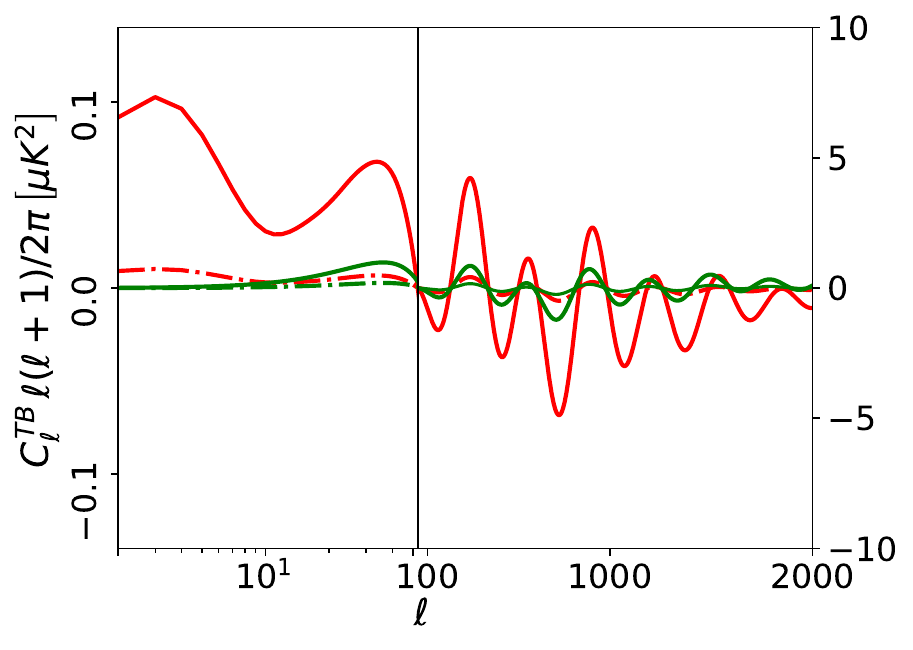}
\includegraphics[width=165pt,height=4cm]{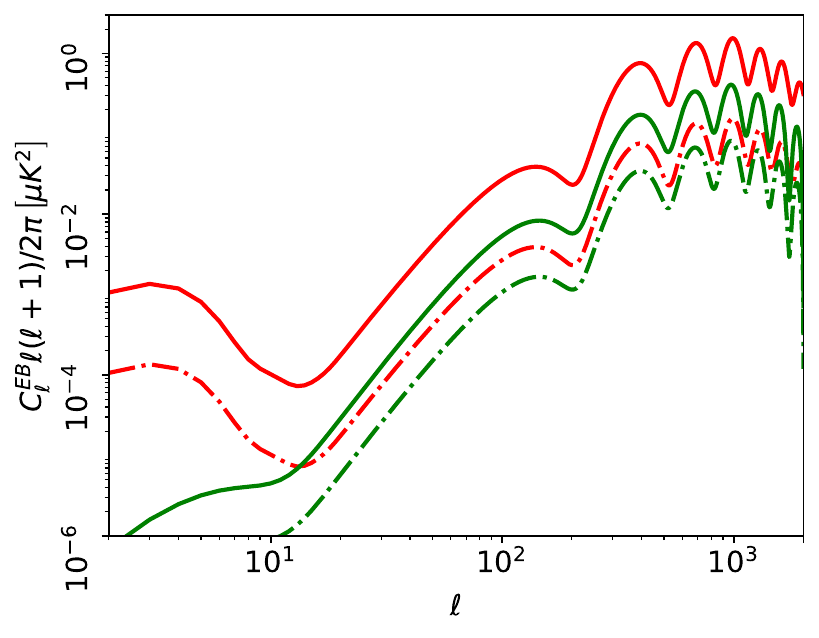}
\caption{\footnotesize\label{Fig:axionDM_m22_g23} Axion-like dark matter: (a) evolution of $g_\phi \phi(\eta)/2$ is plotted in green
as a function of conformal time $\left(\eta-\eta_{\rm rec}\right)/\left(\eta_0-\eta_{\rm rec}\right)$ fixed $m=10^{-22}\mathtt{eV}$ and $g_\phi=10^{-14}\mathtt{GeV}^{-1}$ (green continuous line);
the \texttt{CAMB} visibility function $g_\mathrm{CAMB}$ is plotted in red (on a different scale).
Assuming $g_\phi=10^{-14}\mathtt{GeV}^{-1}$ (green continuous line),
$g_\phi=2\times10^{-15}\mathtt{GeV}^{-1}$  (light-green dashed line),
we plot the angular power spectra for:  
(b) $C_\ell^{EE} - C_\ell^{EE}(\bar{\alpha}=0)$, (c) $C_\ell^{BB}$,
(d)  $C_\ell^{TE} - C_\ell^{TE}(\bar{\alpha}=0)$, (e) $C_\ell^{TB}$, (f) $C_\ell^{EB}$.
}
\end{figure*}

Since the Boltzmann \texttt{CAMB} code works in conformal time
we fit numerically the relation between cosmic and conformal time
from recombination to today; 
for the $\Lambda$CDM model cosmological model with \textit{Planck} 2018 estimates of cosmological parameters we obtain \cite{Planck:2018vyg}:
\begin{equation}
t\simeq\frac{\eta_0}{3.5041}\left(\frac{\eta}{\eta_0}\right)^{3.09358}\,.    
\end{equation}
As in the Early dark Energy case, the field quickly oscillates at $z=0$ therefore we can assume $\phi(\eta_0)\simeq 0$.
We plot  $g_\phi\phi(\eta)/2$
as a function of time chosen a particular value for $m=10^{-22}\mathtt{eV}$ and $g_\phi=10^{-14}\mathtt{GeV}^{-1}$,
see Fig.~\ref{Fig:axionDM_m22_g23}a.
Once the source terms for scalar perturbations in the Boltzmann code
are modified inserting the new terms proportional to $\alpha (\eta)-\alpha(\eta_{0})$, see Eqs.~\eqref{source:E}-\eqref{source:B},
the rotated power spectra are obtained.
In Fig.~\ref{Fig:axionDM_m22_g23} we plot $C_\ell^{EE} - C_\ell^{EE}(\bar{\alpha}=0)$,  $C_\ell^{BB}$,
$C_\ell^{TE} - C_\ell^{TE}(\bar{\alpha}=0)$, $C_\ell^{TB}$, and $C_\ell^{EB}$ for
$m=10^{-22}\mathtt{eV}$ and for two different values of the coupling constant:
$g_\phi=10^{-14}\mathtt{GeV}^{-1}$ - corresponding to a total rotation angle $\alpha(\eta_0)=0.52$ deg,
and 
$g_\phi=2\times10^{-15}\mathtt{GeV}^{-1}$ - corresponding to a total rotation angle $\alpha(\eta_0)=0.103$ deg.

\section{Current measurements and forecasts for future experiments}
\label{Sect:IV}

In this Section we discuss the status of current measurements and forecast the science capabilities of future experiments in the context of cosmological birefringence. We will analyze various cosmological models with different approximations  by providing effective $\Delta \chi^2$ and posterior probabilities for parameters by Monte Carlo Markov Chain (MCMC) exploration. 

The parity violating nature of the
interaction generates nonzero parity-odd correlators ($C_\ell^{TB}$ and $C_\ell^{EB}$). We 
therefore consider the full theoretical data covariance matrix:
\begin{eqnarray}
\label{C:theor}
&&\mathbf{\bar{C}}_l =\left( 
\begin{array}{ccc}
  \bar{C}_\ell^{TT} & \bar{C}_\ell^{TE} & \bar{C}_\ell^{TB}\\
  \bar{C}_\ell^{TE} & \bar{C}_\ell^{EE} & \bar{C}_\ell^{EB}\\
  \bar{C}_\ell^{TB} & \bar{C}_\ell^{EB} & \bar{C}_\ell^{BB}
  \end{array}
\right)\\
&=& \left( 
\begin{array}{ccc}
  C_\ell^{TT}+N_\ell^{TT} & C_\ell^{TE} & C_\ell^{TB}\\
  C_\ell^{TE} &  C_\ell^{EE}+N_\ell^{EE} & C_\ell^{EB}\\
  C_\ell^{TB} & C_\ell^{EB} &  C_\ell^{BB}+N_\ell^{BB}
  \end{array}
  \right)\nonumber\,.
\end{eqnarray}
The noise power spectra are obtained by considering 
an inverse-variance weighted sum of the noise sensitivity 
convolved with a Gaussian beam window function for each frequency channel $\nu$ 
\cite{Finelli:2016cyd}:
\begin{equation}
\label{noise:XX}
N_\ell^{XX}=\left[\sum_\nu\frac{1}{N_{l\nu}^{XX}}\right]^{-1}\,,    
\end{equation}
with:
\begin{equation}
N_{l\nu}^{XX}=\Delta^2_{X\nu} 
\exp\left[l\left(l+1\right)\frac{\theta_{\rm FWHM\,\nu}^2}{8\ln 2}\right]\,,
\end{equation}
here $X=\left\{T,E,B\right\}$, $\Delta_{X\nu}$ is the detector noise level,
and $\theta_{\rm FWHM\,\nu}$ is the full width half maximum (FWHM) for a given frequency channel $\nu$. 

\begin{table}[htbp]
\centering
\begin{tabular}{|c|c|c|c|}
\hline
$\nu$ [GHz]&$\theta_{\rm FWHM\,\nu}$[arcmin] & $\Delta_{T\nu}$ [$\mu K\,$arcm.] &
$\Delta_{P\nu}$ [$\mu K\,$arcm.] \\
\hline
$78$  & 39 &9.56 & 13.5 \\
$89$ & 35 & 8.27&11.7\\
$100$ & 29 &6.50 &9.2\\
$119$ & 25 &5.37 &7.6\\
$140$ & 23 &4.17 &5.9\\
$166$ & 21 & 4.60&6.5\\
$195$ & 20 & 4.10&5.8\\
\hline
\end{tabular}
\caption{\footnotesize\label{TableLiteBIRD} Experimental specification for LiteBIRD:
the full width half maximum ($\theta_{\rm FWHM\,\nu}$) and  the detector noise levels
for different frequency channels
\cite{Hazumi:2019lys,Paoletti:2019pdi}.} 
\end{table}

\begin{figure}[!htb]
\includegraphics[width=\columnwidth]{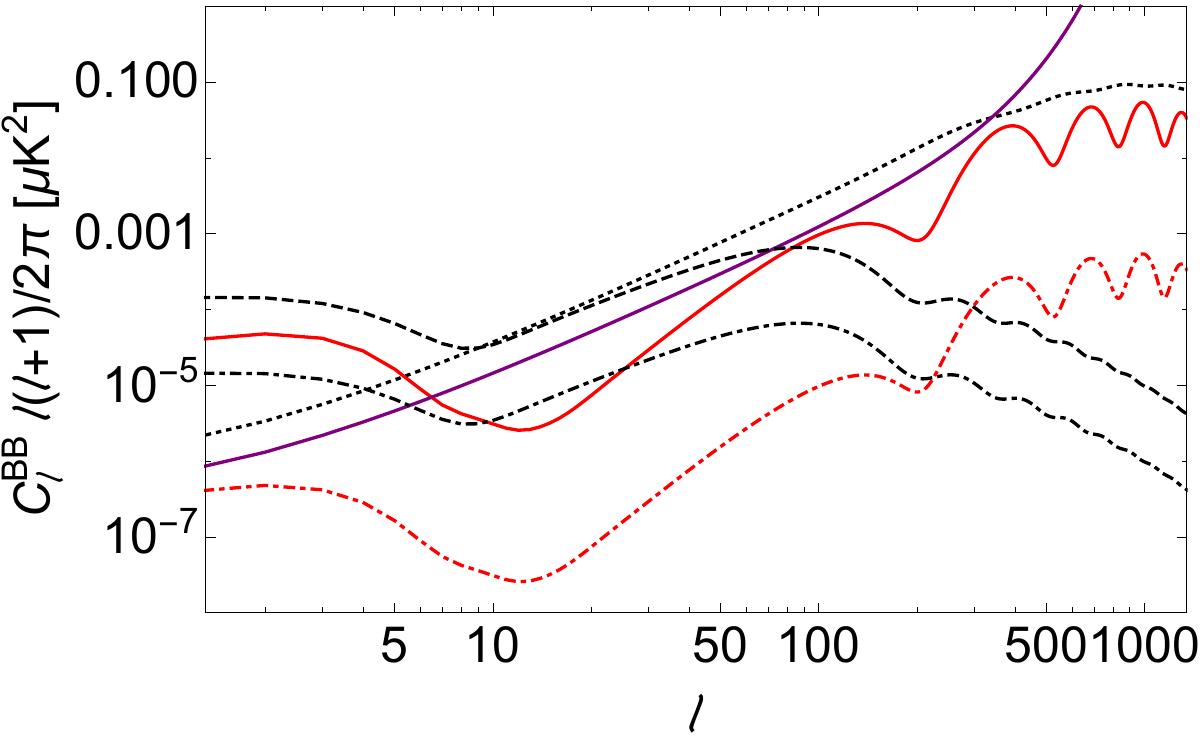}
\caption{\footnotesize\label{plot::noiseBB} 
Estimated noise power spectrum $N_{l\nu}^{BB}$ for LiteBIRD (purple continuous line), 
compared with the signal induced by gravitational lensing (black dotted line), 
primordial signal for $r=10^{-2}$ (black dashed line),
primordial signal for $r=10^{-3}$ (black dot-dashed line),
signal induced by cosmic birefringence for $\bar{\alpha}=\pm 1$~deg (red continuous line),
signal induced by cosmic birefringence for $\bar{\alpha}=\pm 0.1$~deg (red dot-dashed line).
}
\end{figure}

Following \cite{Easther:2004vq,Xia:2007gz,Xia:2007qs},we consider a Wishart likelihood and introduce the effective $\chi^2_\mathrm{eff}$: 
\begin{equation}
\label{chi2}
\chi^2_\mathrm{eff} = \sum_\ell \left(2 \ell + 1\right) f_\mathrm{sky}\left(\frac{A}{\left|\bar{C} \right|}
+\ln\frac{|\bar{C}|}{|\hat{C}|}-3 \right)\,, 
\end{equation}
where $f_\mathrm{sky}$ denotes the observed fraction of the sky,
$A$ is defined as:
\begin{widetext}
\begin{eqnarray}
A &=& \hat{C}_\ell^{TT}\left(\bar{C}_\ell^{EE}\bar{C}_\ell^{BB}-(\bar{C}_\ell^{EB})^2\right)
+\hat{C}_\ell^{TE}\left(\bar{C}_\ell^{TB}\bar{C}_\ell^{EB}-\bar{C}_\ell^{TE} \bar{C}_\ell^{BB}\right)
+\hat{C}_\ell^{TB}\left(\bar{C}_\ell^{TE}\bar{C}_\ell^{EB}-\bar{C}_\ell^{TB} \bar{C}_\ell^{EE}\right)
\nonumber\\
&+&\hat{C}_\ell^{TE}\left(\bar{C}_\ell^{TB}\bar{C}_\ell^{EB}-\bar{C}_\ell^{TE} \bar{C}_\ell^{BB}\right)
+\hat{C}_\ell^{EE}\left(\bar{C}_\ell^{TT}\bar{C}_\ell^{BB}-(\bar{C}_\ell^{TB})^2\right)
+\hat{C}_\ell^{EB}\left(\bar{C}_\ell^{TE}\bar{C}_\ell^{TB}-\bar{C}_\ell^{TT} \bar{C}_\ell^{EB}\right)
\nonumber\\
&+&\hat{C}_\ell^{TB}\left(\bar{C}_\ell^{TE}\bar{C}_\ell^{EB}-\bar{C}_\ell^{EE} \bar{C}_\ell^{TB}\right)
+\hat{C}_\ell^{EB}\left(\bar{C}_\ell^{TE}\bar{C}_\ell^{TB}-\bar{C}_\ell^{TT} \bar{C}_\ell^{EB}\right)
+\hat{C}_\ell^{BB}\left(\bar{C}_\ell^{TT}\bar{C}_\ell^{EE}-(\bar{C}_\ell^{TE})^2\right)\,,
\end{eqnarray}
$|\bar{C}|$ is the determinant of the theoretical covariance matrix, see Eq.~\eqref{C:theor}:
\begin{equation}
|\bar{C}|=\bar{C}_\ell^{TT}\bar{C}_\ell^{EE}\bar{C}_\ell^{BB}
+2\bar{C}_\ell^{TE}\bar{C}_\ell^{TB}\bar{C}_\ell^{EB}
-\bar{C}_\ell^{TT}\left(\bar{C}_\ell^{EB}\right)^2
-\bar{C}_\ell^{EE}\left(\bar{C}_\ell^{TB}\right)^2
-\bar{C}_\ell^{BB}\left(\bar{C}_\ell^{TE}\right)^2\,, 
\end{equation}
and $|\hat{C}|$ is the determinant of the observed covariance matrix:
\begin{equation}
|\hat{C}|=\hat{C}_\ell^{TT}\hat{C}_\ell^{EE}\hat{C}_\ell^{BB}
+2\hat{C}_\ell^{TE}\hat{C}_\ell^{TB}\hat{C}_\ell^{EB}
-\hat{C}_\ell^{TT}\left(\hat{C}_\ell^{EB}\right)^2
-\hat{C}_\ell^{EE}\left(\hat{C}_\ell^{TB}\right)^2
-\hat{C}_\ell^{BB}\left(\hat{C}_\ell^{TE}\right)^2\,.    
\end{equation}
\end{widetext}

As a representative example for the next generation of CMB polarization experiments we consider Lite
(Light) satellite for the study of B-mode polarization and Inflation from cosmic background
Lite Background Radiation Detection (LiteBIRD) \cite{Hazumi:2019lys,LiteBIRD:2022cnt}, selected by the Japan Aerospace Exploration Agency (JAXA) as a strategic large class mission. 
In Tab.~\ref{TableLiteBIRD} we report the LiteBIRD-like experimental specifications that we use for our forecasts. 
We produce simulated data for $T,E$ by considering the inverse noise weighting of the central frequency channels in Tab.~\ref{TableLiteBIRD} and by assuming that the lowest  and highest frequencies are used to separate the foreground emission as done in \cite{LiteBIRD:2022cnt} (see also \cite{CORE:2016ymi}). For the $B$-mode polarization (in addition to the instrumental noise)  we include the following two sources of confusion: the lensing signal and a contribution which mimics the foreground residuals, as also done in \cite{Paoletti:2022anb}.
We compare these LiteBIRD-like noise power spectrum $N_\ell^{BB}$ with the signal induced by cosmic birefringence
in Fig.~\ref{plot::noiseBB}. With these settings we consider $f_{\mathrm sky}=0.7$ and $\ell_{max}=1350$.

As first step we consider which constant birefringence angle $\bar\alpha$ could be detected with this LiteBIRD-like configuration.
We consider a 
covariance matrix $\bar{C}$ 
obtained 
with $\bar{\alpha}=\left\{1,0.5,0.35,0.2,0.1,0.01\right\}$ deg 
and using the power spectra obtained from of Eqs.~\eqref{C_ll_TE_constant} - \eqref{C_ll_EB_constant}
we estimated for some values of $\chi^2_\mathrm{eff}$, see Tab.~\ref{TableCHI:const} .
The observed power spectra $\hat{C}_\ell$ correspond to the case without cosmic birefringence ($\alpha=0$).
We add, both to the theoretical and to the observed power spectra, the noise power spectra $N_\ell^{TT}$, $N_\ell^{EE}$ and $N_\ell^{BB}$ from Eq.~\eqref{noise:XX};
for $C_\ell^{BB}$ we consider also the contribution of lensing and foregrounds.

\begin{table}[htbp]
\centering
\begin{tabular}{|c|c||c|}
\hline
$\bar{C}_\ell$ theoretical ($\bar{\alpha}$)+$N_\ell$ & $\hat{C}_\ell$ observed+$N_\ell$ & $\chi^2_\mathrm{eff}$  \\
\hline
\hline
$C_\ell (\bar{\alpha}=1 \,{\rm deg})$ & $C_\ell (\alpha=0 \,{\rm deg})$ & $3.03\times 10^{4}$ \\
\hline
$C_\ell (\bar{\alpha}=0.5 \,{\rm deg})$ & $C_\ell (\alpha=0 \,{\rm deg})$ & $7.57\times 10^{3}$\\
\hline
$C_\ell (\bar{\alpha}=0.35 \,{\rm deg})$  & $C_\ell (\alpha=0 \,{\rm deg})$ &$3.71\times 10^{3}$\\
\hline
$C_\ell (\bar{\alpha}=0.2 \,{\rm deg})$ & $C_\ell (\alpha=0 \,{\rm deg})$ & $1.12\times 10^{3}$ \\
\hline
$C_\ell (\bar{\alpha}=0.1 \,{\rm deg})$ & $C_\ell (\alpha=0 \,{\rm deg})$ & $3.03\times 10^{2}$ \\
\hline
$C_\ell (\bar{\alpha}=0.01 \,{\rm deg})$ & $C_\ell (\alpha=0 \,{\rm deg})$ & $3.03$\\
\hline
\end{tabular}
\caption{\footnotesize\label{TableCHI:const} $\chi^2_\mathrm{eff}$, see Eq. \eqref{chi2}, for different values of $\bar{\alpha}$: 
the theoretical power spectra are obtained using the analytic approximation
 of Eqs.~\eqref{C_ll_TE_constant} - \eqref{C_ll_EB_constant}
and they are compared to the un-rotated case ($\alpha=0$). 
We assume $f_\mathrm{sky}=0.7$ and $\ell_{max}=1350$ for LiteBIRD.}
\end{table}

\subsection{Limits for axion-like as Early Dark Energy}

We find that an axion-like field acting as Early Dark Energy (EDE) could produce
a signal similar to the detection of $\bar \alpha = 0.35$~deg \cite{Minami:2020odp}
by taking into account the redshift dependence of the scalar field with:
\begin{equation} 
 g_\phi \sim 1.65 \times 10^{-18}\mathtt{GeV}^{-1}\,,   
\end{equation}
assuming, as in Section~\ref{Sect:III:EDE}, $n=2$, $\Lambda=0.417$ eV, $f=0.05\, M_\mathrm{pl}$,
$\Theta_i=1$ and $\dot{\Theta}_i=0$. We determine this value of $g_\phi$ by finding the $C_\ell$ obtained when taking into account the
redshift dependence of the axion-like field acting as EDE which best mimics the
$C_\ell^\mathrm{ const} (\bar \alpha = 0.35\,{\rm deg})$ in Eq.~(\ref{C_ll_TE_constant})-(\ref{C_ll_EB_constant}) by considering a minimization of $\Delta \chi^2_\mathrm{eff}$ in presence
of the lensing BB.

Let us now turn to future experiments such as LiteBIRD.
It is important to note that a LiteBIRD-like experiment can in principle distinguish between the
EDE signal induced by $g_\phi \sim 1.65 \times 10^{-18}\mathtt{GeV}^{-1}$ 
and $C_\ell^\mathrm{ const}  (\bar \alpha = 0.35\,{\rm deg})$ at very high statistical significance, with
$\chi^2_\mathrm{eff}=67.3$ 
according to Tab.~\ref{TableCHI:EDE}.
This capability of future experiments
opens up the possibility to understand the physical mechanism of cosmological birefringence.

From Tab.~\ref{TableCHI:EDE} 
we retrieve other two important information.
We report a value of $g_\phi \sim 1.4 \times 10^{-19}\mathtt{GeV}^{-1}$ as the smallest value of the coupling which can be distinguished
by a $C_\ell (\bar{\alpha})$ with $\bar{\alpha} = \alpha (\eta_\mathrm{rec})-\alpha(\eta_{0})$ 
in Eq. (\ref{eq:baralpha});
and 
$g_\phi \sim 6.0 \times 10^{-20}\mathtt{GeV}^{-1}$ as
the 95 \% upper bound which a LiteBIRD-like experiment as the one we adopt can achieve.

Our results improve those obtained in Tab.~I of \cite{Capparelli:2019rtn} for LiteBIRD:
$g_\phi\simeq1.45\times10^{-16}\mathtt{GeV}^{-1}$
(considering the power spectrum of the rotation angle $C_\ell^{\alpha\alpha}$)
and $g_\phi\simeq7.8\times10^{-17}\mathtt{GeV}^{-1}$
(considering the cross-correlation between the rotation angle and the temperature  $C_\ell^{\alpha T}$).
See also the constraints  for axionlike particles acting as Early Dark Energy discussed in \cite{Fujita:2020ecn}.

\begin{table}[htbp]
\centering
\begin{tabular}{|c|c||c|}
\hline
$\bar{C}_\ell$ theoretical (EDE) +$N_\ell$ & $\hat{C}_\ell$ observed +$N_\ell$ & $\chi^2_\mathrm{eff}$  \\
\hline
\hline
$C_\ell ( g_\phi= 8.17\times10^{-18}\mathtt{GeV}^{-1})$ & $C_\ell (\alpha=0 \,{\rm deg})$ 
& $1.10\times 10^{5}$\\
\hline
$C_\ell (g_\phi=1.51\times10^{-18}\textbf{}\mathtt{GeV}^{-1})$ & $C_\ell (\alpha=0 \,{\rm deg})$ 
& $3.81\times 10^{3}$\\
\hline
$C_\ell ( g_\phi=4.35\times10^{-19}\mathtt{GeV}^{-1})$ & $C_\ell (\alpha=0 \,{\rm deg})$ 
& $3.21\times 10^{2}$\\
\hline
$C_\ell ( g_\phi=3.5\times10^{-19}\mathtt{GeV}^{-1})$ & $C_\ell (\alpha=0 \,{\rm deg})$ 
& $2.09\times 10^{2}$\\
\hline
$C_\ell ( g_\phi=6.0\times10^{-20}\mathtt{GeV}^{-1})$ & $C_\ell (\alpha=0 \,{\rm deg})$ 
& $10.5$\\
\hline
\hline
$C_\ell ( g_\phi=1.65\times10^{-18}\mathtt{GeV}^{-1})$ & $C_\ell (\bar{\alpha}=0.35 \,{\rm deg})$ 
& $67.3$\\
\hline
$C_\ell ( g_\phi=1.4\times10^{-19}\mathtt{GeV}^{-1})$ & $C_\ell (\bar{\alpha}=0.02 \,{\rm deg})$ 
& $9.62$\\
\hline
\end{tabular}
\caption{\footnotesize\label{TableCHI:EDE} 
Early Dark energy: $\chi^2_\mathrm{eff}$, see Eq. \eqref{chi2}, for different values of $g_\phi$, 
fixed $n=2$, $\Lambda=0.417$ eV, $f=0.05\, M_\mathrm{ pl}=1.22\times 10^{17}$ GeV,
$\Theta_i=1$ and $\dot{\Theta}_i=0$.
The theoretical power spectra are obtained using the modified version of \texttt{CAMB}.
The observed power spectra correspond to the case without cosmic birefringence ($\alpha=0$),
except the last two lines where we consider a rotation $\bar{\alpha}$, see  Eqs.~\eqref{C_ll_TE_constant}-\eqref{C_ll_EB_constant}.
We assume $f_\mathrm{sky}=0.7$ and $\ell_{max}=1350$ for LiteBIRD.}
\end{table}

\subsection{Limits for axion-like for dark energy}

In the case of axion-like field acting as dark energy
we find that we should consider a coupling constant of the order:
\begin{equation}
g_\phi\sim-1.8\times10^{-20}\mathtt{GeV}^{-1}\,,    
\end{equation}
in order to best mimic a birefringence signal
$C_\ell^\mathrm{ const} (\bar \alpha = 0.35\,{\rm deg})$ \cite{Minami:2020odp},
assuming $M=1.95\times10^{-3}$ eV, $f=0.265 M_\mathrm{pl}$,
$\Theta_i=0.25$ and $\dot{\Theta}_i=0$, 
as in Section~\ref{Sect:III:DE}.
We always consider a minimization of $\Delta \chi^2_\mathrm{eff}$ in presence of the lensing BB.

In this dark energy case the pseudoscalar field $\phi$ becomes dynamic at late times and therefore 
the linear polarization angle rotates at low redshift. 
The power spectra are quite similar to those obtained using  
the analytic approximation of Eqs.~\eqref{C_ll_TE_constant} - \eqref{C_ll_EB_constant}.
The smallest coupling $g_\phi$ that a LiteBIRD-like experiment can distinguish 
from $C_\ell^\mathrm{ const} (\bar \alpha)$ is 
$\sim8.0\times10^{-19}\mathtt{GeV}^{-1}$,
always assuming  $\bar{\alpha} = \alpha (\eta_\mathrm{rec})-\alpha(\eta_{0})$. 
As 95\% upper bound we find 
$g_\phi\sim9.0\times10^{-22}\mathtt{GeV}^{-1}$
(see Tab.~\ref{TableCHI:DE}).

\begin{table}[htbp]
\centering
\begin{tabular}{|c|c||c|}
\hline
$\bar{C}_\ell$ theoretical (DE) +$N_\ell$ & $\hat{C}_\ell$ observed +$N_\ell$ & $\chi^2_\mathrm{eff}$ \\
\hline
\hline
$C_\ell (g_\phi=1.8\times10^{-20}\mathtt{GeV}^{-1})$ & $C_\ell (\alpha=0 \,{\rm deg})$ 
& $3.78\times 10^{3}$\\
\hline
$C_\ell (g_\phi=5.2\times10^{-21}\mathtt{GeV}^{-1})$ & $C_\ell (\alpha=0 \,{\rm deg})$ 
& $3.15\times 10^{2}$\\
\hline
$C_\ell (g_\phi=7.2\times10^{-21}\mathtt{GeV}^{-1})$ & $C_\ell (\alpha=0 \,{\rm deg})$ 
& $6.04\times 10^{2}$\\
\hline
$C_\ell (g_\phi=9.0\times10^{-22}\mathtt{GeV}^{-1})$ & $C_\ell (\alpha=0 \,{\rm deg})$ 
& $9.4$\\
\hline
\hline
$C_\ell (g_\phi=8.0\times10^{-19}\mathtt{GeV}^{-1})$ & $C_\ell (\bar{\alpha}=-15.7 \,{\rm deg})$ 
& $9.8$\\
\hline
$C_\ell (g_\phi=1.8\times10^{-20}\mathtt{GeV}^{-1})$ & $C_\ell (\bar{\alpha}=-0.35 \,{\rm deg})$ 
& $0.30$\\
\hline
\end{tabular}
\caption{\footnotesize\label{TableCHI:DE} 
Axion like dark energy: $\chi^2_\mathrm{eff}$, see Eq. \eqref{chi2},  for different values of $g_\phi$, 
fixed $M=1.95\times10^{-3}$ eV, $f=0.265 M_\mathrm{pl}$,
$\Theta_i=0.25$ and $\dot{\Theta}_i=0$.
The theoretical power spectra are obtained using the modified version of \texttt{CAMB}.
The observed power spectra correspond to the case without cosmic birefringence ($\alpha=0$),
except the last two lines where we consider a rotation $\bar{\alpha}$, see  Eqs.~\eqref{C_ll_TE_constant}-\eqref{C_ll_EB_constant}.
We assume $f_\mathrm{sky}=0.7$ and $\ell_{max}=1350$ for LiteBIRD.}
\end{table}

\subsection{Limits for axion-like for dark matter}

For a pseudoscalar field acting as dark matter 
we find that 
\begin{equation}
\label{limit:DM22}
 g_\phi\sim1.37\times10^{-14}\mathtt{GeV}^{-1}\,,  
\end{equation}
is needed to reproduce a signal similar to $\bar \alpha = 0.35$~deg \cite{Minami:2020odp} for $m=10^{-22}\mathtt{eV}$,
as in Section~\ref{Sect:III:DM}, 
by considering a minimization of $\Delta \chi^2_\mathrm{eff}$ in presence
of the lensing BB.

A LiteBIRD-like experiment can easily distinguish between a birefringence signal induced by dark matter
with a coupling constant of this order of magnitude and  $C_\ell^\mathrm{ const}  (\bar \alpha = 0.35\,{\rm deg})$ at very high statistical significance
$\chi^2_\mathrm{eff}=69.8$,
see Tab.~\ref{TableCHI:DM}.

From this Table we report also a value of 
$g_\phi \sim 7.5 \times 10^{-15}\mathtt{GeV}^{-1}$ as the smallest value of the coupling which can be distinguished
by a $C_\ell (\bar{\alpha})$ with $\bar{\alpha} = \alpha (\eta_\mathrm{rec})-\alpha(\eta_{0})$ 
in Eq.~(\ref{eq:baralpha});
and 
$g_\phi \sim 8.1 \times 10^{-16}\mathtt{GeV}^{-1}$ as
the 95 \% upper bound which a LiteBIRD-like experiment as the one we adopt can achieve.

\begin{figure*}[!htb]
\centering
\includegraphics[width=300pt]{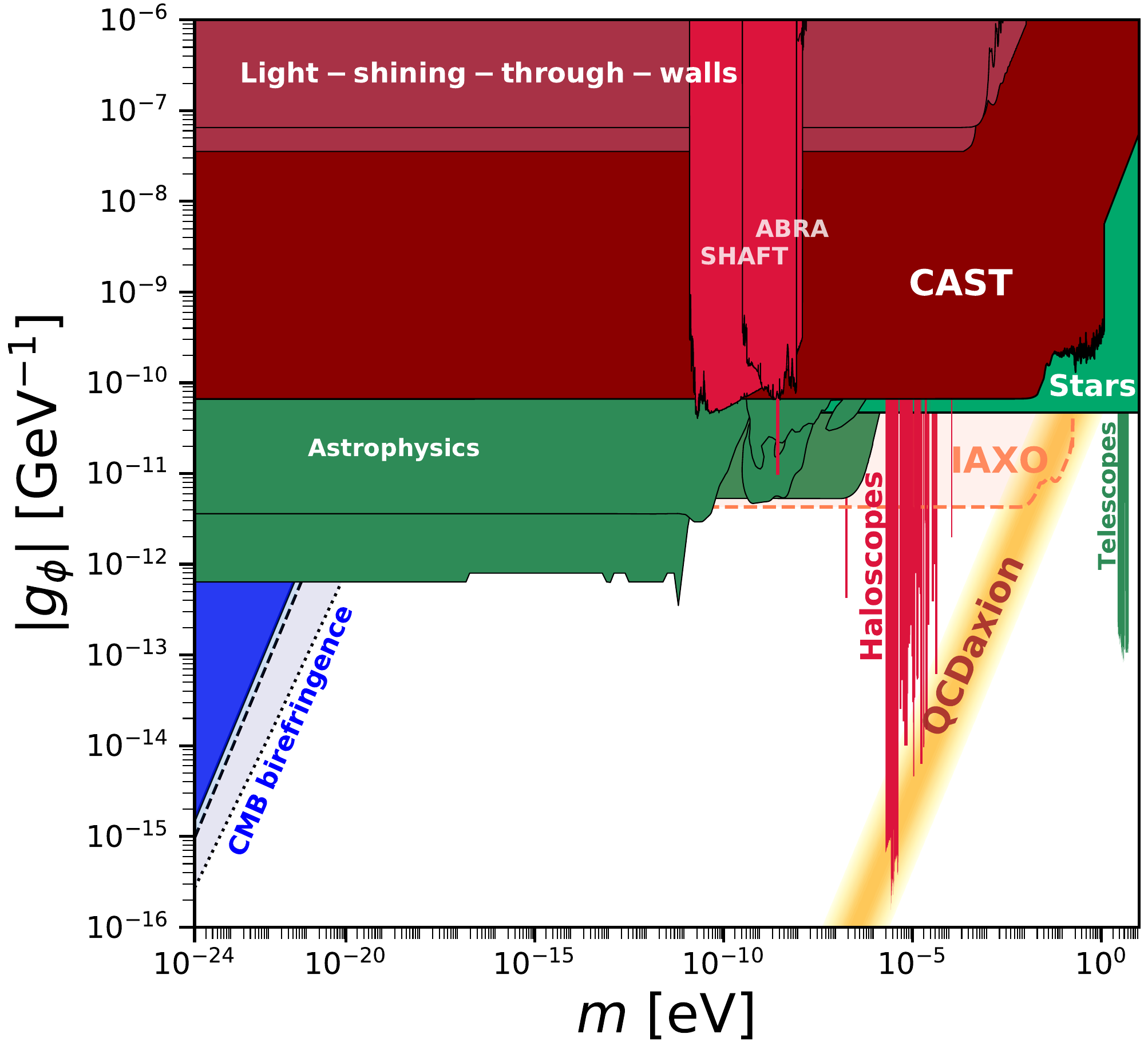}
\caption{\footnotesize\label{Fig:axionLimits} 
Limits on the coupling constant with photons $g_\phi$ 
for pseudoscalar field acting as dark matter as a function of mass $m$ 
(coloured regions are excluded). 
The light blue dotted line corresponds to CMB birefringence of the order of 0.35~deg \cite{Minami:2020odp} obtained by taking into account the redshift dependency of the birefringence angle, compared to CMB birefringence limits presented in \cite{Finelli:2008jv} (dark blue region) and in \cite{Fedderke:2019ajk} (blue dashed line).
Plot created with the \texttt{AxionLimits} code \cite{AxionLimits}, we refer to online documentation for references on the other constraints.
}
\end{figure*}

The above results depend on the mass of the axion. If we consider a heavier mass for the pseudoscalar field, as $m=10^{-20}$ eV, 
the coupling that mimics  $C_\ell^\mathrm{ const} (\bar \alpha = 0.35\,{\rm deg})$ 
is smaller than the one reported in Eq.~\eqref{limit:DM22} and is:
\begin{equation}
\label{limit:DM20}
 g_\phi\sim8.0\times10^{-13}\mathtt{GeV}^{-1}\,,  
\end{equation}
On the contrary, for smaller masses (e.g. $m=10^{-20}$ eV), we have to consider a smaller 
coupling constant of the order:
\begin{equation}
\label{limit:DM24}
 g_\phi\sim2.7\times10^{-16}\mathtt{GeV}^{-1}\,.  
\end{equation}
In Fig.~\ref{Fig:axionLimits}  we compare the limits on the axion-photon coupling 
obtained from isotropic cosmic birefringence with the other limits present in literature \cite{ParticleDataGroup:2022pth,AxionLimits}. 
CMB cosmic birefringence nicely complements 
other experimental/astrophysical tests \cite{Galaverni:2022eoy}. 

\begin{table}[htbp]
\centering
\begin{tabular}{|c|c||c|}
\hline
$\bar{C}_\ell$ theoretical (DM) +$N_\ell$ & $\hat{C}_\ell$ observed +$N_\ell$ & $\chi^2_\mathrm{eff}$  \\
\hline
\hline
$C_\ell (g_\phi=1.5\times 10^{-14}\mathtt{GeV}^{-1})$ & $C_\ell (\alpha=0 \,{\rm deg})$ 
& $3.75\times 10^{3}$\\
\hline
$C_\ell (g_\phi=1.0\times 10^{-14}\mathtt{GeV}^{-1})$ & $C_\ell (\alpha=0 \,{\rm deg})$ 
& $1.70\times 10^{4}$\\
\hline
$C_\ell (g_\phi=4.3\times10^{-15}\mathtt{GeV}^{-1})$ & $C_\ell (\alpha=0 \,{\rm deg})$
& $3.05\times 10^{2}$\\
\hline
$C_\ell (g_\phi=2.0\times10^{-15}\mathtt{GeV}^{-1} )$ & $C_\ell (\alpha=0 \,{\rm deg})$ 
& $66.0$\\
\hline
$C_\ell (g_\phi=8.1\times10^{-16}\mathtt{GeV}^{-1} )$ & $C_\ell (\alpha=0 \,{\rm deg})$ 
& $10.4$\\
\hline
\hline
$C_\ell (g_\phi=1.37\times10^{-14}\mathtt{GeV}^{-1})$ &$C_\ell (\bar{\alpha}=0.35 \,{\rm deg})$
& $69.8$\\
\hline
$C_\ell (g_\phi=7.5\times10^{-15}\mathtt{GeV}^{-1})$ &$C_\ell (\bar{\alpha}=0.17 \,{\rm deg})$
& $10.4$ \\ 
\hline
\end{tabular}
\caption{\footnotesize\label{TableCHI:DM} 
Axion like Dark Matter: $\chi^2_\mathrm{eff}$, see Eq. \eqref{chi2},  for different values of $g_\phi$, 
fixed $m=10^{-22}\mathtt{eV}$.
The theoretical power spectra are obtained using the modified version of \texttt{CAMB}.
The observed power spectra correspond to the case without cosmic birefringence ($\alpha=0$),
except the last two lines where we consider a rotation $\bar{\alpha}$, , see  Eqs.~\eqref{C_ll_TE_constant}-\eqref{C_ll_EB_constant}.
We assume $f_\mathrm{sky}=0.7$ and $\ell_{max}=1350$ for LiteBIRD.}
\end{table}

\subsection{ Markov Chain MonteCarlo results}

\begin{figure}[htb]
\includegraphics[width=\columnwidth]{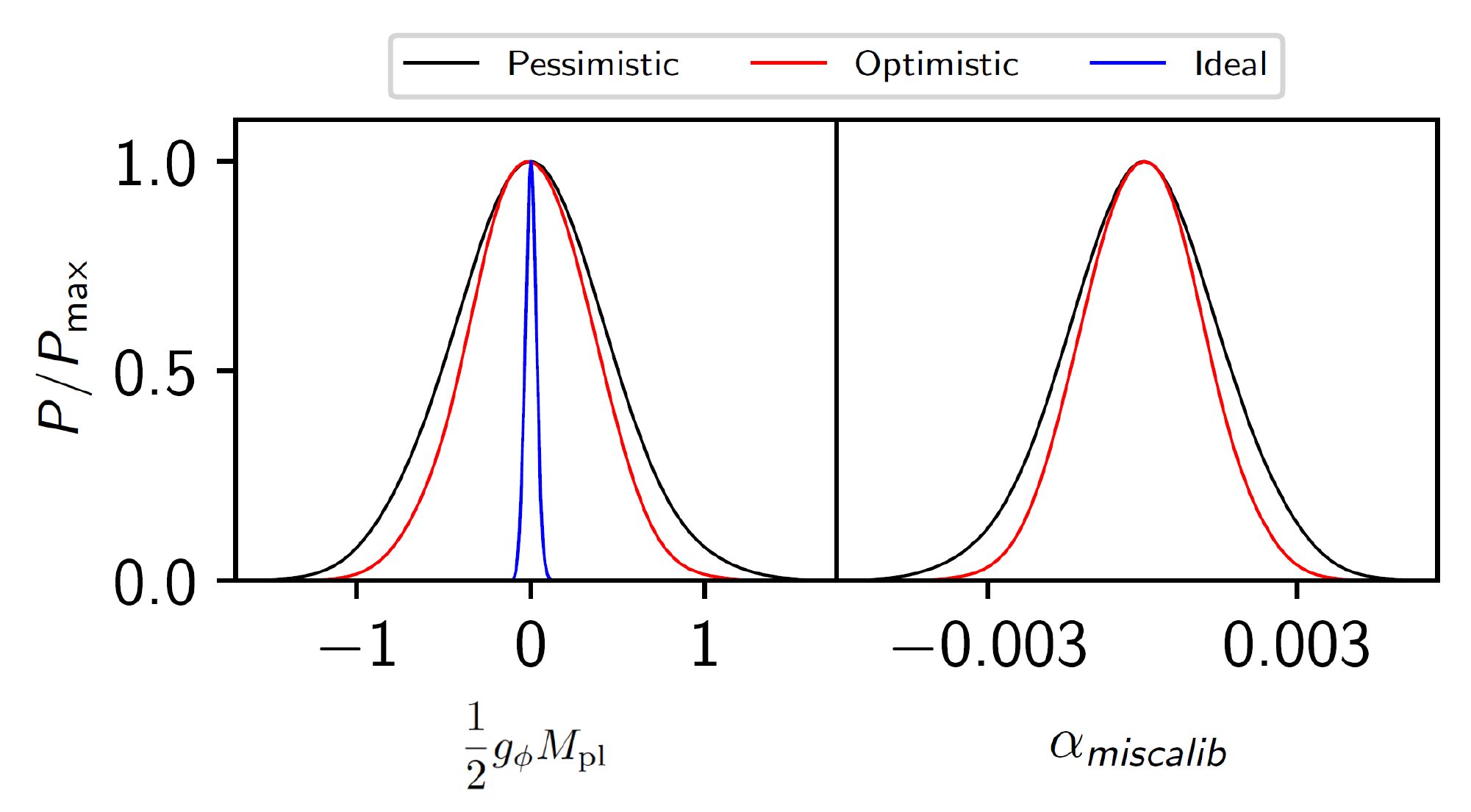}
\caption{\footnotesize\label{Fig:CosmoMC_1} One dimensional posterior distribution for the coupling and the miscalibration angle. In blue the ideal case where only the coupling is considered. In red the optimistic case with the miscalibration angle and in black the pessimistic case. 
}
\end{figure}

We perform few exploratory runs exploring the whole cosmological and birefringence parameter space using the MCMC code \texttt{cosmomc} \cite{Lewis:2002ah,Lewis:2013hha}. 
We use the exact Wishart likelihood with mock data generated following Eq. (\ref{chi2}), with an effective sky fraction of 70\% for all channels T, E, B and including the same instrumental noise (see Fig.~\ref{plot::noiseBB}) and foreground residuals in BB. In order to include the full contribution of cosmic birefringence we extended the standard exact likelihood to include also the odd cross correlators TB and EB. 
We perform first an idealistic case where we vary only the coupling together with the six standard parameters of the $\Lambda$CDM:dark matter density $\Omega_c h^2$,  baryon density $\Omega_{\mathrm B} h^2$, angular diameter distance to the last scattering surface $\theta$, optical depth $\tau$, the scalar spectral index $n_s$ and the amplitude of primordial fluctuations $A_s$. The resulting posterior probability distribution is shown in blue Fig.~\ref{Fig:CosmoMC_1} and the 68\% error bar is
$\sigma(g_\phi M_{\mathrm{pl}}/2)=0.032$ with a fiducial $g_\phi=0$
(corresponding to $|g_\phi|\lesssim 5.3\times10^{-20}\mathtt{GeV}^{-1}$ at 2$\sigma$).
 Note that the degradation of a factor of few in the constraints in $g_\phi$ with respect to those quoted in Section IV is due to the variation of all the cosmological parameters in the MCMC exploration. 

The result above mentioned represents an ideal case because we have not included the uncertainty due to the miscalibration angle related to the uncertainty in the calibration of polarization angles. In order to account for such uncertainty we add an isotropic rotation of the spectra due to an isotropic angle that we call $\alpha_{miscalib}$. 
This mimics the confusion created by not knowing the calibration angle when the birefringence signal arrives at the detectors. We vary this additional parameter assuming a Gaussian prior. We consider two cases: optimistic with a width of the prior of 
0.00175~rad  ($=0.1$~deg= 6~arcmin) 
and pessimistic with a width of the prior of 0.0035~rad ($=0.2$~deg= 12~arcmin). 
The resulting one dimensional posteriors are presented in red and black in Fig.~\ref{Fig:CosmoMC_1}, we note how with respect to the blue ideal curve we have a slight degradation of the constraints on the coupling whose uncertainty increase by roughly one order of magnitude to 
$\sigma(g_\phi M_{\mathrm{pl}}/2)=0.35$
(corresponding to $|g_\phi|\lesssim 5.7\times10^{-19}\mathtt{GeV}^{-1}$ at 2$\sigma$)
for the optimistic and 
$\sigma(g_\phi M_{\mathrm{pl}}/2)=0.41$
(corresponding to $|g_\phi|\lesssim 6.7\times10^{-19}\mathtt{GeV}^{-1}$ at 2$\sigma$)
for the pessimistic case. 
Note that the constraints on $g_\phi$ obtained by taking into account the miscalibration angle degrade by roughly one order of magnitude.

\begin{figure}[htb]
\includegraphics[width=\columnwidth]{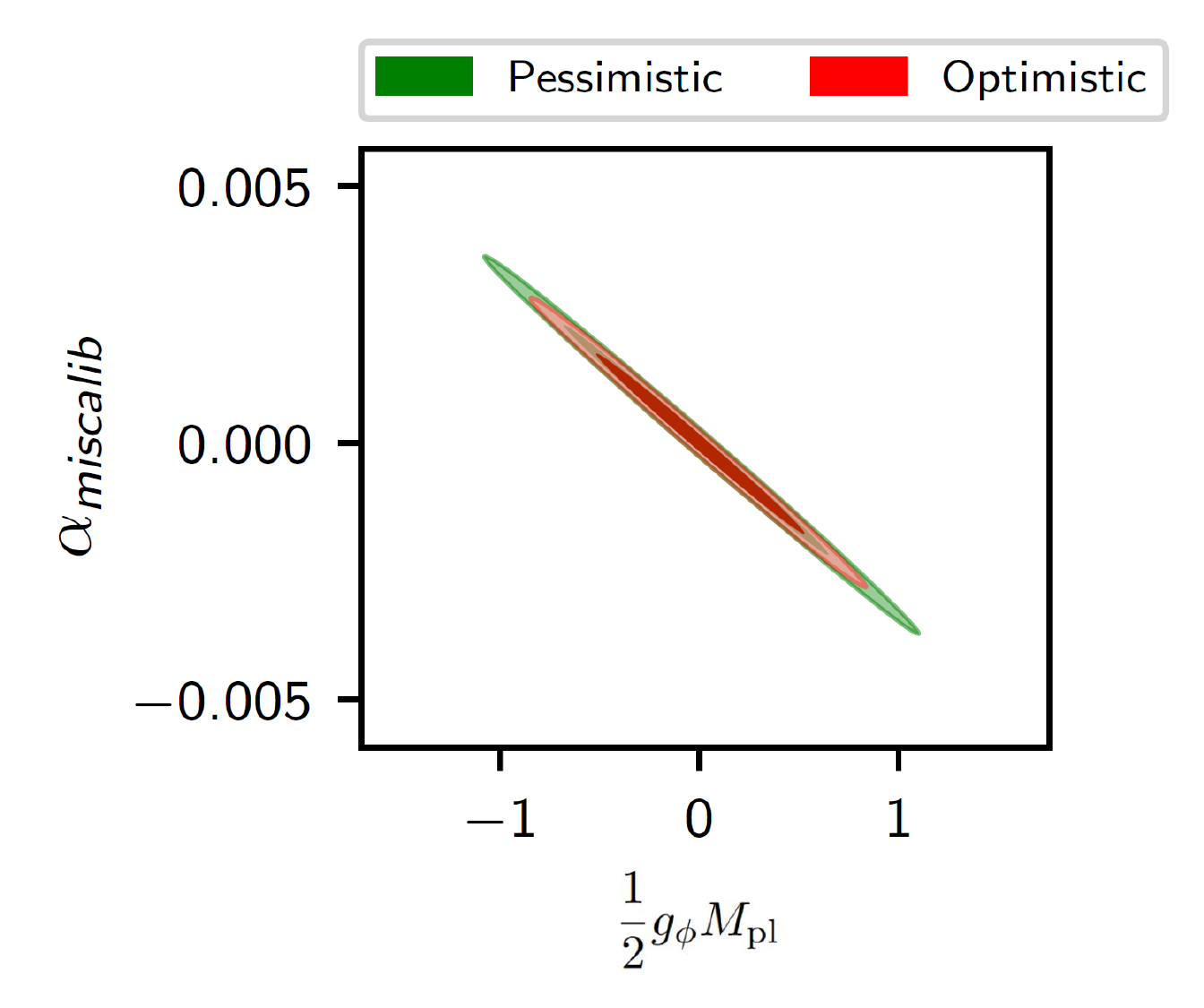}
\caption{\footnotesize\label{Fig:CosmoMC_2}Two dimensional posterior distribution for the miscalibration angle vs the coupling. In green the pessimistic case and in red the optimistic one. 
}
\end{figure}

In Fig.~\ref{Fig:CosmoMC_2} we show the correlation between the miscalibration angle and the coupling. 
Note that the degeneracy the miscalibration angle and the coupling is not exact as would be with the birefringence angle when the redshift dependence of the rotation angle is neglected.

\section{Conclusions}
\label{Sect:V}

We studied isotropic cosmological birefringence induced by a cosmological redshift-dependent pseudoscalar field 
with a coupling $ g_\phi \phi F^{\mu\nu} \tilde F_{\mu \nu}$. 
We showed how time evolution of the background pseudoscalar field
imprints in general a non-trivial multipole dependence in the observed CMB angular power spectra which is not captured by the
widely adopted approximation in which the redshift dependence of the rotation angle is neglected. 
This effect could be important in interpreting reported hints 
of birefringence in the Planck CMB spectra \cite{Minami:2020odp,Diego-Palazuelos:2022dsq,Eskilt:2022cff,Komatsu:2022nvu}.
Beyond considering phenomenological
redshift evolution for the rotation angle induced by a pseudo-scalar field, we also considered the theoretical prediction 
for Early Dark Energy, Quintessence and  axion-like matter.

As consequences, not only the total rotation
$\bar{\alpha}=\alpha (\eta_\mathrm{rec})-\alpha(\eta_{0})$ determines the final CMB spectra, but also when the rotation occurs compared to
the main changes in the visibility function, i.e. recombination and reionization. 
Moreover, non-vanishing parity violating effects occur 
also in the particular case $\alpha (\eta_\mathrm{rec})=\alpha(\eta_{0})$.

Due to the non-trivial multipole dependence induced by the redshift evolution of the pseudo-scalar field, the resulting isotropic birefringence 
is not degenerate with a polarization rotation angle independent on the multipoles, 
which is connected to a systematic calibration angle uncertainty.                                         

For the theoretical models of EDE, DE, and axion-like DM we estimated the size of the couplings
which will be detected by a LiteBIRD-like experiment by a $\chi^2$ calculation. Moreover, always for these models and by a $\chi^2$ calculation, 
we also computed at which level
our theoretical predictions can be distinguished by
the widely adopted approximation in which the redshift dependence of the rotation angle is neglected for a LiteBIRD-like experiment.
Finally, we have explicitly shown by MCMC the reduction of the degeneracy between the isotropic birefringence effect for Early Dark Energy and the miscalibration angle by allowing 
all the cosmological parameters to vary, always for a LiteBIRD-like experiment.

As a next step, we will add the effects due inhomogeneities in the pseudoscalar field, i.e. anisotropic birefringence, to complete the theoretical predictions 
of interesting models with a pseudo-scalar field, such as 
Early Dark Energy, Quintessence and  axion-like matter.

\begin{acknowledgments}
We would like to thank A.~Gruppuso and E.~Komatsu for useful discussions. FF and DP acknowledge financial support by ASI Grant 2016-24-H.0 and the agreement n. 2020-9-HH.0 ASI-UniRM2. 
We acknowledge the use of the INAF-OAS HPC cluster. 
\end{acknowledgments}

\appendix
\section{Additional phenomenological power spectra}
\label{App_1}

In this Appendix we discuss other phenomenological examples of redshift dependence of the linear polarization angle.

\begin{figure*}[!htb]
\includegraphics[width=168pt,height=4.5cm]{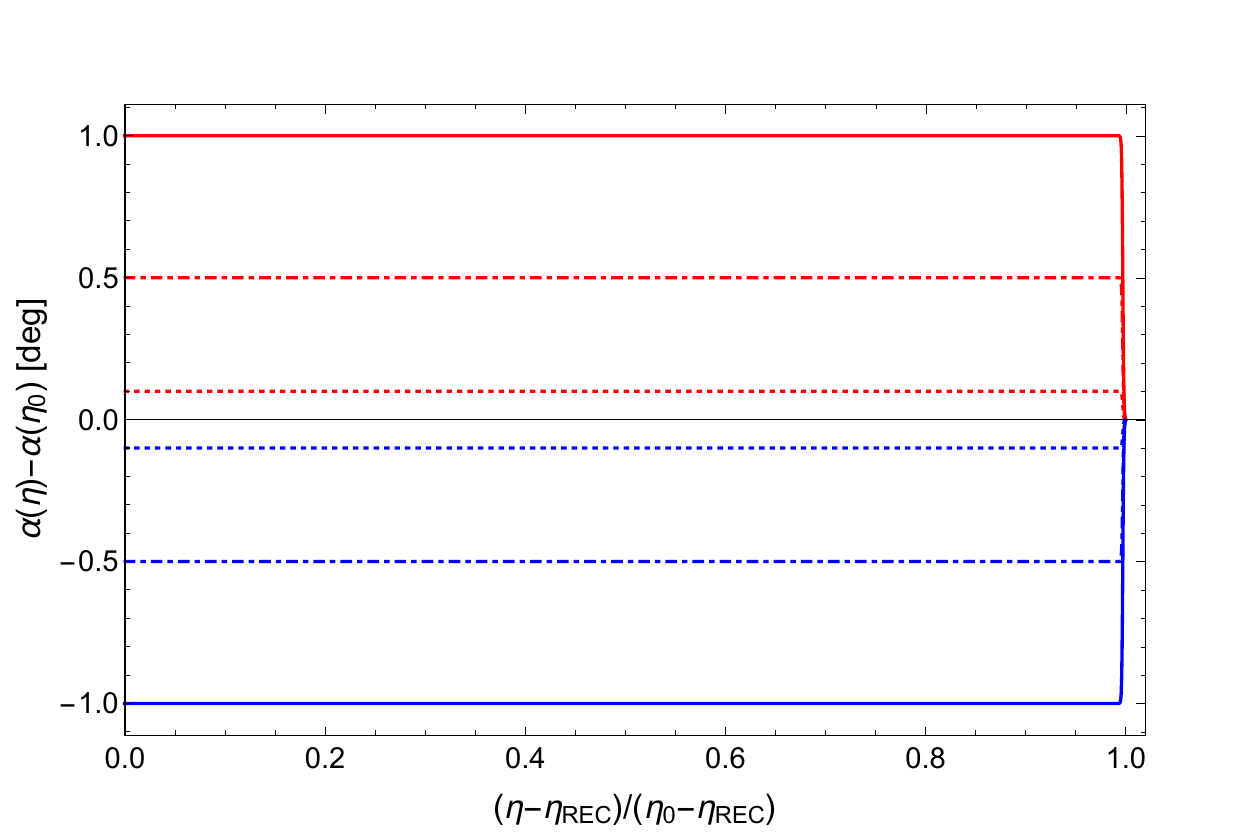}
\includegraphics[width=168pt,height=4cm]{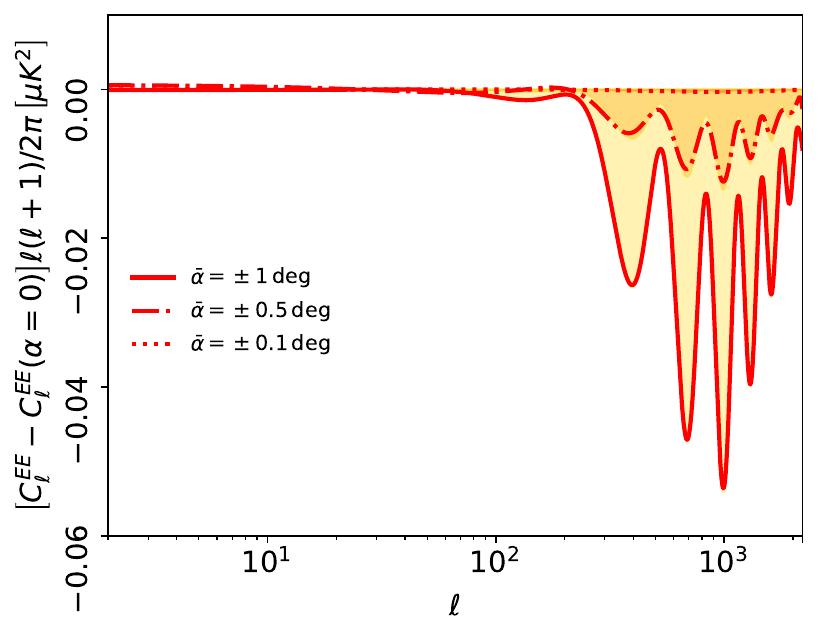}
\includegraphics[width=168pt,height=4cm]{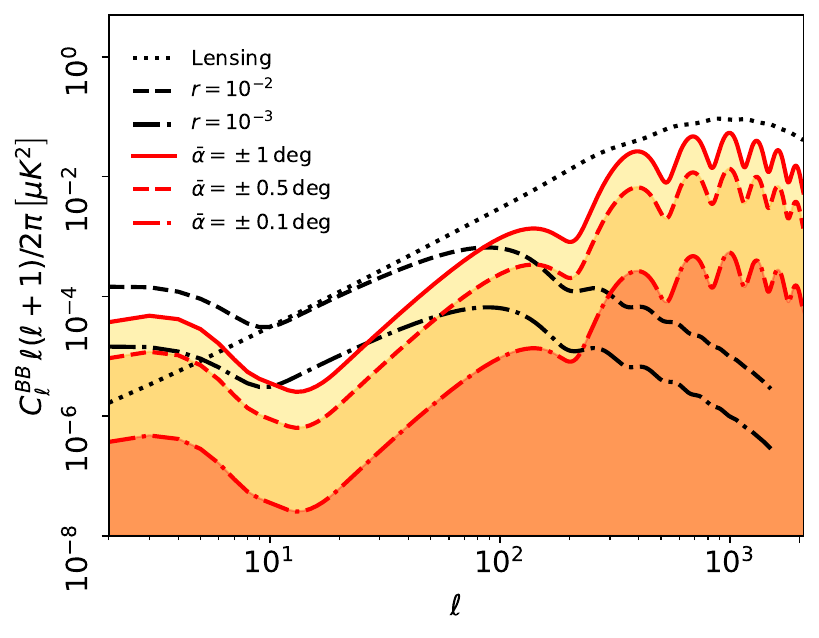}
\includegraphics[width=168pt,height=4cm]{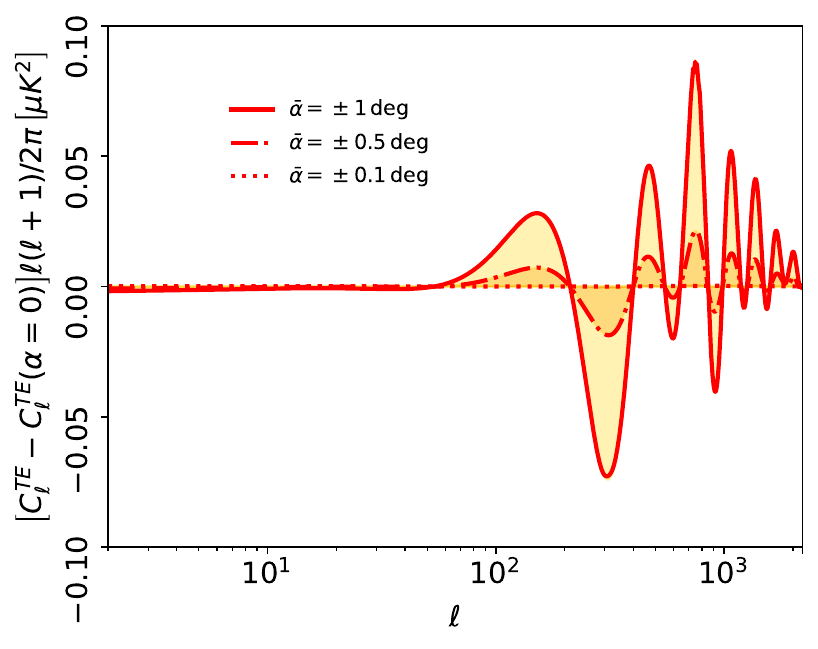}
\includegraphics[width=168pt,height=4cm]{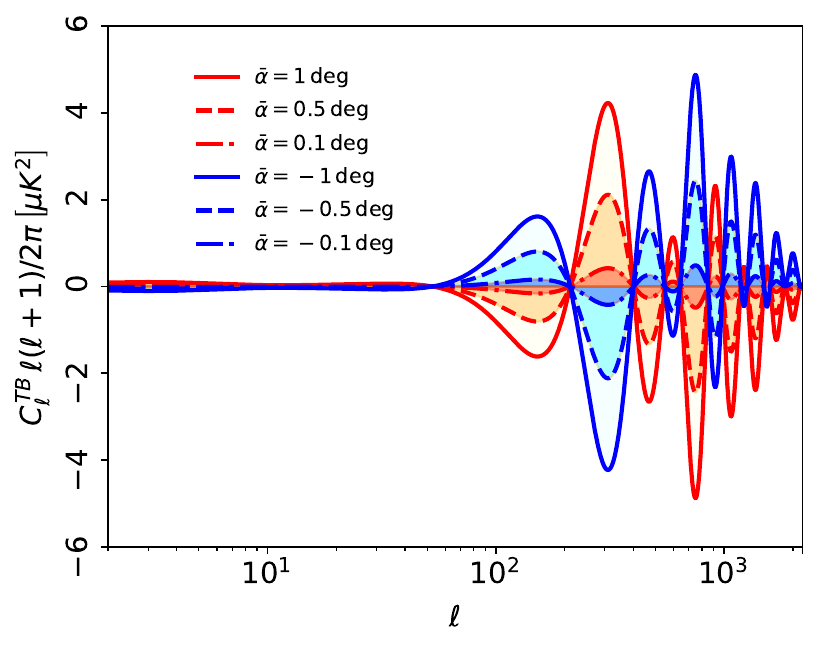}
\includegraphics[width=168pt,height=4cm]{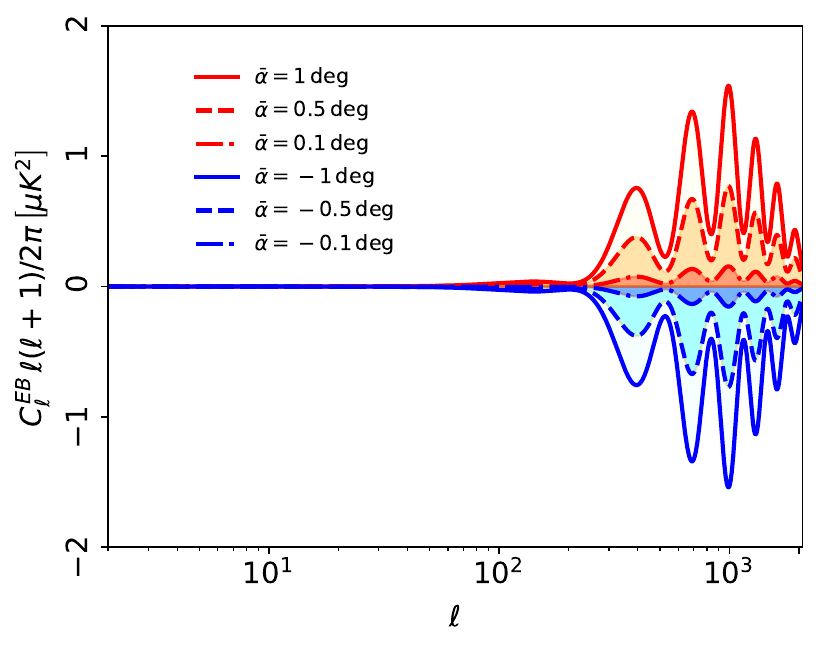}
\caption{\footnotesize\label{Fig:constant}
(a) Evolution of $\alpha(\eta)-\alpha(\eta_0)$ as a function of conformal time: 
during propagation $\bar\alpha=+/- 1$ deg (continue red/blue line), $+/- 0.5$~deg (dashed red/blue line),
$+/- 0.1$ deg (dotted red/blue line).
In this case the output of the modified \texttt{CAMB} code (coloured lines) coincides with 
the analytic approximations of Eqs.~(\ref{C_ll_TE_constant})-(\ref{C_ll_EB_constant}) (coloured regions):
(b) $C_\ell^{EE} - C_\ell^{EE}(\bar{\alpha}=0)$, (c) $C_\ell^{BB}$,
(d)  $C_\ell^{TE} - C_\ell^{TE}(\bar{\alpha}=0)$, (e) $C_\ell^{TB}$, (f) $C_\ell^{EB}$.
}
\end{figure*}

First we consider ``instantaneous rotation at present time'' (see Fig.~\ref{Fig:constant}).
In this case $\bar{\alpha}=\alpha (\eta_\mathrm{rec})-\alpha(\eta_{0})$ is exactly constant during
integration along the line-of-sight
therefore the power spectra obtained using the modified \texttt{CAMB} code (coloured lines) exactly coincide 
with the $C_\ell^\mathrm{obs}$ given by the analytic expressions 
of Eqs.~\eqref{C_ll_TE_constant}-\eqref{C_ll_EB_constant} (coloured regions).

\begin{figure*}[!htb]
\includegraphics[width=168pt,height=4.5cm]{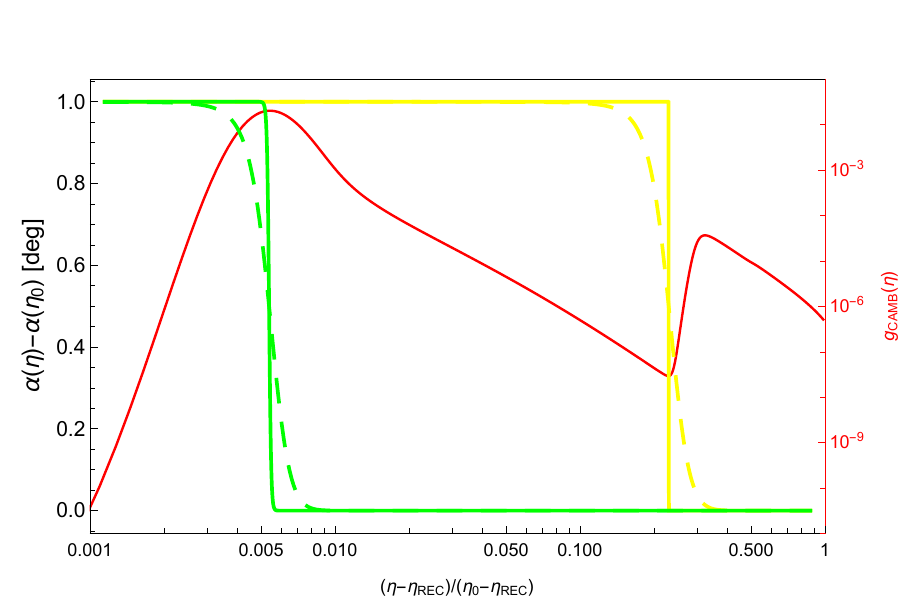}
\includegraphics[width=168pt,height=4cm]{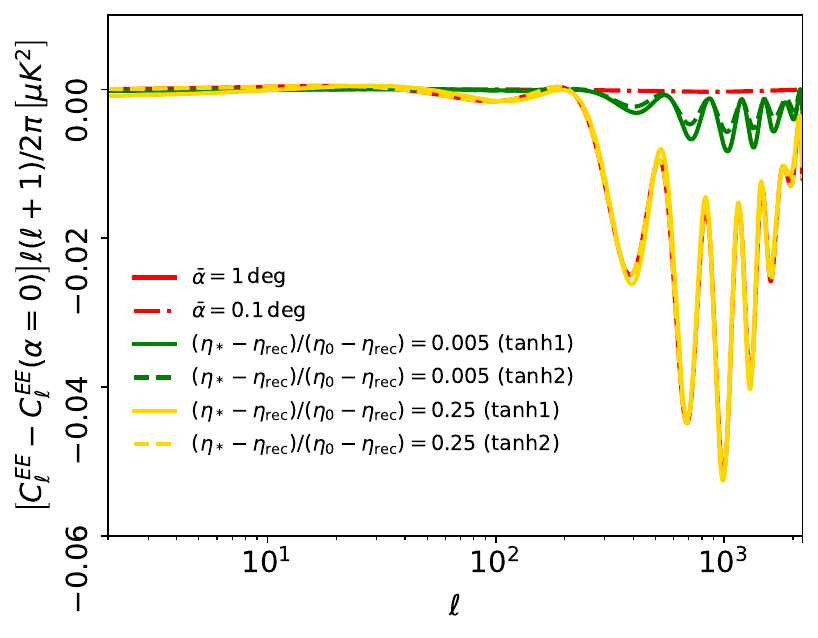}
\includegraphics[width=168pt,height=4cm]{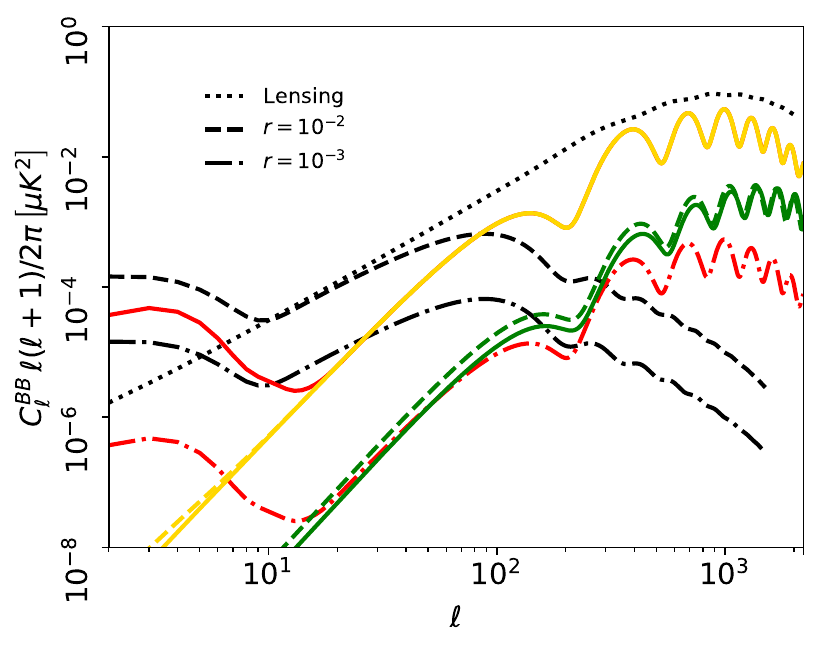}
\includegraphics[width=168pt,height=4cm]{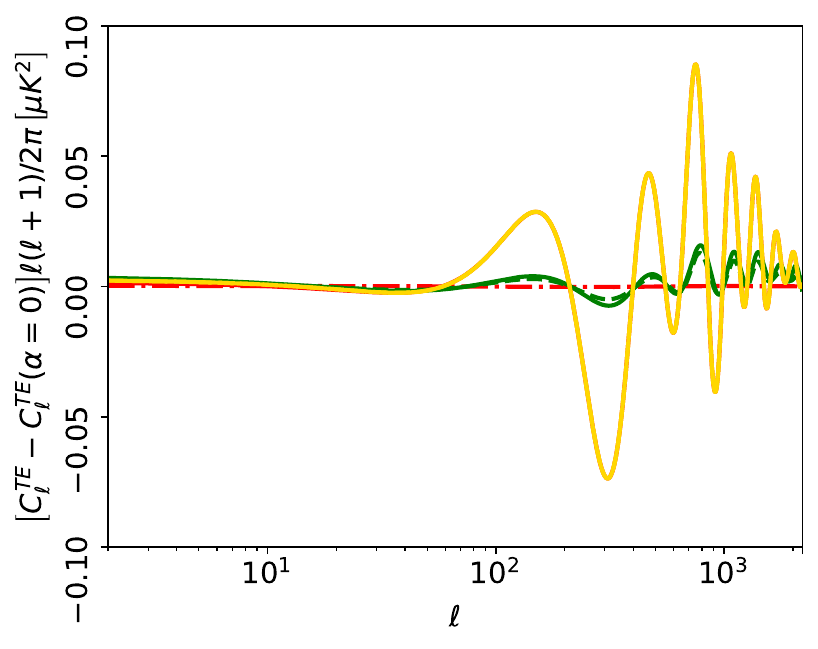}
\includegraphics[width=168pt,height=4cm]{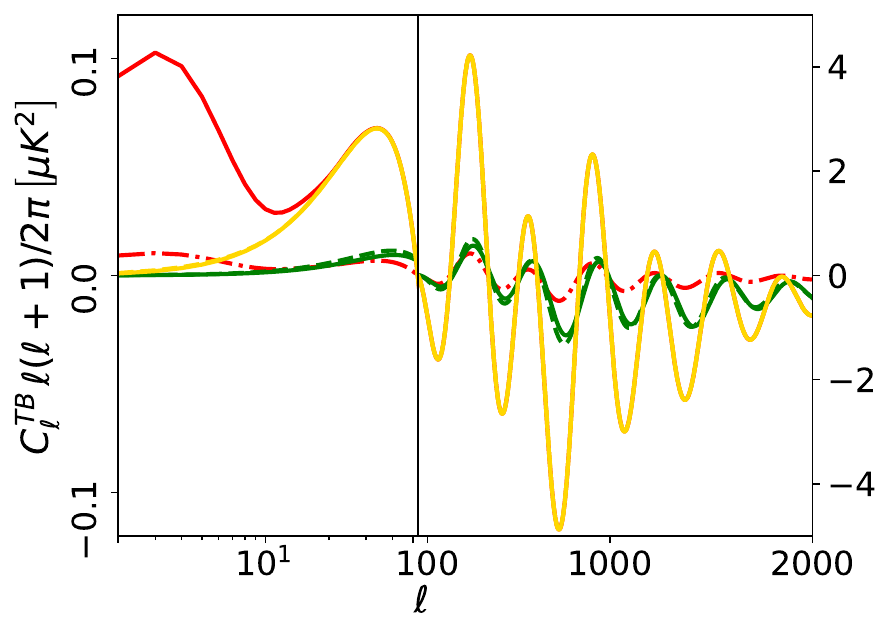}
\includegraphics[width=168pt,height=4cm]{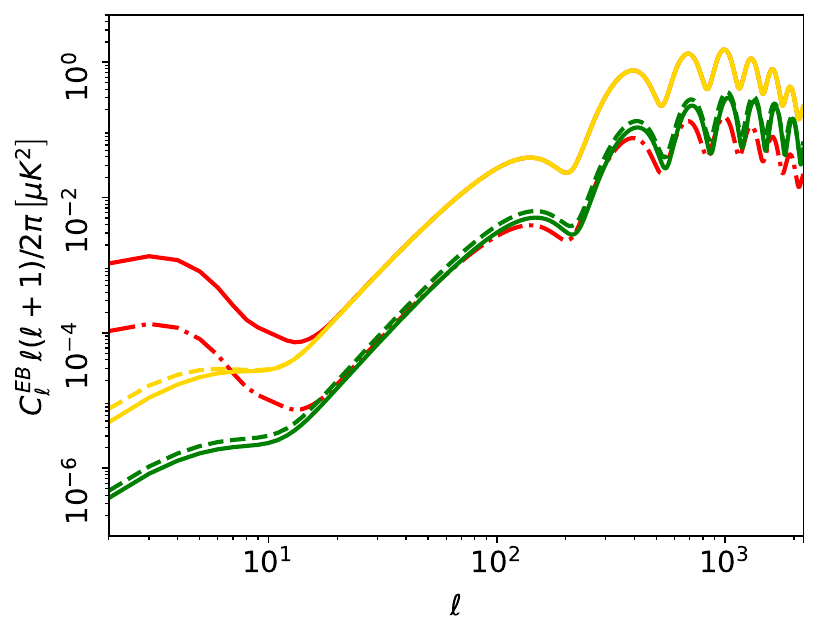}
\caption{\footnotesize\label{Fig:test_tanhevol:theta:120}  
Comparison between different hyperbolic tangents:
(a) evolution of $\alpha(\eta)-\alpha(\eta_0)$ as a function of conformal time $x\equiv\left(\eta-\eta_{\rm rec}\right)/\left(\eta_0-\eta_{\rm rec}\right)$:
$\alpha(\eta)=- 1/2\left\{1+\tanh\left[10^4 (x-0.005)\right]\right\}$ (green continuous line),
$\alpha(\eta)=- 1/2\left\{1+\tanh\left[10^3 (x-0.005)\right]\right\}$ (green dotted line),
$\alpha(\eta)=- 1/2\left\{ 1+\tanh\left[10^4 (x-0.25)\right]\right\}$ (yellow dotted line),
$\alpha(\eta)=- 1/2\left\{ 1+\tanh\left[20 (x-0.25)\right]\right\}$ (yellow dotted line)
- the \texttt{CAMB} visibility function $g_\mathrm{CAMB}$ is plotted in red (on a different scale); 
angular power spectra obtained with the modified version of \texttt{CAMB} are compared in
(b) $C_\ell^{EE} - C_\ell^{EE}(\bar{\alpha}=0)$, (c) $C_\ell^{BB}$,
(d)  $C_\ell^{TE} - C_\ell^{TE}(\bar{\alpha}=0)$, (e) $C_\ell^{TB}$, (f) $C_\ell^{EB}$.
}
\end{figure*}

The comparison between Fig.~\ref{Fig:tanhevol:theta:120} and  Fig.~\ref{Fig:test_tanhevol:theta:120} 
assures us that the effects on the power spectra
are not dominated by the slope of the tanh function describing the transition between two different values of the linear polarization angle. Instead it is very important when this transition occurs: earlier in time the rotation happens, 
smaller are the effects on the power spectra.

\begin{figure*}[!htb]
\includegraphics[width=178pt,height=4.5cm]{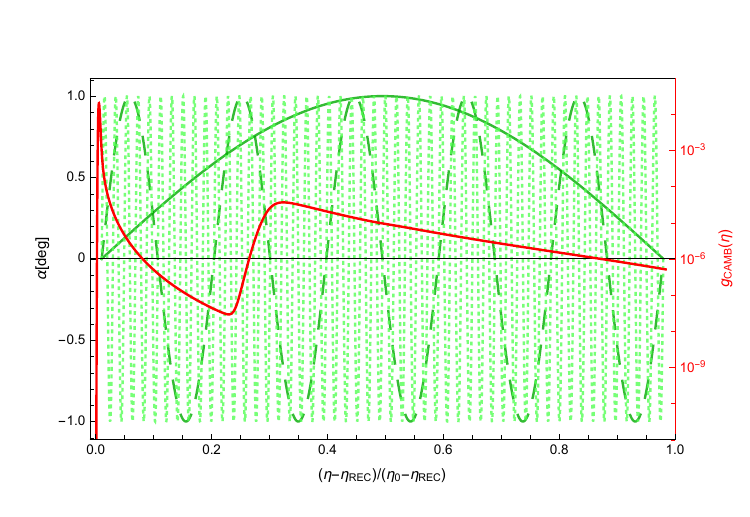}
\includegraphics[width=160pt,height=4cm]{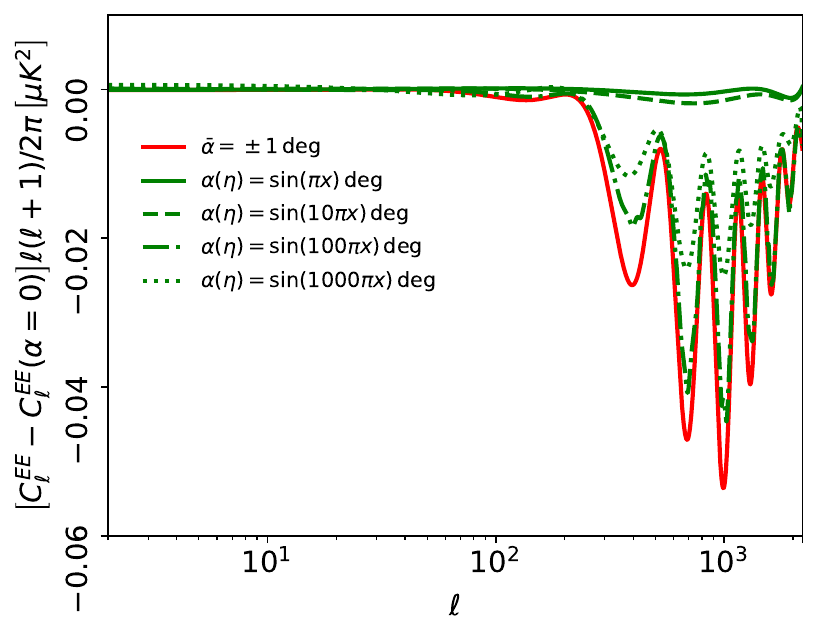}
\includegraphics[width=160pt,height=4cm]{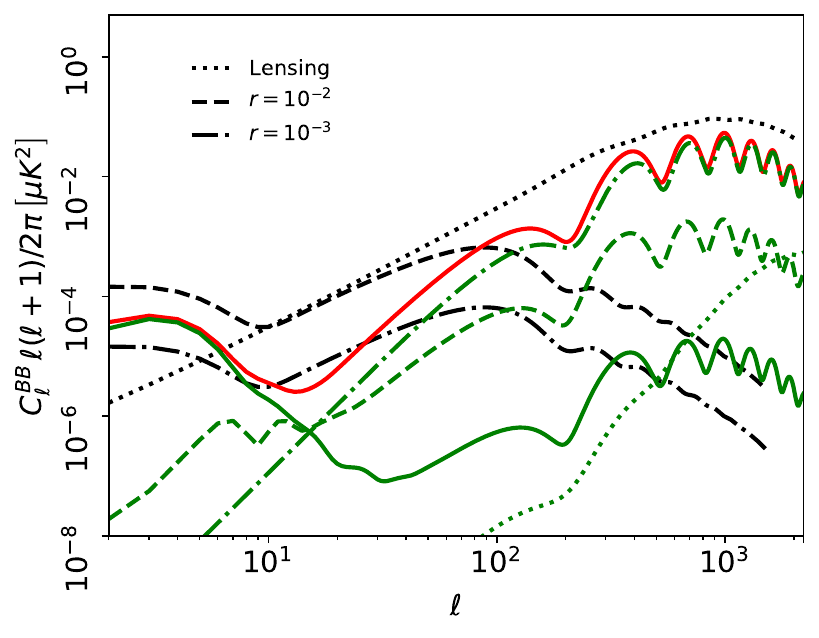}
\includegraphics[width=160pt,height=4cm]{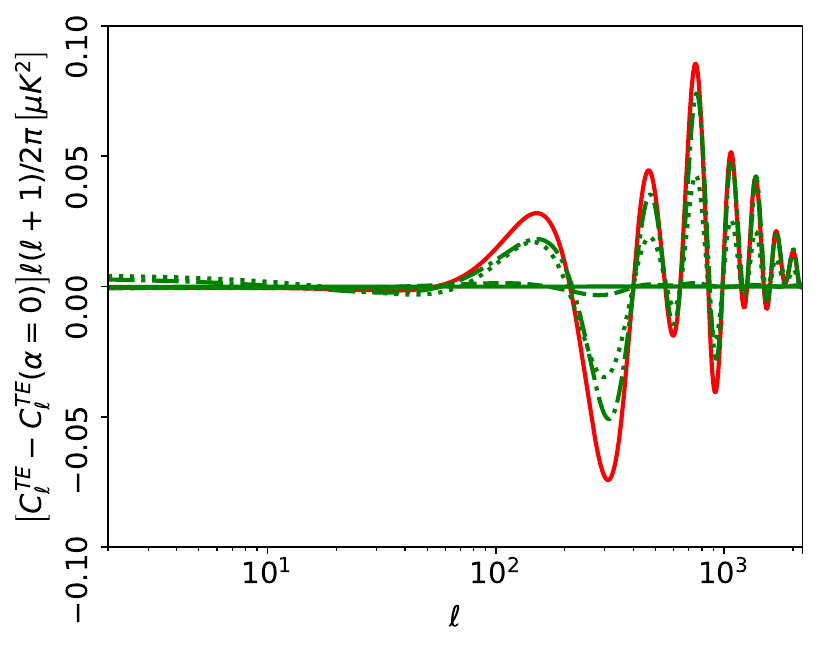}
\includegraphics[width=160pt,height=4cm]{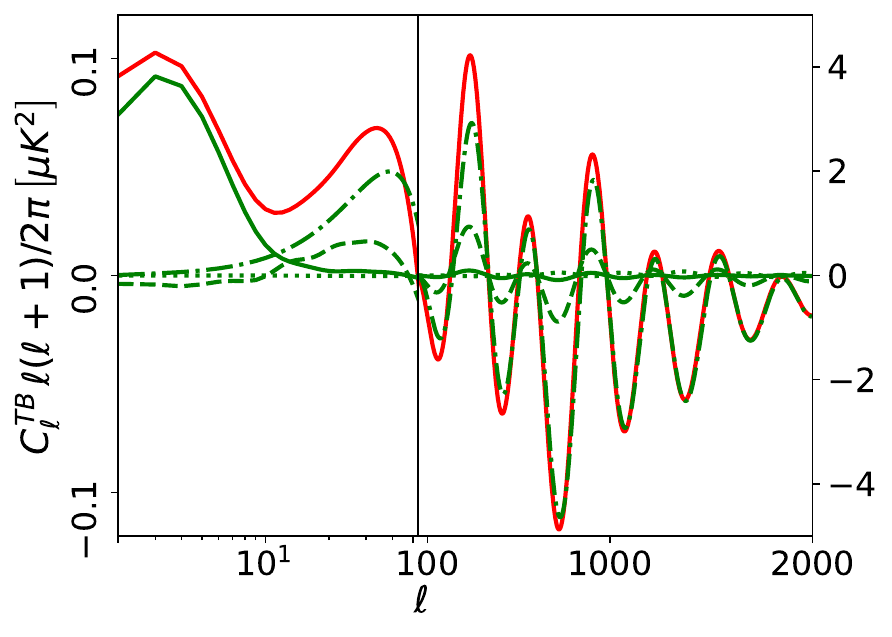}
\includegraphics[width=160pt,height=4cm]{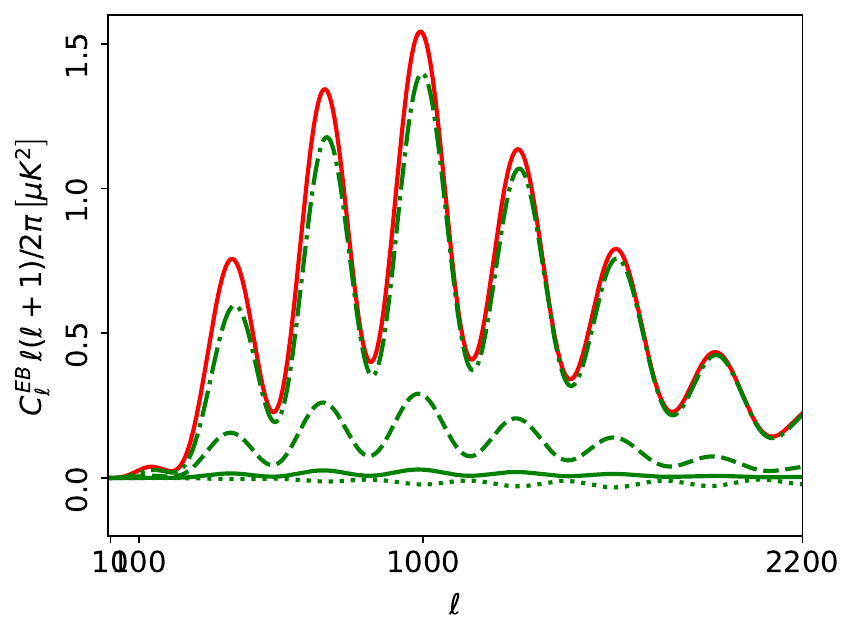}
\caption{\footnotesize\label{Fig:sin20:freq} 
(a) Evolution of the birefringence angle $\alpha$ as a function of conformal time $x\equiv\left(\eta-\eta_{\rm rec}\right)/\left(\eta_0-\eta_{\rm rec}\right)$:
$\alpha(\eta)= \sin(\pi x)$ (green continuous line),
$\alpha(\eta)= \sin(10\pi x)$ (green dotted line),
$\alpha(\eta)= \sin(100\pi x)$ (green dot-dashed line),
$\alpha(\eta)= \sin(1000\pi x)$ (green dotted line),
note that in all tree cases
$\alpha (\eta_\mathrm{rec})=\alpha(\eta_{0})$ ($\bar{\alpha}=0$~deg)
- the \texttt{CAMB} visibility function $g_\mathrm{CAMB}$ is plotted in red (on a different scale);
(b) $C_\ell^{EE} - C_\ell^{EE}(\bar{\alpha}=0)$, (c) $C_\ell^{BB}$,
(d)  $C_\ell^{TE} - C_\ell^{TE}(\bar{\alpha}=0)$, (e) $C_\ell^{TB}$, (f) $C_\ell^{EB}$.}
\end{figure*}

In Fig.~\ref{Fig:sin20} we already showed that power spectra are influenced by cosmic birefringence also when 
$\bar{\alpha}=\alpha (\eta_\mathrm{rec})-\alpha(\eta_{0})=0$. 
In Fig.~\ref{Fig:sin20:freq} we always focus on an oscillating behaviour with $\bar{\alpha}=0$, but we compare different frequencies.
Effects on the power spectra at high-$\ell$ seem to increase at higher oscillating frequencies 
($\alpha(\eta)= \sin(10\pi x)$ and $\alpha(\eta)= \sin(100\pi x)$),
but for a linear polarization angle oscillating extremely quickly (e.g. $\alpha(\eta)= \sin(1000\pi x)$ ) the effects cancel out.
Since the visibility function reaches its maximum at recombination and a second peak at reionization 
- see Section~\ref{Sect:II} for more details -
the overall effects highly depend on the value of $\alpha(\eta)-\alpha(\eta_0)$ at these two epochs.
In the case $\alpha(\eta)= \sin(\pi x)$ the birefringence angle is too small at recombination to modify the source terms of Eqs.~\eqref{source:E}-\eqref{source:B}, but effects are visible at reionization (lower-$\ell$).
On the contrary for $\alpha(\eta)= \sin(10 \pi x)$ and $\alpha(\eta)= \sin(100 \pi x)$
the effects at recombination (high-$\ell$) are quite important,
while the effects at low-$\ell$ (reionization) are wiped out by the rapid oscillations of the birefringence angle;
for even faster oscillations, $\alpha(\eta)= \sin(1000 \pi x)$, the effects are deleted also at recombination.

\begin{figure*}[!htb]
\includegraphics[width=168pt,height=4.5cm]{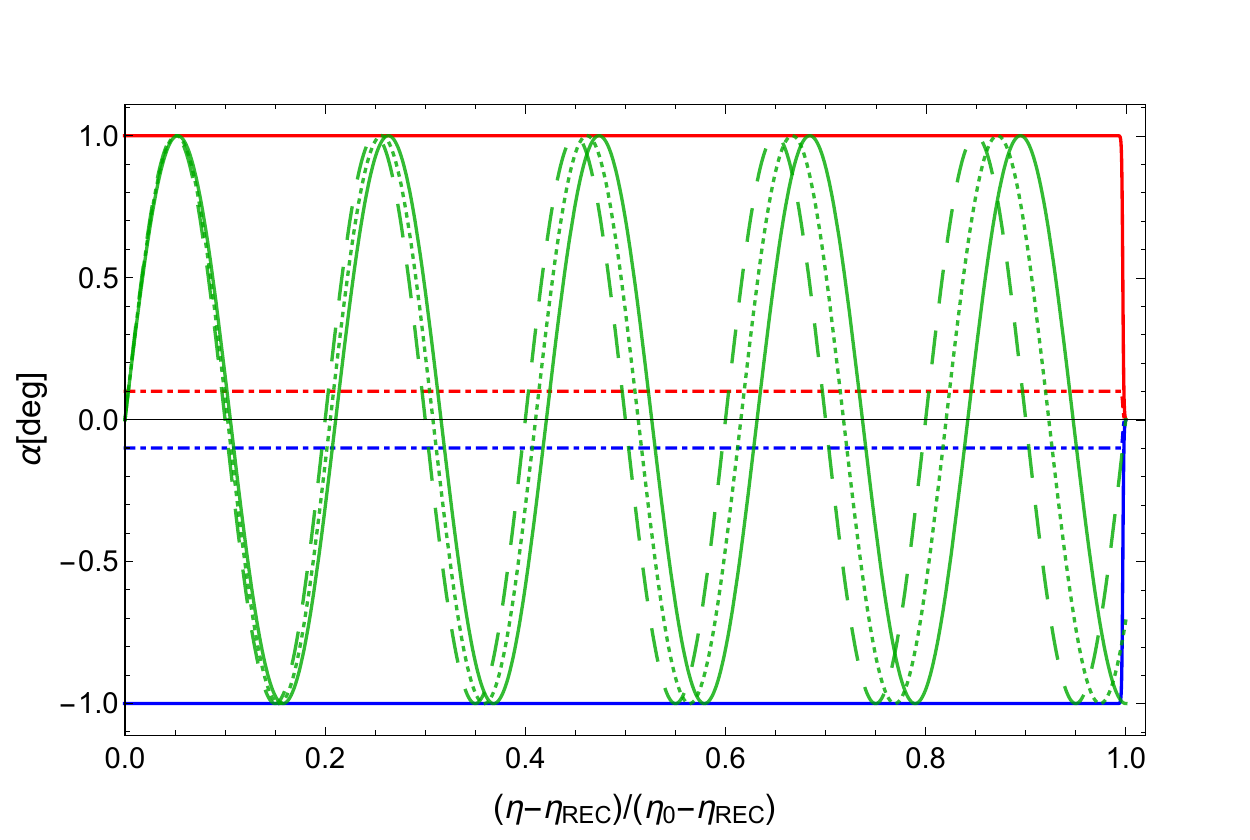}
\includegraphics[width=168pt,height=4cm]{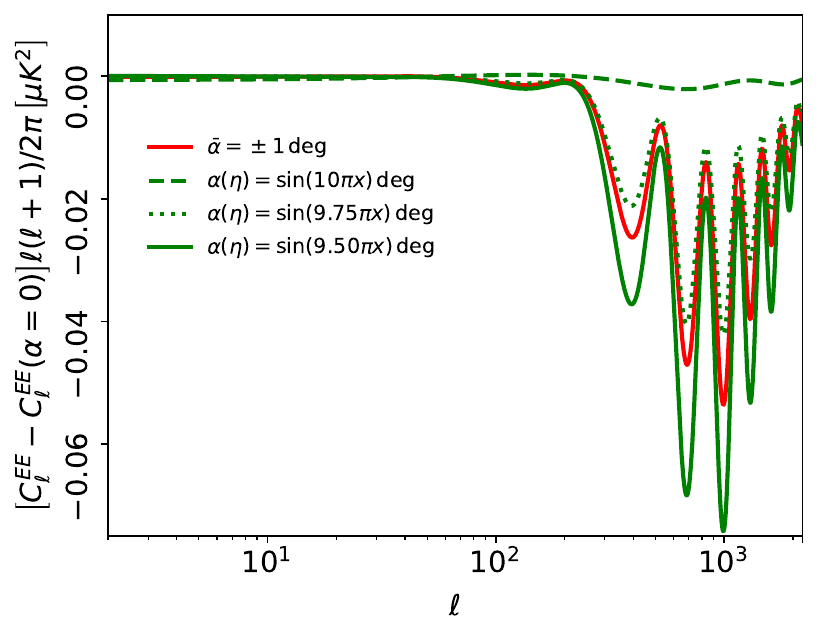}
\includegraphics[width=168pt,height=4cm]{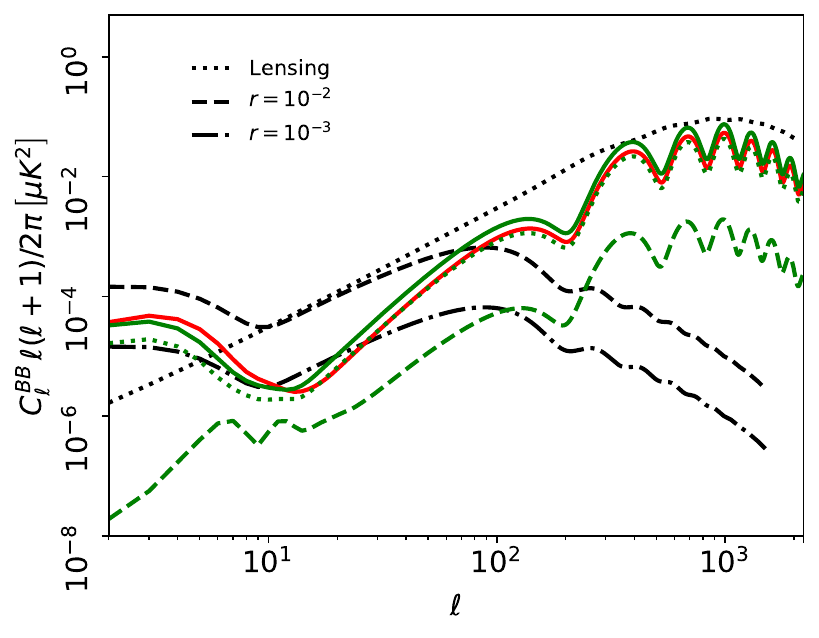}
\includegraphics[width=168pt,height=4cm]{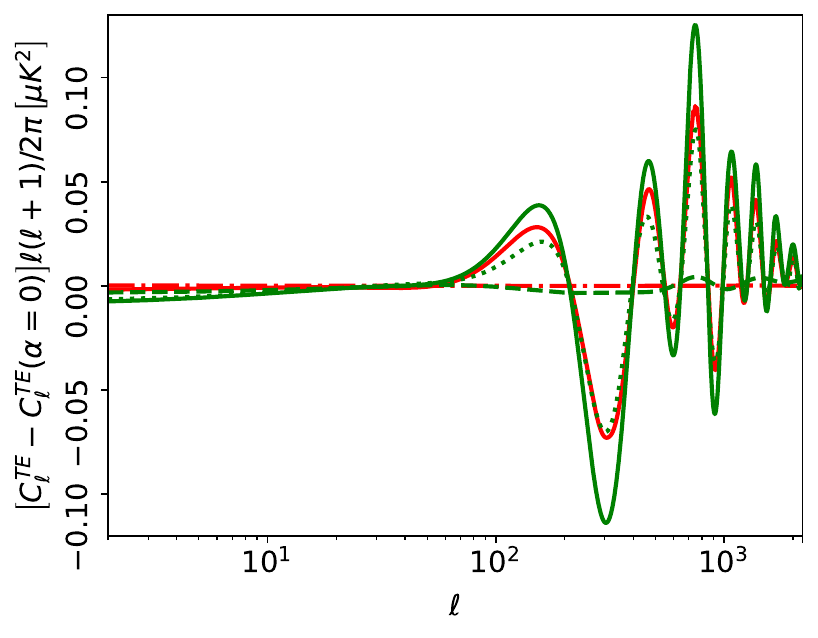}
\includegraphics[width=168pt,height=4cm]{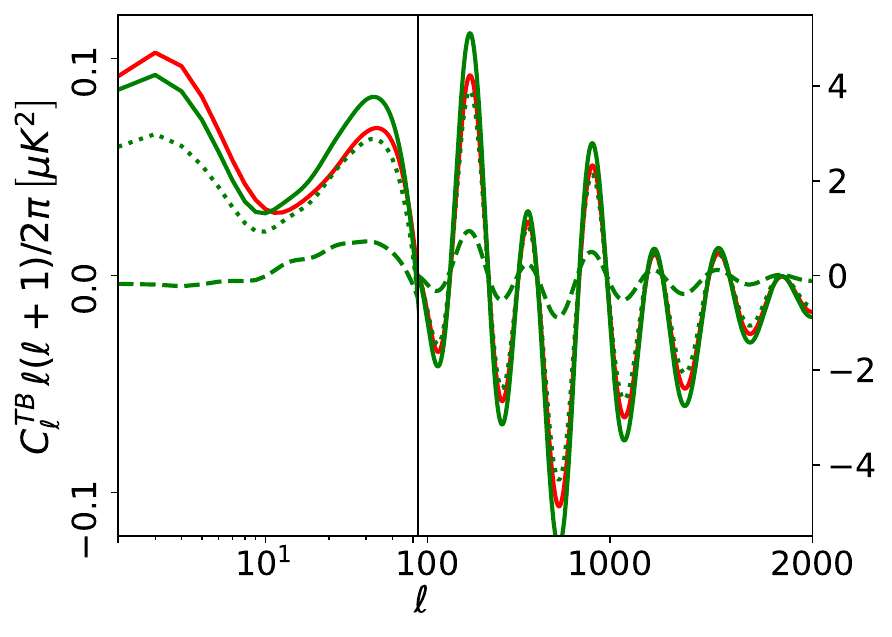}
\includegraphics[width=168pt,height=4cm]{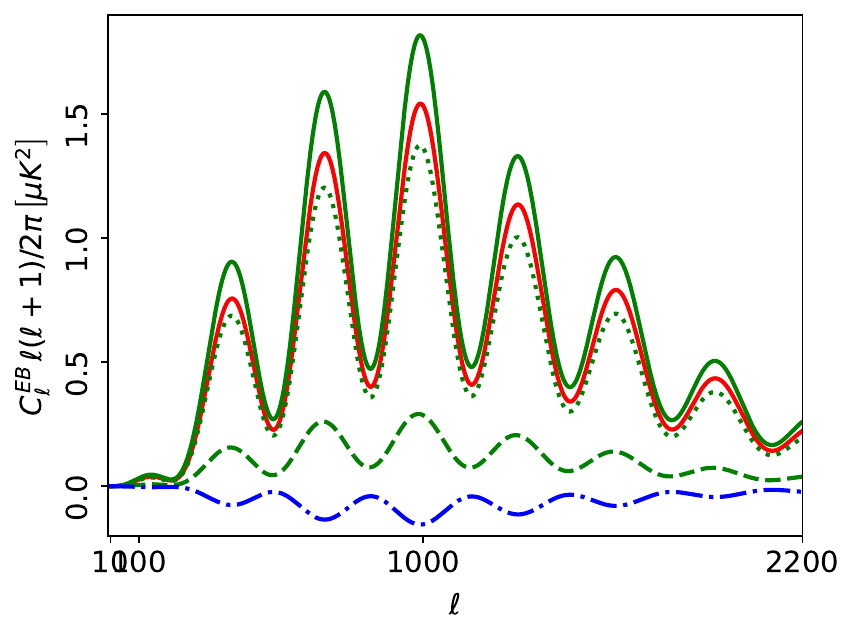}
\caption{\footnotesize\label{Fig:sin20:delta}
(a) Evolution of the birefringence angle $\alpha$ as a function of conformal time $x\equiv\left(\eta-\eta_{\rm rec}\right)/\left(\eta_0-\eta_{\rm rec}\right)$:
$\alpha(\eta)= \sin(9.5 \pi x)$ (green continuous line, $\bar{\alpha}=1$ deg),
$\alpha(\eta)= \sin(9.75\pi x)$ (green dotted line, $\bar{\alpha}=0.71$ deg),
$\alpha(\eta)= \sin(10\pi x)$ (green dashed  line, $\bar{\alpha}=0$ deg) - 
we plot for comparison also a the case of a sudden rotation of $+/-1$~deg ($+/-0.1$~deg) occurring at present time, 
see continuous (dot-dashed) red/blue line;
(b) $C_\ell^{EE} - C_\ell^{EE}(\bar{\alpha}=0)$, (c) $C_\ell^{BB}$,
(d)  $C_\ell^{TE} - C_\ell^{TE}(\bar{\alpha}=0)$, (e) $C_\ell^{TB}$, (f) $C_\ell^{EB}$.}
\end{figure*}

Finally, see Fig.~\ref{Fig:sin20:delta}, we compare three different oscillating behaviours of the linear polarization angle.
In one case $\alpha (\eta_\mathrm{rec})=\alpha(\eta_{0})$, 
while in the other two cases $\alpha (\eta_\mathrm{rec})\neq\alpha(\eta_{0})$.
When there is a difference between the value of $\alpha$ at recombination and today 
this clearly dominates the effects on the power spectra.

\begin{thebibliography}{99}
\bibitem{EllipticSine}
In order to insert the elliptic sine $\mathrm{sn}\left(x,\frac{1}{\sqrt{2}}\right)$ in the \texttt{CAMB} code
we applied twice descending Landen transformation \cite[Sect. 16.12]{Abramowitz}:
$$
\mathrm{sn}\left(x,m\right)=\frac{\left(1+\mu_1^{1/2}\right)\mathrm{sn}\left(\frac{x}{1+\mu_1^{1/2}},\mu_1\right)}{1+\mu_1^{1/2}\mathrm{sn}^2\left(\frac{x}{1+\mu_1^{1/2}},\mu_1\right)}
$$
where:
$$
\mu_1\equiv \left(\frac{1-\sqrt{1-m}}{1+\sqrt{1-m}}\right)^2
$$
Starting from $m=1/\sqrt{2}\simeq0.707$ we obtained first $\mu_1\simeq0.089$, 
and applying the Landen transformation a second time  $\mu_2\simeq0.00053$.
Finally, since we reduced to a case where $m\ll1$, we used the approximation
of the elliptic sine in terms of circular functions \cite[Sect. 16.12]{Abramowitz}:
$$
\mathrm{sn}\left(x,m\right)\simeq \sin x
-\frac{m}{4}\left(x-\sin x \cos x \right)\cos x\,.
$$
\end{thebibliography}
\end{document}